\documentclass[11pt]{article}
\usepackage{amsmath,amsthm,comment}
\usepackage{algorithm}
\usepackage{algorithmic}
\usepackage{amsfonts}
\usepackage{graphicx}
\usepackage{multirow}
\usepackage{booktabs}
\usepackage[top=1in, bottom=1in, left=1in, right=1in]{geometry}
\usepackage{graphicx}
\usepackage{rotating}
\usepackage{enumerate}
\usepackage{dsfont}
\usepackage{bm}
\usepackage{bbm}
\usepackage{natbib}
\usepackage{color}
\usepackage{subcaption}
\usepackage{caption}
\usepackage[colorlinks=true,citecolor=blue,linkcolor=red,urlcolor=blue]{hyperref}
\usepackage{float}
\usepackage{booktabs}
\usepackage{multirow}
\allowdisplaybreaks
\numberwithin{equation}{section}
\newtheorem{theorem}{Theorem}[section]

\newtheorem{lemma}{Lemma}[section]
\newtheorem{proposition}{Proposition}[section]
\theoremstyle{definition}

\theoremstyle{remark}

\newtheorem{assumption}{\textbf{Assumption}}[section]

\newtheorem{example}{Example}[section]

\newcommand{\one}{\mathbbm{1}}

\begin{document}
\title{Tail Risk Equivalent Level Transition and Its Application for Estimating Extreme $L_p$-quantiles}
\author{
            Qingzhao Zhong\footnote{ School of Data Science, Fudan University, Shanghai, China.},\quad
			Yanxi Hou\footnote{Corresponding author: yxhou@fudan.edu.cn. School of Data Science, Fudan University, Shanghai, China.}\\
		}
\date{\today}
\maketitle
\begin{abstract}
$L_p$-quantile has recently been receiving growing attention in risk management since it has desirable properties as a risk measure and is a generalization of two widely applied risk measures, Value-at-Risk and Expectile. The statistical methodology for $L_p$-quantile is not only feasible but also straightforward to implement as it represents a specific form of M-quantile using $p$-power loss function. In this paper, we introduce the concept of Tail Risk Equivalent Level Transition (TRELT) to capture changes in tail risk when we make a risk transition between two $L_p$-quantiles. TRELT is motivated by PELVE in \cite{Li2023} but for tail risk. As it remains unknown in theory how this transition works, we investigate the existence, uniqueness, and asymptotic properties of TRELT (as well as dual TRELT) for $L_p$-quantiles. In addition, we study the inference methods for TRELT and extreme $L_p$-quantiles by using this risk transition, which turns out to be a novel extrapolation method in extreme value theory. The asymptotic properties of the proposed estimators are established, and both simulation studies and real data analysis are conducted to demonstrate their empirical performance. 
\end{abstract}

\section{Introduction}

In risk management, regulators and practitioners commonly utilize risk measures to ascertain the necessary capital reserves to withstand potential losses or risk events. These risk measures are of considerable importance in risk modeling and prediction, and they are extensively studied across various disciplines, including actuarial science, economic statistics, and financial engineering. Particularly for tail risk, accurately measuring extreme losses is crucial to enhance the precision of risk assessment, which in turn directly influences a company's approach to decision-making.

Numerous risk measures have been proposed and examined theoretically across various scenarios. For instance, \cite{Breckling1988} generalized the quantile to the M-quantile within the framework of the M-estimation. The generalized quantile was proposed in \cite{Bellini2014}, which is the minimizer of a suitable convex asymmetric loss function. Under the $p$-power loss, both the generalized quantile and the M-quantile reduce to the $L_p$-quantile, which was initially explored by \cite{Chen1996}. Furthermore, from an economic perspective, two extensions were made based on Rank Dependent Expected Utility (RDEU) in \cite{Bellini2014} and Cumulative Prospect Theory (CPT) in \cite{Mao2018}, respectively. We refer to \cite{Haezendonck1982,Artzner1999,Landsman2001,Goovaerts2004,Heras2012,Cai2020} for further insights into diverse types of risk measures.

Among them, VaR (Value-at-Risk, see \cite{Christoffersen2009}) and ES (Expected Shortfall, see \cite{Tasche2002,Acerbi2002a,Acerbi2002b}) are two risk measures widely applied in the industry. Their mathematical definitions are given by,
\begin{equation*}
  {\rm VaR}_{\tau}(X) = \inf_{x \in \mathbb{R}} \left\{ ~x ~\big|~ \mathbb{P}(X \leq x) \ge \tau \right\} \quad\text{and}\quad
  {\rm ES}_{\tau}(X) = \frac{1}{1-\tau} \int_{\tau}^{1} {\rm VaR}_{q}(X) \,dq,
\end{equation*}
where $X$ is a random loss, and  $\tau \in (0,1)$ is a risk level. The Basel Committee on Banking Supervision, as documented in \cite{BCBS2019}, has confirmed that banks should replace the $\rm VaR_{0.99}$ by $\rm ES_{0.975}$ for quantifying market risks. This shift is due to ES's superior ability to capture tail risk. Under this transition, the capital requirement remains unchanged if $\rm ES_{0.975} \approx VaR_{0.99}$ while it will increase to better withstand potential risks if $\rm ES_{0.975} > VaR_{0.99}$. To characterize the equivalent transition between VaR and ES, \cite{Li2023} introduced a new concept of Probability Equivalent Level of VaR and ES (PELVE), which identifies the balancing point for this transition. 
For a (tail) risk level $\varepsilon \in (0,1)$, PELVE is essentially a constant multiplier $c \in [1,1/\varepsilon]$ such that,
\begin{equation}\label{eq:pelve}
  {\rm ES}_{1-c\varepsilon}(X) = {\rm VaR}_{1-\varepsilon}(X).
\end{equation}
Its formal definition $\Pi(\varepsilon)$ is given by
\begin{equation}\label{eq:PELVE}
  \Pi(\varepsilon) = \inf_{c \in [1,1/\varepsilon]} \left\{ ~c~ \big|~ {\rm ES}_{1-c\varepsilon}(X) \leq {\rm VaR}_{1-\varepsilon}(X) \right\}.
\end{equation}
PELVE enjoys many desirable theoretical properties and can roughly distinguish heavy-tailed distributions from light-tailed ones through a threshold $e \approx 2.72$. \cite{Li2023} also reported that the existence and uniqueness of $\Pi(\varepsilon) = c$ in \eqref{eq:pelve} can be guaranteed under only some mild conditions. 
Further work on model calibration from PELVE has been considered in \cite{Assa2024} as well.

The tail risk transition between a couple of risk measures presents a novel theoretical challenge for tail risk measurement. Although some studies have already adopted a certain transition mechanism between risk measures, the theoretical underpinnings of the transition have not been explored well. In addition to PELVE, \cite{Xu2022} has considered another transition between VaR and expectile within a regression framework. It is noted that VaR and expectile indeed correspond to the $L_p$-quantiles with $p = 1$ and 2 respectively. However, the conditions of the tail risk transition are not justified; for example, the existence and uniqueness of this transition are unknown, and thus the statistical methods proposed based on extreme value theory suffer some biased problems. Therefore, motivated by the idea of PELVE, this paper aims to study the tail risk transition mechanism between the $L_p$-qauntiles with a more general $p$ and establishes the concept of Tail Risk Equivalent Level Transition (TRELT). TRELT builds a bridge between two levels given tail equivalent risk measures, paving the way for intriguing inference problems for tail risk transition. Furthermore, it appears reasonable to treat it as an implication of stability. Unlike PELVE, TRELT is only established on the tail region rather than the global uncertainty, and several explicit estimators for TRELT are also proposed later.

We now give a review on $L_p$-quantile first. Suppose $X$ is a random variable with a distribution $F$. Given any order $p\ge 1$ and a risk level $\tau\in(0,1)$, the $L_p$-quantile $\theta_{p}(\tau)$ of $F$ is defined as
\begin{equation}\label{eq:lp-quantile1}
\theta_{p}(\tau):= \mathop{\arg\min}\limits_{u \in \mathbb{R}} \mathbb{E}[ \rho_{p,\tau}(X-u)],
\end{equation}
where the check function $\rho_{p,\tau}(s)$ is a $p$-power loss function such that
\begin{equation*}
\rho_{p,\tau}(s) = \left|  \tau -  1_{\{ s \leq 0 \}} \right|\cdot \left| s \right|^p = \tau s_{+}^p + (1-\tau)s_{-}^p,
\end{equation*}
with $s_{+} = \max\{ s,0 \}$ and $s_{-} = \max\{ -s,0 \}$. Note that the minimizer $\theta_{p}(\tau)$ exists since $\mathbb{E}[ \rho_{p,\tau}(X-u) ]$ is convex and tends to infinity as $u \to \pm \infty$, and $\theta_{p}(\tau)$ is unique when $p>1$ since $\mathbb{E}[ \rho_{p,\tau}(X-u) ]$ is strictly convex in this case. There are some interesting properties for $L_p$-quantile, such as monotonicity and continuity in $\tau$. Both \cite{Bellini2014} and \cite{Ziegel2016} imply that the $L_p$-quantile is a consistent risk measure when $p=2$. Substantial literature, such as \cite{Bellini2017,Daouia2018,Daouia2020,Girard2021a,Girard2022,Daouia2024,Danilevicz2024}, have already studied the expectile as a risk measure. These works focused on theoretical properties and further predicted tail risk via expectile in view of extreme value theory.

By \cite{Chen1996} and \eqref{eq:lp-quantile1}, the $L_p$-quantile for $p \ge 1$ satisfies the following scale equation,
\begin{equation}\label{eq:lp-quantile2}
\tau \mathbb{E}[(X - \theta_{p}(\tau))_{+}^{p-1}] = (1-\tau)\mathbb{E}[(X - \theta_{p}(\tau))_{-}^{p-1}].
\end{equation}
Especially for $p=1$, \eqref{eq:lp-quantile2} reduces to $\tau \mathbb{E}[\one_{\{X \ge \theta_{1}(\tau)\}}] = (1-\tau)\mathbb{E}[\one_{\{X \leq \theta_{1}(\tau)\}}]$ and it yields quantile or VaR. As explained in \cite{Daouia2019}, the existence of $L_p$-quanitle in \eqref{eq:lp-quantile2} will be guaranteed as long as $\mathbb{E}[|X|^{p-1}]<\infty$. Moreover, \eqref{eq:lp-quantile2} leads to a transformation for the risk level $\tau$ (see \cite{Jones1994}) such that
\begin{equation}\label{eqn:level_tran}
\tau = \frac{\mathbb{E}[(X - \theta_{p}(\tau))_{-}^{p-1}]}{\mathbb{E}[|X - \theta_{p}(\tau)|^{p-1}]},
\end{equation}
from which $\theta_{p}(\tau)$ can be seen as a quantile of a transformed distribution
\begin{equation*}
G(x) = \frac{\mathbb{E}[ (X - x)_{-}^{p-1}]}{\mathbb{E}[|X - x|^{p-1}]}.
\end{equation*}
Given a set of samples $X_1, X_2,...,X_n$, at a fixed or an intermediate level, \cite{Daouia2019} proposed an empirical estimator for $\theta_p(\tau)$ via solving the empirical form of \eqref{eq:lp-quantile1},
\begin{equation}\label{eq:inter_2019}
\hat{\theta}_p(\tau) = \mathop{\arg\min}\limits_{u \in \mathbb{R}} \frac{1}{n} \sum_{i=1}^{n} \rho_{p,\tau}(X_i-u) = \mathop{\arg\min}\limits_{u \in \mathbb{R}} \frac{1}{n} \sum_{i=1}^{n} |\tau- \one_{\{X_i \leq u\}}|\cdot|X_i -u|^p.
\end{equation}
When we consider an extreme level, due to the lack of sufficient samples on the tail, \eqref{eq:inter_2019} may lead to an ineffective estimator and thus extrapolative technique must be employed. \cite{Daouia2019} further put up a standard extrapolative estimator for $\theta_p(1 - \varepsilon'_n)$, which was given by
\begin{equation}\label{eq:exm1_2019}
  \tilde\theta^{\rm sta}_p(1-\varepsilon'_n) = \left(\frac{\varepsilon'_n}{\varepsilon_n}\right)^{-\hat\gamma} \hat\theta_p(1-\varepsilon_n),
\end{equation}
where $\varepsilon_n$ and $\varepsilon_n'$ were the intermediate and extreme (tail) levels, and $\hat\gamma$ was a suitable estimator for extreme value index $\gamma$ (see Assumption \ref{ass:forv} below). For $p=1$, it could be taken $X_{n-[n\varepsilon_n],n}$ as $\hat{\theta}_1(1-\varepsilon_n)$. Hence, \cite{Daouia2019} provided the other extrapolative estimator,
\begin{equation}\label{eq:exm2_2019}
  \tilde{\theta}^{\rm qua}_p(1-\varepsilon'_n) = \left[ \frac{\hat{\gamma}}{B(p, \hat{\gamma}^{-1}-p+1)} \right]^{-\hat{\gamma}} \left(\frac{\varepsilon'_n}{\varepsilon_n}\right)^{-\hat\gamma} X_{n-[n\varepsilon_n],n}.
\end{equation}
We use superscripts ``sta'' and ``qua'' to mark these two estimators. Notice that our proposed estimator \eqref{eq:thetap_extest1} will reduce to \eqref{eq:exm2_2019} when $q=1$. We thus will conduct \eqref{eq:exm1_2019} as a benchmark method in the following empirical studies. We refer to \cite{Usseglio-Carleve2018,Girard2021b,Bignozzi2022} for other relevant works.

Then, TRELT addresses the tail risk transition between $L_p$- and $L_q$-quantiles by considering
\begin{equation}\label{eq:trelt}
  \theta_p(1-c\varepsilon)=\theta_q(1-\varepsilon), \quad\varepsilon\in(0,1-\tau_0),
\end{equation}
where $c$ is a certain constant determined by $p,q,\varepsilon$, and $\tau_0$ is a threshold for tail region. One contribution of this paper is to establish the mathematical conditions and properties of \eqref{eq:trelt}. It is important to note that \eqref{eq:trelt} may only be meaningful on the one-side tail region, since only when $\tau \in (\tau_0,1)$ does $\theta_p(\tau) > \theta_q(\tau)$ hold, as detailed in Proposition \ref{pro:rank_lpquant} below, in which case, definition \eqref{eq:trelt} makes sense. Although TRELT appears to have some limitations in its definition, it is sufficient for studying extreme risks. Thus, our study is established on the extreme value theory for heavy-tailed distributions.

Actually, TRELT gives rise to several additional approaches to estimate extreme $L_p$-quantile, which turn out to be more innovative and capable of describing the uncertainty than classic ones. Another contribution of this paper is to propose new extrapolation methods of the extreme estimators via TRELT and $\theta_q(1-\varepsilon_n)$ when risk measure transitions from $L_q$-quantile to $L_p$-quantile. To be specific, we study a TRELT-based extrapolation of $\theta_p(1-\varepsilon'_n)$ from $\theta_q(1-\varepsilon_n)$, adhering two principles of tail risk transition: first, we estimate $\theta_q(1-\varepsilon_n)$ at an intermediate level $\varepsilon_n$ via \eqref{eq:inter_2019}; second, we apply the method of TRELT to extrapolate $\theta_p(1-\varepsilon'_n)$ at an extreme level. Compared to \eqref{eq:exm1_2019} and \eqref{eq:exm2_2019}, there are several improvements. Firstly, our methods are all embedded in the estimators of TRELT, potentially providing a better characterization of risk uncertainty when risk measures vary. Secondly, within our framework, the extreme $L_p$-quantile $\theta_p(1-\varepsilon'_n)$ can be estimated effectively by any $\theta_q(1-\varepsilon_n)$ via TRELT under $1-q < p - \frac{1}{2\gamma} < 1$, not only $\theta_p(1-\varepsilon_n)$ or $\theta_1(1-\varepsilon_n)$. This suggests that, we can get some information of $\theta_p(1-\varepsilon'_n)$ from $\theta_q(1-\varepsilon_n)$ by the method of tail risk transition, while both \eqref{eq:exm1_2019} and \eqref{eq:exm2_2019} fail. Thirdly, simulation evidence has shown that our methods indeed report superior empirical performance.

The rest of the paper is organized as follows. The construction of TRELT for $\theta_p(\tau)$ and $\theta_q(\tau)$ is organized in Section \ref{sec2}, where we also discuss the existence and uniqueness, and the limiting behavior of TRELT. Section \ref{sec3} and \ref{sec4} provide the estimators for the coefficient of TRELT $\Pi_{p,q}$ and the extreme $L_p$-quantiles $\theta_p(\tau)$ via TRELT, respectively. Finally, we conduct a series of simulation studies and real data analysis to illustrate the performance of our proposed estimators in Section \ref{sec5} and Section \ref{sec6}.

\section{Tail Risk Equivalent Level Transition Between $L_p$-quantiles}\label{sec2}

Let the two orders $p,q$ satisfy $p > q \ge 1$. We aim to construct a risk equivalent level transition between $L_p$- and $L_q$-quanitles via a coefficient $c$ such that,
\begin{equation}\label{eq:trelt0}
  \theta_p(1-c\varepsilon) = \theta_q(1-\varepsilon).
\end{equation}
However, it might fail to achieve such a transition for all $\varepsilon \in (0,1)$. The main reason is that it is not clear about the size of $\theta_p(\tau)$ and $\theta_q(\tau)$; consequently, it is infeasible to find the range of the coefficient $c$ because we are uncertain whether \eqref{eq:trelt0} has a solution $c$, or even if it does, its uniqueness is unknown. Thus, it needs more information about the relative sizes between $\theta_p(\tau)$ and $\theta_q(\tau)$, and the study of the existence and uniqueness of $c$ is absolutely necessary. Fortunately, Proposition \ref{pro:rank_lpquant} provides a helpful interpretation through a limit relation over them. We can claim that $\theta_p(\tau) > \theta_q(\tau)$ (or $\theta_p(\tau) < \theta_q(\tau)$) always holds on the tail region $(\tau_0,1)$ (or $(\tau'_0,1)$) for a certain threshold $\tau_0$. This suggests that the risk equivalent level transition could be established on the tail, which is both sufficient and feasible for predicting extreme risks in risk management.

Recall that $X$ is a random variable with a distribution function $F$, then we denote $\overline{F} = 1-F$ and $U(\cdot)$ as the survival function of $F$ and the left-continuous inverse of $1/\overline{F}$, (\emph{i.e.} tail quantile function) respectively. In this paper, we study the right tail of $F$ for $\theta_{p}(\tau)$ with a risk level $\tau$ close to 1. The following assumption is a first-order regular variation condition for the right tail of $F$.

\begin{assumption}[First-order regular variation]\label{ass:forv}
The survival function $\overline{F}$ satisfies a (first-order) regular variation condition with an extreme value index $\gamma >0$, \textit{i.e.}, for all $x>0$,
\begin{equation}\label{eq:forv_F}
\lim_{t \to \infty} \frac{\overline{F}(tx)}{\overline{F}(t)} = x^{-\frac{1}{\gamma}}.
\end{equation}
Equivalently, this can be reformulated in terms of $U$ by
\begin{equation}\label{eq:forv_U}
\lim_{t \to \infty} \frac{U(tx)}{U(t)} = x^{\gamma}.
\end{equation}
\end{assumption}

The next assumption is a necessary moment condition for the left tail of $F$.

\begin{assumption}\label{ass:mom_negapart}
$\mathbb{E}[X_-^{p-1}]<\infty$.
\end{assumption}

\begin{proposition}\label{pro:rank_lpquant}
Suppose $F$ satisfies both Assumption \ref{ass:forv} for some $\gamma > 0$ and Assumption \ref{ass:mom_negapart} with an order $p \in (1, 1+1/\gamma)$. Then, for all $q \in [1,p)$, we have that,
  \begin{equation}\label{eq:rank_lpquant}
    \lim_{\varepsilon \downarrow 0} \frac{\theta_p(1-\varepsilon)}{\theta_q(1-\varepsilon)} = \left[ \frac{B(p, \gamma^{-1}-p+1)}{B(q, \gamma^{-1}-q+1)} \right]^{\gamma} := \mathcal{L}(\gamma, p ,q),
  \end{equation}
  where $B(a,b) = \int_{0}^{1} t^{a-1}(1-t)^{b-1}\,dt$ is the Beta function. Furthermore, let $\delta = p - 1/\gamma$, and we have the following three statements:
  \begin{enumerate}
    \item[(a)] \label{rank1} If $\delta = 1-q$, then $\mathcal{L}(\gamma, p ,q)=1$;
    \item[(b)] \label{rank2} If $1-q<\delta < 1$, then $\mathcal{L}(\gamma, p ,q)>1$, there exists $\tau_0 \in [0,1)$, such that $\theta_{p}(1-\varepsilon) > \theta_{q}(1-\varepsilon)$ for $\varepsilon \in (0, 1-\tau_0)$;
    \item[(c)] \label{rank3} If $\delta < 1-q$, then $\mathcal{L}(\gamma, p ,q)<1$, there exists $\tau'_0 \in [0,1)$, such that $\theta_{p}(1-\varepsilon) < \theta_{q}(1-\varepsilon)$ for $\varepsilon \in (0,1-\tau'_0)$.
  \end{enumerate}
\end{proposition}

Note that \eqref{eq:rank_lpquant} is a straightforward consequence of Corollary 1 in \cite{Daouia2019}. Note also that the condition $p-1 < 1/\gamma$ is not only indispensable to make $B(p,\gamma^{-1} -p+1)$ work, but also implies $\mathbb{E}[X_{+}^{p-1}] < \infty$ (see Lemma \ref{lem:finite_mom}). When combined the condition $\mathbb{E}[X_-^{p-1}] < \infty$, it implies that $\mathbb{E}[|X|^{p-1}] < \infty$, and thus, the $L_p$-quantile is indeed well-defined. It is also worth noting that we only need the existence of threshold $\tau_0$. Taking the second case $(1-q < \delta < 1)$ as an example, it is readily to see that $\tau_0$ is essentially threshold such that $\theta_p(\tau) > \theta_q(\tau)$ for all $\tau > \tau_0$. From monotonicity and continuity of $\theta_p(\tau)$ (see Proposition \ref{pro:mp_lpquant}), one alternative definition can be given by
\begin{equation}\label{eq:tau0}
\tau_0 =
  \begin{cases}
    \sup_{\tau \in (0,1)} \left\{ ~\tau ~\big| ~\theta_p(\tau) \leq \theta_q(\tau)\right\}, & \mbox{if } \left\{ ~\tau ~\big| ~\theta_p(\tau) \leq \theta_q(\tau)\right\} \neq \emptyset, \\
    0, & \mbox{if } \left\{ ~\tau ~\big| ~\theta_p(\tau) \leq \theta_q(\tau)\right\} = \emptyset.
  \end{cases}
\end{equation}
Obviously, the value of $\tau_0$ depends on $p,q$ and $F$. The value of $\tau'_0$ can be defined similarly and we lay it in \eqref{eq:another_tau0}.

\begin{example}[Expectile and VaR, 1]
  The $L_p$-quantile will reduce to expectile and VaR when $p=2$ and $q=1$ correspondingly. According to Proposition \ref{pro:rank_lpquant}, we have that,
  \begin{equation}\label{eq:exp-var}
    \lim_{\varepsilon \downarrow 0} \frac{\theta_{2}(1-\varepsilon)}{\theta_{1}(1-\varepsilon)} = \left[ \frac{\gamma}{1-\gamma} \right]^{\gamma} =: \mathcal{L}(\gamma, 2 ,1).
  \end{equation}
  Then, the following statements hold with different $\gamma$,
  \begin{itemize}
    \item If $\frac{1}{2}< \gamma <1$, there exists $\tau_0 \in (0,1)$, then $\theta_{2}(1-\varepsilon) > \theta_{1}(1-\varepsilon)$ for $\varepsilon \in (0, 1-\tau_0)$;
    \item If $0<\gamma <\frac{1}{2}$, there exists $\tau'_0 \in (0,1)$, then $\theta_{2}(1-\varepsilon) < \theta_{1}(1-\varepsilon)$ for $\varepsilon \in (0, 1-\tau'_0)$.
  \end{itemize}
  Similar arguments can be found in Theorem 11 of \cite{Bellini2014} and Proposition 2.2 of \cite{Bellini2017}.
\end{example}

Below, we focus on the case of $1-q < p - 1/\gamma < 1$ first and redefine the positive constant $c:=c(\varepsilon)$ the coefficient of TRELT (CTRELT) between $\theta_p(\tau)$ and $\theta_q(\tau)$ with the tail probability $\varepsilon:=1-\tau$ and reformulate \eqref{eq:trelt0} as follows,
\begin{equation}\label{eq:ctrelt}
\theta_p(1-c\varepsilon) = \theta_q(1-\varepsilon), \quad \varepsilon \in (0,1-\tau_0),
\end{equation}
where $\tau_0$ is given in Proposition \ref{pro:rank_lpquant}. Given the orders $p,q$ and threshold $\tau_0$, the coefficient $c$ varies with the value of $\varepsilon$ and its range can be determined as $\left[1,\frac{1-\tau_0}{\varepsilon} \right]$ by $\tau_0 \leq 1-c\varepsilon < 1$ and $1 -c\varepsilon \leq 1-\varepsilon$. We also provide a formal definition for CTRELT as follows,
\begin{equation}\label{eq:ctrelt_form}
\Pi_{p,q,\tau_0}(\varepsilon) = \inf_{c \in \left[1,\frac{1-\tau_0}{\varepsilon}\right]} \left\{ ~c~\big|~\theta_p(1-c\varepsilon) \leq \theta_q(1-\varepsilon) \right\}. 
\end{equation}

We define the CTRELT $\Pi_{p,q,\tau_0}(\varepsilon)$ in \eqref{eq:ctrelt_form} by an infimum rather than a definite point to prevent some infrequent situations from occurring, such as several values or even no value of $c \in \left[1,\frac{1-\tau_0}{\varepsilon}\right]$ satisfying \eqref{eq:ctrelt}. It is worth noting that the threshold $\tau_0$ can be determined as soon as $p,q$ and $F$ are known. Consequently, the relevance of CTRELT to $\tau_0$ is essentially due to $p,q$ and $F$; moreover, the focus of this article is not to investigate the relationship between $\Pi_{p,q,\tau_0}(\varepsilon)$ and $\tau_0$ at all. Therefore, we omit the dependence of $\Pi_{p,q,\tau_0}(\varepsilon)$ on $\tau_0$ and write $\Pi_{p,q,\tau_0}(\varepsilon)$ as $\Pi_{p,q}(\varepsilon)$ in the rest of the paper.

\begin{example}[\cite{Koenker1993}]
  This example describes what happens when $c$ is a boundary point. On the one hand, if $c=1$ for all $\varepsilon$, then the equality $\theta_p(1-\varepsilon) = \theta_q(1-\varepsilon)$ holds. It has been proven for expectile and quantile that the quantile function should meet with,
  \begin{equation*}
    \theta_1(1-\varepsilon) = \frac{1-2\varepsilon}{\sqrt{\varepsilon (1-\varepsilon)}},
  \end{equation*}
  or any affine transformations of it. Consequently, the distribution function can be written as
  \begin{equation*}
  F(x)=
  \begin{cases}
    \frac{1}{2} \left( 1+ \sqrt{1- \frac{4}{(4+x^2)}} \right), & x \ge 0, \\
    \frac{1}{2} \left( 1- \sqrt{1- \frac{4}{(4+x^2)}} \right), & x<0,
  \end{cases}
  \end{equation*}
  with extreme value index $\gamma = \frac{1}{2}$.  On the other hand, if $c = \frac{1-\tau_0}{\varepsilon}$, then $\theta_p \left( 1- \varepsilon \, \frac{1-\tau_0}{\varepsilon} \right) = \theta_p(\tau_0) = \theta_q(1-\varepsilon)$, which may not always hold. As seen below, only by satisfying $\theta_p(\tau_0) \leq \theta_q(1-\varepsilon)$ can we ensure the CTRELT is both existent and unique.
\end{example}

Next, we present an analysis of the existence and uniqueness for CTRELT.

\begin{assumption}\label{ass:euas} For all $p > q \ge 1$, there exists a threshold $\tau_0$ such that
\begin{enumerate}
  \item[(a)] $\theta_p(\tau_0) \leq \theta_q(1-\varepsilon)$ for all $\varepsilon \in (0,1-\tau_0)$;
  \item[(b)] both $\theta_p(\tau)$ and $\theta_q(\tau)$ are not constants on $[\tau_0,1]$.
\end{enumerate}
\end{assumption}

\begin{proposition}[Existence and Uniqueness of CTRELT]\label{pro:exist-unique}
  Suppose $F$ satisfies both Assumption \ref{ass:forv} for some $\gamma > 0$ and Assumption \ref{ass:mom_negapart} with $p,q$ satisfying $1 \leq q < p$, $1-q < p - \frac{1}{\gamma} < 1$.
  Then, for all $\varepsilon \in (0,1-\tau_0)$, there exists $c \in \left[1,\frac{1-\tau_0}{\varepsilon}\right]$ such that \eqref{eq:ctrelt} holds for $c = \Pi_{p,q}(\varepsilon)$ if and only if Assumption \ref{ass:euas} (a) holds. Moreover, if Assumption \ref{ass:euas} (b) also holds, then the $c \in \left[1,\frac{1-\tau_0}{\varepsilon}\right]$ in \eqref{eq:ctrelt} is unique.
\end{proposition}

Assumption \ref{ass:euas} is a mild condition that many common distributions meet, including continuous heavy-tailed distributions with non-constant $L_p$-quantile. Proposition \ref{pro:exist-unique} indicates that there always exists a finite solution $\Pi_{p,q}(\varepsilon)$ to \eqref{eq:ctrelt} such that $\Pi_{p,q}(\varepsilon) = c \in \left[1,\frac{1-\tau_0}{\varepsilon}\right]$ as long as $\theta_p(\tau_0) \leq \theta_q(1-\varepsilon)$. Assumption \ref{ass:euas} (b) accounts for the strict monotonicity, which is essential for uniqueness. This condition can also be rephrased to state that the quantile function is not constant on $[\tau_0,1]$, since a not-constant quantile implies a non-constant $L_p$-quantile as well. The existence and uniqueness of  $c=\Pi_{p,q}(\varepsilon)$ provide the foundation for the inference problems in Sections \ref{sec3} and \ref{sec4}, especially for the extrapolation method in \eqref{eq:uni_form}.

One critical feature of CTRELT is \emph{location-scale invariance}, that is, $\Pi_{p,q}(\varepsilon;\lambda X + \mu) = \Pi_{p,q}(\varepsilon;X)$ for all $\lambda > 0$ and $\mu \in \mathbb{R}$. This property can be easily verified by \emph{location-scale invariance} of $L_p$-quantile. A reasonable interpretation for this is that the value of CTRELT remains unchanged when a portfolio is scaled by a constant, or shifted by a constant loss or gain. CTRELT characterizes the shape of the distribution under risk transition without considering its location and scale. Hence, CTRELT may show some merits in measuring the variability of risk for an asset assessment, especially compared with some non-scale-free measures, such as variance.

Alternatively, the dual CTRELT is going to be defined when we move the multiplier in CTRELT from the $\theta_p$ side to the $\theta_q$ side, more precisely,
\begin{equation}\label{eq:dual_ctrelt}
\theta_p(1-\varepsilon) = \theta_q\left(1-\frac{\varepsilon}{d}\right), \quad \varepsilon \in (0,1-\tau_0).
\end{equation}
Its formal definition can also be formulated similarly,
\begin{equation}\label{eq:dual_ctrelt_form}
\pi_{p,q}(\varepsilon) = \inf_{d \in [1,\infty)} \left\{ ~d ~\bigg|~\theta_p(1-\varepsilon) \leq \theta_q\left(1-\frac{\varepsilon}{d}\right) \right\}.
\end{equation}
Compared to CTRELT, an advantage of using dual CTRELT is that we do not require the condition $\theta_p(\tau_0) \leq \theta_q(1-\varepsilon)$ anymore and $\pi_{p,q}(\varepsilon)$ is always finite for all $\varepsilon \in (0,1-\tau_0)$ assuredly. 

\begin{proposition}[Existence and Uniqueness of dual CTRELT]\label{pro:exist_unique_dualctrelt}
Suppose $F$ satisfies both Assumption \ref{ass:forv} for some $\gamma > 0$ and Assumption \ref{ass:mom_negapart} with $p,q$ satisfying $1 \leq q < p$, $1-q < p - \frac{1}{\gamma} < 1$. Then, for all $\varepsilon \in (0, 1-\tau_0)$, there exists $d \in [1,\infty)$ such that \eqref{eq:dual_ctrelt} holds for $d = \pi_{p,q}(\varepsilon)$. Moreover, if Assumption \ref{ass:euas} (b) holds, then the $d \in [1,\infty)$ in \eqref{eq:dual_ctrelt} is unique.
\end{proposition}

The asymptotic relationship for levels between expectile and quantile has been discussed in Proposition 1 of \cite{Xu2022}, and a similar argument for PELVE is also presented in Theorem 3 in \cite{Li2023}. Similarly, the limit behavior for both CTRELT $\Pi_{p,q}(\varepsilon)$ and dual CTRELT $\pi_{p,q}(\varepsilon)$ can be described definitely as $\varepsilon$ tends to zero.

\begin{proposition}\label{pro:lim_behavior}
  Suppose the conditions of Proposition \ref{pro:exist-unique} and Assumption \ref{ass:euas} hold. Then, we have that
  \begin{equation}\label{eq:lim_ctrelt}
    \lim_{\varepsilon \downarrow 0} \Pi_{p,q}(\varepsilon) = \frac{B(p,\gamma^{-1}-p+1)}{B(q,\gamma^{-1}-q+1)} := \ell(\gamma, p, q),
  \end{equation}
  as well as
  \begin{equation}\label{eq:lim_dual_ctrelt}
  \lim_{\varepsilon \downarrow 0} \pi_{p,q}(\varepsilon) =  \frac{B(p,\gamma^{-1}-p+1)}{B(q,\gamma^{-1}-q+1)} := \ell(\gamma, p, q),
  \end{equation}
  where $\ell(\gamma, p, q) = (\mathcal{L}(\gamma, p,q))^{1/\gamma}$.
\end{proposition}

Note that both $\Pi_{p,q}(\varepsilon)$ and $\pi_{p,q}(\varepsilon)$ converge to one same limit that depends on $p,q$ and $\gamma$. This convergence suggests an approach to estimate $\Pi_{p,q}(\varepsilon)$ or $\pi_{p,q}(\varepsilon)$ by substituting an estimator of $\gamma$ into this limit. Although this estimator is computationally tractable, it suffers some limitations and we discuss it in Section \ref{subsec3.1}.

\begin{example}[Expectile and VaR, 2]
  If we take $q=1$ and $p=2$, then the limit \eqref{eq:lim_ctrelt} becomes $\frac{\gamma}{1-\gamma}$ which coincides with the results in Proposition 1 of \cite{Xu2022}.
\end{example}

The following proposition provides a straightforward relationship between CTRELT $\Pi_{p,q}(\varepsilon)$ and dual CTRELT $\pi_{p,q}(\varepsilon)$. We remark that the condition $\theta_p(\tau_0) \leq \theta_q(1-\varepsilon)$ is sufficient to make $\pi_{p,q}(\Pi_{p,q}(\varepsilon)\varepsilon)$ sense in \eqref{eq:correlation1}, while it is unnecessary for \eqref{eq:correlation2}, since $\pi_{p,q}(\varepsilon)$ is always finite.

\begin{proposition}\label{pro:correlation}
  Suppose the conditions of Proposition \ref{pro:exist-unique} hold. For all $\varepsilon \in (0,1-\tau_0)$, we have that
  \begin{equation}\label{eq:correlation2}
    \Pi_{p,q}(\varepsilon/\pi_{p,q}(\varepsilon)) = \pi_{p,q}(\varepsilon).
    \end{equation}
  Moreover, if Assumption \ref{ass:euas}(a) also holds, then we have that
  \begin{equation}\label{eq:correlation1}
  \pi_{p,q}(\Pi_{p,q}(\varepsilon)\varepsilon) = \Pi_{p,q}(\varepsilon).
  \end{equation}
\end{proposition}

We conclude this section with a discussion below on the case of $p-\frac{1}{\gamma} < 1-q$, which corresponds to the third statement in Proposition \ref{pro:rank_lpquant}. The definitions of the CTRELT $\Pi'_{p,q}(\varepsilon)$ and the dual CTRELT $\pi'_{p,q}(\varepsilon)$ with $\tau'_0$ can be defined in a similar way to the case of $1-q < p - \frac{1}{\gamma} < 1$. Hence, we only exhibit these definitions without further elaboration. On the one hand, the CTRELT $\Pi'_{p,q}(\varepsilon)$ is a multiplier $c' \in \left[1, \frac{1-\tau'_0}{\varepsilon} \right]$ solving
\begin{equation}\label{eq:ctrelt_prime}
  \theta_q(1-c'\varepsilon) = \theta_p(1-\varepsilon), \quad \varepsilon \in (0,1-\tau'_0),
\end{equation}
where $\tau'_0$ is defined by
\begin{equation}\label{eq:another_tau0}
\tau'_0 =
\begin{cases}
\sup_{\tau \in (0,1)} \left\{ ~\tau ~\big| ~\theta_q(\tau) \leq \theta_p(\tau)\right\}, & \mbox{if } \left\{ ~\tau ~\big| ~\theta_q(\tau) \leq \theta_p(\tau)\right\} \neq \emptyset, \\
0, & \mbox{if } \left\{ ~\tau ~\big| ~\theta_q(\tau) \leq \theta_p(\tau)\right\} = \emptyset.
\end{cases}
\end{equation}
Its formal definition is given by
\begin{equation}\label{eq:form_ctrelt_prime}
\Pi'_{p,q}(\varepsilon) = \inf_{c' \in \left[1,\frac{1-\tau'_0}{\varepsilon}\right]} \left\{ ~c' ~\big| ~\theta_q(1-c'\varepsilon) \leq \theta_p(1-\varepsilon) \right\}. 
\end{equation}
On the other hand, the dual CTRELT $\pi'_{p,q}(\varepsilon)$ is a multiplier $d' \in [1,\infty)$ solving
\begin{equation}\label{eq:dual_ctrelt_prime}
  \theta_q(1-\varepsilon) = \theta_p\left(1-\frac{\varepsilon}{d'}\right), \quad \varepsilon \in (0,1-\tau'_0),
\end{equation}
with formal definition given by
\begin{equation}\label{eq:form_dual_ctrelt_prime}
\pi'_{p,q}(\varepsilon) = \inf_{d' \in [1,\infty)} \left\{ ~d' ~\bigg| ~\theta_q(1-\varepsilon) \leq \theta_p\left(1-\frac{\varepsilon}{d'}\right) \right\}. 
\end{equation}

The analysis of the existence and uniqueness for $\Pi'_{p,q}(\varepsilon)$ and $\pi'_{p,q}(\varepsilon)$ is similar to the scenario when $1-q < p - \frac{1}{\gamma} < 1$ and is therefore omitted here. In the following sections, we will concentrate solely on the case of $1-q < p-\frac{1}{\gamma}<1$ to discuss statistical inference problems.

\section{Estimation of the CTRELT}\label{sec3}

Another contribution of this paper is to develop inference methods for extreme $L_p$-quantiles under the tail risk equivalent level transition between $L_p$- and $L_q$-quantiles. Suppose that the samples $X_1, X_2,...,X_n$ are independent and identically distributed from a distribution function $F$. Let $1 - \varepsilon'_n \uparrow 1$ be extreme such that $n \varepsilon'_n \to a \in [0,\infty)$ and $1 - \varepsilon_n \uparrow 1$ be intermediate such that $n\varepsilon_n \to \infty$. We aim to estimate the extreme $L_p$-quantile $\theta_p(1-\varepsilon_n')$ through an estimator of intermediate $L_q$-quantile $\theta_q(1-\varepsilon_n)$, which can be achieved by a novel TRELT-based extrapolation method. Different from the classical extrapolation methods in previous works, for example, the one in \eqref{eq:exm1_2019}, the proposed TRELT-based extrapolation methods have some additional elements to estimate due to the transition between the two risk measures. The key issue is to combine extrapolative technique with \eqref{eq:ctrelt} and then to plug in good estimators of CTRELT with intermediate and extreme levels. We first sketch two approaches to extrapolate intermediate $L_q$-quantile to extreme $L_p$-quantile.

Firstly, by Proposition \ref{pro:rank_lpquant}, the asymptotic relationship between $\theta_p(1-\varepsilon'_n)$ and $\theta_p(1-c(\varepsilon_n)\varepsilon_n)$ follows from,
\begin{equation}\label{eq:extrapolation1}
  \frac{\theta_p(1-\varepsilon'_n)}{\theta_p(1-c(\varepsilon_n)\varepsilon_n)} = \frac{\frac{\theta_p(1-\varepsilon'_n)}{\theta_1(1-\varepsilon'_n)}}{\frac{\theta_p(1-c(\varepsilon_n)\varepsilon_n)}{\theta_1(1-c(\varepsilon_n)\varepsilon_n)}}\times \frac{\theta_1(1-\varepsilon'_n)}{\theta_1(1-c(\varepsilon_n)\varepsilon_n)} \sim \left( \frac{c(\varepsilon_n) \varepsilon_n}{\varepsilon'_n} \right)^{\gamma}, \quad as~\varepsilon'_n,~\varepsilon_n \downarrow 0.
\end{equation}
Then, \eqref{eq:ctrelt} motivates us to reformulate \eqref{eq:extrapolation1} as,
\begin{equation}\label{eq:trans_extra1}
  \theta_p(1-\varepsilon'_n) \sim \left( \frac{c(\varepsilon_n) \varepsilon_n}{\varepsilon'_n} \right)^{\gamma} \theta_q(1-\varepsilon_n).
\end{equation}

Secondly, the other path can be given by considering $\theta_p(1-\varepsilon'_n)$ and $\theta_p(1-c(\varepsilon'_n)\varepsilon'_n)$,
\begin{equation}\label{eq:extrapolation2}
  \frac{\theta_p(1-\varepsilon'_n)}{\theta_p(1-c(\varepsilon'_n)\varepsilon'_n)} = \frac{\frac{\theta_p(1-\varepsilon'_n)}{\theta_1(1-\varepsilon'_n)}}{\frac{\theta_p(1-c(\varepsilon'_n)\varepsilon'_n)}{\theta_1(1-c(\varepsilon'_n)\varepsilon'_n)}}\times \frac{\theta_1(1-\varepsilon'_n)}{\theta_1(1-c(\varepsilon'_n)\varepsilon'_n)} \sim [c(\varepsilon'_n)]^{\gamma}, \quad as~\varepsilon'_n \downarrow 0.
\end{equation}
Then, by \eqref{eq:ctrelt} and the extrapolation \eqref{eq:exm1_2019},  
\eqref{eq:extrapolation2} can be rewritten as,
\begin{equation}\label{eq:trans_extra2}
  \theta_p(1-\varepsilon'_n) \sim [c(\varepsilon'_n)]^{\gamma} \theta_q(1-\varepsilon'_n) = \left( \frac{c(\varepsilon'_n) \varepsilon_n}{\varepsilon'_n} \right)^{\gamma} \theta_q(1-\varepsilon_n).
\end{equation}


From \eqref{eq:trans_extra1} and \eqref{eq:trans_extra2}, it is evident that the estimation of $\theta_p(1-\varepsilon'_n)$ is intricately linked to that of $\gamma$, $c(\varepsilon_n)$ or $c(\varepsilon'_n)$, and $\theta_q(1-\varepsilon_n)$. Therefore, we need to estimate each component in the above extrapolation formulations. First, we always apply \eqref{eq:inter_2019} as the estimator for $\theta_q(1-\varepsilon_n)$, whose asymptotic property has been studied well. Second, several estimators for $\gamma$ have been summarized in Chapter 3 of \cite{Haan2006}, including the Hill estimator (\cite{Hill1975}), the Pickands estimator (\cite{Pickands1975}), the moment estimator (\cite{Dekkers1989}) and the maximum likelihood estimator. For a positive $\gamma$, a widely-used one is the Hill estimator,
\begin{equation}\label{eq:hill}
\hat{\gamma}_H = \frac{1}{k} \sum_{i=0}^{k-1} \log X_{n-i,n} - \log X_{n-k,n},
\end{equation}
where $X_{1,n} \leq X_{2,n} \leq \cdots \leq X_{n,n}$ are the order statistics and $k:= k(n)$ is an intermediate sequence satisfying $k = k(n) \to \infty$ and $k/n \to 0$ as $n \to \infty$. Throughout this paper, we adopt \eqref{eq:hill} as the estimator for $\gamma$. Thirdly, we propose three estimators \eqref{eq:plug_in}, \eqref{eq:pi_inter_est} and \eqref{eq:pi_ext_est} for CTRELT, which are motivated by the limit relation \eqref{eq:lim_ctrelt} and \eqref{eqn:level_tran} respectively. 

Before that, it's necessary to put the second-order regular variation condition here.

\begin{assumption}[Second-order regular variation]\label{ass:sorv}
The survival function $\overline{F}$ satisfies a second-order regular variation condition with an extreme value index $\gamma >0$, \textit{i.e.}, for all $x>0$,
\begin{equation}\label{eq:sorv_F}
\lim_{t \to \infty} \frac{1}{A\left( 1/\overline{F}(t)   \right)}\left[\frac{\overline{F}(tx)}{\overline{F}(t)}-x^{-1/\gamma}\right] = x^{-1/\gamma} \frac{x^{\rho / \gamma}-1}{\gamma \rho},
\end{equation}
where $\rho \leq 0$ and $A$ is a positive or negative auxiliary function with $\lim_{t \to \infty} A(t) = 0$. Equivalently, this can also be reformulated in terms of $U$ (recall $U$ is the tail quantile function of $F$),
\begin{equation}\label{eq:sorv_U}
\lim_{t \to \infty} \frac{1}{A(t)}\left[\frac{U(tx)}{U(t)}-x^{\gamma}\right] = x^{\gamma} \frac{x^{\rho}-1}{\rho}.
\end{equation}
\end{assumption}

Note that Assumption \ref{ass:sorv} implies Assumption \ref{ass:forv} and further controls the rate of convergence in Assumption \ref{ass:forv}. In addition, the auxiliary function $A$ is also regularly varying with index $\rho \leq 0$ (see Theorem 2.3.3 in \cite{Haan2006}).

\subsection{Estimation of $\Pi_{p,q}(\varepsilon)$ via \eqref{eq:lim_ctrelt}}\label{subsec3.1}

As previously discussed, the limit relation \eqref{eq:lim_ctrelt} provides a good approximation for $\Pi_{p,q}(\varepsilon)$ as $\varepsilon \to 0$, which inspires us to put up an estimator for $\Pi_{p,q}(\varepsilon)$ by plugging in $\hat{\gamma}_H$ directly, without considering what level $\varepsilon$ is. It seems reasonable since $\varepsilon$ always tends to 0 for both intermediate or extreme cases when the sample size is sufficiently large. This plug-in estimator is given by
\begin{equation}\label{eq:plug_in}
  \widehat{\Pi}_{p,q} = \ell (\hat{\gamma}_H,p,q) = \frac{B(p,\hat{\gamma}_H^{-1} - p +1)}{B(q,\hat{\gamma}_H^{-1} - q+ 1)}.
\end{equation}
Indeed, this estimator is level-free and its uncertainty depends completely on $\hat{\gamma}_H$. Its asymptotic normality can be derived immediately by that of $\hat{\gamma}_H$ under second-order regular variation condition via Delta-method.

\begin{theorem}\label{th:asynorm_plugin}
  Suppose $F$ satisfies Assumptions \ref{ass:mom_negapart}, \ref{ass:euas} and \ref{ass:sorv} with $p,q$ satisfying $1 \leq q < p$, $1-q < p - \frac{1}{\gamma} < 1$. Then, for all $\varepsilon_n, \varepsilon'_n \in (0,1-\tau_0)$, we have that as $n \to \infty$,
  \begin{equation}\label{eq:asynorm_plugin}
    \sqrt{k} \left(\widehat{\Pi}_{p,q} - \ell(\gamma,p,q)\right) \xrightarrow{d}  \mathcal{N} \left( \frac{\partial}{\partial \gamma} \ell(\gamma,p,q) \frac{\lambda}{1-\rho} , \left(\frac{\partial}{\partial \gamma} \ell(\gamma,p,q) \right)^2 \gamma^2  \right),
  \end{equation}
  provided the sequence $k: = k(n)$ satisfy $k \to \infty$ and $k/n \to 0$, and
  \begin{equation*}
    \lim_{n \to \infty} \sqrt{k} A\left( \frac{n}{k} \right) = \lambda < \infty.
  \end{equation*}
  The derivatives in the limit are given by
  \begin{equation*}
     \frac{\partial}{\partial t} \ell(t,p,q) = \frac{\Gamma(p)}{\Gamma(q)} \cdot \frac{\Gamma(t^{-1}-q+1) \left(\frac{\partial}{\partial t}\Gamma(t^{-1}-p+1)\right)-\Gamma(t^{-1}-p+1) \left(\frac{\partial}{\partial t} \Gamma(t^{-1}-q+1)\right)}{[\Gamma(t^{-1}-q+1)]^2},
  \end{equation*}
  and
  \begin{equation*}
    \frac{\partial}{\partial t}\Gamma(t^{-1}-p+1) = -\frac{1}{t^2} \int_{0}^{\infty} s^{t^{-1}-p} e^{-s} \log s \,ds.
  \end{equation*}
\end{theorem}

Although this estimator can work as a simple approximation to the CTRELT in practice, it suffers from several drawbacks under rigorous scrutiny. First, the CTRELT is a mapping from $\varepsilon$ to $\Pi_{p,q}(\varepsilon)$ so that $\Pi_{p,q}(\varepsilon)$ may differ for different values $\varepsilon$. However, from the essence of this estimator, it makes no sense that $\widehat\Pi_{p,q}$ does not correspond to the value $\varepsilon$. Thus, the estimator \eqref{eq:plug_in} may only perform well when $\varepsilon$ is sufficiently small and may show poor performance when it deviates far from 0 or the sample size is small. Second, the level $\tau=1-\varepsilon$ admits the transformation \eqref{eqn:level_tran}, but the limit in \eqref{eq:lim_ctrelt} fails to admit it. Thus, the estimator \eqref{eq:plug_in} does not truly estimate the tail risk transition given large risk levels. Third, the estimator \eqref{eq:plug_in} may be a bad statistical model when one considers the rates of convergence for an intermediate $\varepsilon_n \to 0$ and an extreme $\varepsilon'_n\to 0$ at the same time. It is because the rate of convergence of $\widehat\Pi_{p,q}$ is completely determined by the Hill estimator $\hat\gamma_H$. However, as $n\to\infty$, there is no difference between the convergence rates for $\varepsilon_n \to 0$ and $\varepsilon'_n\to 0$. 


\subsection{Estimation of $\Pi_{p,q}(\varepsilon_n)$ at intermediate level}\label{subsec3.2}

We now propose other two empirical methods for estimating $\Pi_{p,q}(\varepsilon)$ via transformation \eqref{eqn:level_tran} by considering the intermediate and extreme levels. Using \eqref{eq:lp-quantile2}, \eqref{eqn:level_tran} and \eqref{eq:ctrelt} yields that
\begin{equation*}
  1-c \varepsilon  = \frac{\mathbb{E}[(\theta_p(1-c\varepsilon)-X)^{p-1}_+]}{\mathbb{E}[|X-\theta_p(1-c\varepsilon)|^{p-1}]} = \frac{\mathbb{E}[(\theta_q(1-\varepsilon)-X)^{p-1}_+]}{\mathbb{E}[|X-\theta_q(1-\varepsilon)|^{p-1}]},
\end{equation*}
and
\begin{equation*}
  1-\varepsilon = \frac{\mathbb{E}[(\theta_q(1-\varepsilon)-X)^{q-1}_+]}{\mathbb{E}[|X-\theta_q(1-\varepsilon)|^{q-1}]}.
\end{equation*}
Then, we can derive $\Pi_{p,q}(\varepsilon)$ explicitly by
\begin{equation}\label{eq:pi_rep}
   \Pi_{p,q}(\varepsilon)= \frac{\mathbb{E}[(X - \theta_q(1-\varepsilon))^{p-1}_+] }{\mathbb{E}[(X - \theta_q(1-\varepsilon))^{q-1}_+]} \times \frac{\mathbb{E}[|X-\theta_q(1-\varepsilon)|^{q-1}]}{\mathbb{E}[|X-\theta_q(1-\varepsilon)|^{p-1}]},
\end{equation}
and the corresponding estimator $\widetilde\Pi_{p,q}(\varepsilon)$ can be defined by using its empirical counterpart and plugging in $\hat{\theta}_q(1-\varepsilon)$ (the estimator of $\theta_q(1-\varepsilon)$ given in \eqref{eq:inter_2019}) such that
\begin{equation}\label{eq:pi_emp}
  \widetilde\Pi_{p,q}(\varepsilon) = \frac{\frac{1}{n}\sum_{i=1}^{n}(X_i-\hat{\theta}_q(1-\varepsilon))^{p-1}_+ }{\frac{1}{n}\sum_{i=1}^{n}(X_i-\hat{\theta}_q(1-\varepsilon))^{q-1}_+} \times \frac{\frac{1}{n} \sum_{i=1}^{n} |X_i - \hat{\theta}_q(1-\varepsilon)|^{q-1} }{\frac{1}{n} \sum_{i=1}^{n} |X_i - \hat{\theta}_q(1-\varepsilon)|^{p-1} }.
\end{equation}

The tail level $\varepsilon$ in \eqref{eq:pi_emp} can be either a fixed or an intermediate level, which diverges to $0$. When $\varepsilon$ is intermediate, then the estimator of $\widetilde\Pi_{p,q}(\varepsilon_n)$ can be given as follows,
\begin{equation}\label{eq:pi_inter_est}
  \widetilde\Pi_{p,q}(\varepsilon_n) = \frac{\frac{1}{n}\sum_{i=1}^{n}(X_i-\hat{\theta}_q(1-\varepsilon_n))^{p-1}_+ }{\frac{1}{n}\sum_{i=1}^{n}(X_i-\hat{\theta}_q(1-\varepsilon_n))^{q-1}_+} \times \frac{\frac{1}{n} \sum_{i=1}^{n} |X_i - \hat{\theta}_q(1-\varepsilon_n)|^{q-1} }{\frac{1}{n} \sum_{i=1}^{n} |X_i - \hat{\theta}_q(1-\varepsilon_n)|^{p-1} }.
\end{equation}

\begin{theorem}\label{th:asynorm_pi_inter}
Suppose $F$ satisfies Assumptions \ref{ass:mom_negapart}, \ref{ass:euas} and \ref{ass:sorv} with $p,q$ satisfying $1 \leq q < p$, $1-q < p - \frac{1}{2\gamma} < 1$. Then for all $\varepsilon_n \in (0,1-\tau_0)$, we have that as $n\to\infty$,
\begin{equation*}
\sqrt{n\varepsilon_n} \left( \frac{\widetilde\Pi_{p,q}(\varepsilon_n)}{\Pi_{p,q}(\varepsilon_n)} -1\right) \xrightarrow{d}
\begin{cases}
  \mathcal{N}(\mathcal{E}_1(p,q,\gamma), \mathcal{V}_1(p,q,\gamma)), & \mbox{if } q > 1, \\
  \mathcal{N}(\mathcal{E}_2(p,q,\gamma), \mathcal{V}_2(p,q,\gamma)), & \mbox{if } q = 1,
\end{cases}
\end{equation*}
provided $\varepsilon_n$ satisfies $1 - \varepsilon_n \uparrow 1$ and $n\varepsilon_n \to \infty$, and
\begin{equation*}
  \lim_{n \to \infty}\sqrt{n\varepsilon_n}A\left( \frac{1}{\varepsilon_n} \right) = \lambda < \infty.
\end{equation*}
The expectations and variances $\mathcal{E}_1(p,q,\gamma)$, $\mathcal{E}_2(p,q,\gamma)$, $\mathcal{V}_1(p,q,\gamma)$, and $\mathcal{V}_2(p,q,\gamma)$ are given by
\begin{align*}
\mathcal{E}_1(p,q,\gamma) & = \left[ \frac{(1-p)I(p,\gamma)}{B(p,\gamma^{-1}-p+1)} - \frac{(1-q)I(q,\gamma)}{B(q,\gamma^{-1}-q+1)} + q-p\right] \\
& \times \left[ \frac{\lambda (q-1)}{\gamma^{\rho-1} \rho} \frac{B(q-1,-(\rho-1)/\gamma-q+1)}{[B(q,\gamma^{-1}-q+1)]^{1-\rho}} - \frac{\lambda}{\rho} \right]  \\
 &+ \frac{\lambda}{\gamma^{\rho}\rho} \left\{ \frac{(p-1)[B(q,\gamma^{-1}-q+1)]^{\rho}B(p-1,(1-\rho)/\gamma-p+1)}{B(p,\gamma^{-1}-p+1)} \right.\\
 & \left. - \frac{(q-1)B(q-1,(1-\rho)/\gamma-q+1)}{[B(q,\gamma^{-1}-q+1)]^{1-\rho}} \right\}, \\
\mathcal{V}_1(p,q,\gamma) & = \left\{ \frac{\sqrt{\gamma}}{2} \left[ \frac{(1-p)I(p,\gamma)}{B(p,\gamma^{-1}-p+1)} - \frac{(1-q)I(q,\gamma)}{B(q,\gamma^{-1}-q+1)} + q-p\right] \frac{B(q,(2\gamma)^{-1}-q+1)}{[B(q,\gamma^{-1}-q+1)]^{1/2}} \right.  \\
& + \left. \frac{[B(q,\gamma^{-1}-q+1)]^{1/2}B(p,(2\gamma)^{-1}-p+1)}{2\sqrt{\gamma}B(p,\gamma^{-1}-p+1)} - \frac{B(q,(2\gamma)^{-1}-q+1)}{2\sqrt{\gamma}[B(q,\gamma^{-1}-q+1)]^{1/2}} \right\}^2, \\
\mathcal{E}_2(p,q,\gamma) & = \frac{\lambda (p-1)}{\rho} \frac{B(p-1,-(\rho-1)/\gamma -p+1)}{B(p,\gamma^{-1}-p+1)} - \frac{\lambda}{\gamma \rho}, \\
\mathcal{V}_2(p,q,\gamma) & = \left[ \frac{\gamma(1-p)I(p,\gamma)}{B(p,\gamma^{-1}-p+1)} + \frac{B(p,(2\gamma)^{-1}-p+1)}{2B(p,\gamma^{-1}-p+1)} -(p-1)\gamma -1 \right]^2,
\end{align*}
and $I(p,\gamma) = \gamma (p-2) \int_{0}^{\infty} t^{p-3} (t+1)^{-1/\gamma} \,dt$ ($I(q,\gamma)$ is defined similarly).
\end{theorem}

It should be noted that $\widetilde\Pi_{p,q}(\varepsilon_n)$ is asymptotic unbiased when $\lambda = 0$ and it incorporates $\hat{\theta}_q(1-\varepsilon_n)$, making it impossible to study asymptotic normality without investigating $\hat{\theta}_q(1-\varepsilon_n)$. However, the existing asymptotic results of $\hat{\theta}_q(1-\varepsilon_n)$ (see Theorem 1 in \cite{Daouia2019}), which were established by the Lindeberg-type central limit theorem, may not be sufficient to support our findings here. We hence provide a more reinforced version of the asymptotic normality of $\hat{\theta}_q(1-\varepsilon_n)$ by employing the tail empirical process (see Proposition \ref{pro:daouiath1}). An improvement lies in the relaxation of moment condition $\mathbb{E}\left[X_{-}^{(2+\delta)(q-1)}\right]<\infty$ with $\delta >0$, which was applied for Lyapunov condition in the proof. Particularly, if $2 \leq q < p$, the asymptotic expectations and variances will reduce to
\begin{align*}
\mathcal{E}_1(p,q,\gamma) & = \frac{\lambda}{\gamma^{\rho}\rho} \left\{ \frac{(p-1)[B(q,\gamma^{-1}-q+1)]^{\rho}B(p-1,(1-\rho)/\gamma-p+1)}{B(p,\gamma^{-1}-p+1)} \right.\\
& \left. - \frac{(q-1)B(q-1,(1-\rho)/\gamma-q+1)}{[B(q,\gamma^{-1}-q+1)]^{1-\rho}} \right\}, \\
\mathcal{V}_1(p,q,\gamma) & = \left\{ \frac{[B(q,\gamma^{-1}-q+1)]^{1/2}B(p,(2\gamma)^{-1}-p+1)}{2\sqrt{\gamma}B(p,\gamma^{-1}-p+1)} - \frac{B(q,(2\gamma)^{-1}-q+1)}{2\sqrt{\gamma}[B(q,\gamma^{-1}-q+1)]^{1/2}} \right\}^2, \\
\mathcal{E}_2(p,q,\gamma) & = \frac{\lambda (p-1)}{\rho} \frac{B(p-1,-(\rho-1)/\gamma -p+1)}{B(p,\gamma^{-1}-p+1)} - \frac{\lambda}{\gamma \rho}, \\
\mathcal{V}_2(p,q,\gamma) & = \left[ \frac{B(p,(2\gamma)^{-1}-p+1)}{2B(p,\gamma^{-1}-p+1)} -2 \right]^2.
\end{align*}

\subsection{Estimation of $\Pi_{p,q}(\varepsilon'_n)$ at extremal level}\label{subsec3.3}

Likewise, if we consider extreme level $1 - \varepsilon'_n$, the estimator for $\Pi_{p,q}(\varepsilon'_n)$ can be taken by substituting $\tilde\theta^{\rm sta}_q(1-\varepsilon'_n)$ into \eqref{eq:pi_emp} such that
\begin{equation}\label{eq:pi_ext_est}
  \widetilde\Pi_{p,q}(\varepsilon'_n) = \frac{\frac{1}{n}\sum_{i=1}^{n}(X_i-\tilde\theta^{\rm sta}_q(1-\varepsilon'_n))^{p-1}_+ }{\frac{1}{n}\sum_{i=1}^{n}(X_i-\tilde\theta^{\rm sta}_q(1-\varepsilon'_n))^{q-1}_+} \times \frac{\frac{1}{n} \sum_{i=1}^{n} |X_i - \tilde\theta^{\rm sta}_q(1-\varepsilon'_n)|^{q-1} }{\frac{1}{n} \sum_{i=1}^{n} |X_i - \tilde\theta^{\rm sta}_q(1-\varepsilon'_n)|^{p-1} }.
\end{equation}
Recall that $\tilde{\theta}^{\rm sta}_q(1-\varepsilon'_n)$ is the standard extrapolative estimator for $\theta_q(1-\varepsilon'_n)$ defined in \eqref{eq:exm1_2019} that
\begin{equation}\label{eq:theta_p_ext}
  \tilde{\theta}^{\rm sta}_q(1-\varepsilon'_n) = \left(\frac{\varepsilon'_n}{\varepsilon_n}\right)^{-\hat\gamma_H} \hat\theta_q(1-\varepsilon_n).
\end{equation}

Regrettably, it fails to induce an asymptotic normality for $\widetilde\Pi_{p,q}(\varepsilon'_n)$. It is because the convergence rate of \eqref{eq:theta_p_ext} is much faster than $\sqrt{n\varepsilon'_n}$, which makes it impossible to find a suitable rate to ensure asymptotic normality. As shown in \cite{Daouia2019}, a critical technique to establish asymptotic normality for \eqref{eq:theta_p_ext} involves reformulating the relationship between $\theta_p(\tau)$ and $\theta_1(\tau)$ through a second-order expansion, making the remainder term multiplied by the rate $\sqrt{n\varepsilon_n}/\log [\varepsilon_n / \varepsilon^{\prime}_n]$ be negligible. However, it needs some redundant conditions on the left tail of $F$. To improve this, we provide a more general version of the second-order expansion for $\theta_p(\tau)$ and $\theta_q(\tau)$ in Proposition \ref{pro:2order_expansion}. There are two improvements: we no longer care about whether the left tail is light or heavy, and only retain the moment condition (Assumption \ref{ass:mom_negapart}); it applies to all $p,q$ that satisfy $1 \leq q < p< 1+\frac{1}{\gamma}$.

It is not difficult to verify the asymptotic normality of \eqref{eq:theta_p_ext} still holds under (A.3), provided the remainder can be dominated by the rate $\sqrt{n\varepsilon_n}$. We summarize this in Assumption \ref{ass:cond_2o_expansion}. 

\begin{assumption}\label{ass:cond_2o_expansion}
  $\sqrt{n\varepsilon_n} R(p,1,\gamma, 1-\varepsilon_n) = O(1)$. (See Proposition \ref{pro:2order_expansion} for $R(\cdot)$.)
\end{assumption}

\begin{theorem}\label{th:asy_pi_ext}
Suppose $F$ is strictly increasing and satisfies Assumptions \ref{ass:mom_negapart}, \ref{ass:euas}, \ref{ass:sorv} and \ref{ass:cond_2o_expansion} with $p,q$ satisfying $1 \leq q < p$, $1-q < p - \frac{1}{2\gamma} < 1$. Then, for all $\varepsilon_n, \varepsilon'_n \in (0,1-\tau_0)$, we have that as $n\to\infty$,
\begin{equation*}
\frac{\widetilde\Pi_{p,q}(\varepsilon'_n)}{\Pi_{p,q}(\varepsilon'_n)} \xrightarrow{\mathbb{P}}
\begin{cases}
  \Delta, & \mbox{if } a > 0, \\
  \frac{B(q,\gamma^{-1}-q+1)B(p,(2\gamma)^{-1}-p+1)}{B(p,\gamma^{-1}-p+1)B(q,(2\gamma)^{-1}-q+1)}, & \mbox{if } a = 0,
\end{cases}
\end{equation*}
provided $\varepsilon_n$ and $\varepsilon'_n$ satisfy $1 - \varepsilon_n \uparrow 1$, $n\varepsilon_n \to \infty$ and $1 - \varepsilon'_n \uparrow 1$, $n\varepsilon'_n \to a < \infty$, as well as
\begin{equation*}
  \lim_{n \to \infty} \frac{\sqrt{n\varepsilon_n}}{\log [\varepsilon_n / \varepsilon^{\prime}_n]} = \infty, \quad \lim_{n \to \infty}\sqrt{n\varepsilon_n}A\left( \frac{1}{\varepsilon_n} \right) = \lambda < \infty.
\end{equation*}
Here, $\Delta$ is a random variable with a density function
\begin{equation*}
  f(y) = \frac{\sqrt{a}}{\sqrt{2\pi}}\exp\left\{ - \frac{a(y-1)^2}{2(c_1-c_2y)^2} \right\} \frac{|c_1-c_2|}{(c_1-c_2y)^2},\quad \left(y \neq \frac{c_1}{c_2}\right),
\end{equation*}
where
\begin{equation*}
  \begin{cases}
    c_1 = \frac{[B(q,\gamma^{-1}-q+1)]^{1/2}B(p,(2\gamma)^{-1}-p+1)}{2\sqrt{\gamma}B(p,\gamma^{-1}-p+1)}, \\
    c_2 = \frac{B(q,(2\gamma)^{-1}-q+1)}{2\sqrt{\gamma}[B(q,\gamma^{-1}-q+1)]^{1/2}}.
  \end{cases}
\end{equation*}
\end{theorem}

\section{Estimation of extreme $L_p$-quantiles via TRELT}\label{sec4}

Recall the extrapolation methods we sketched at the beginning of Section \ref{sec3}. A unified form is
\begin{equation}\label{eq:uni_form}
  \theta_p(1-\varepsilon'_n) \sim \left( \frac{c \varepsilon_n}{\varepsilon'_n} \right)^{\gamma} \theta_q(1-\varepsilon_n),\quad\text{as } n\to\infty,
\end{equation}
with $c:=c(\varepsilon_n)$ or $c(\varepsilon'_n)$, where the existence and uniqueness of $c$ was justified in Section \ref{sec2}. For inference problems, \eqref{eq:uni_form} indicates that different combinations of estimators of $\gamma$, $c$ and $\theta_q(1-\varepsilon_n)$ will result in different estimators for $\theta_p(1-\varepsilon'_n)$. So far, all these estimations have been well-established.
Then, the two corresponding extrapolative estimators for $\theta_p(1-\varepsilon'_n)$ can be given by
\begin{equation}\label{eq:thetap_extest2}
  \tilde \theta_p^{\rm int}\left(1-\varepsilon'_n\right) = \left( \frac{\widetilde\Pi_{p,q}(\varepsilon_n) \varepsilon_n}{\varepsilon'_n} \right)^{\hat\gamma_H} \hat \theta_q(1-\varepsilon_n),
\end{equation}
\begin{equation}\label{eq:thetap_extest3}
  \tilde \theta_p^{\rm ext}\left(1-\varepsilon'_n\right) = \left( \frac{\widetilde\Pi_{p,q}(\varepsilon'_n) \varepsilon_n}{\varepsilon'_n} \right)^{\hat\gamma_H} \hat \theta_q(1-\varepsilon_n).
\end{equation}
Given that their extrapolation formulation are the same and the only difference lies in $\widetilde\Pi_{p,q}(\varepsilon_n)$ or $\widetilde\Pi_{p,q}(\varepsilon'_n)$, the third one can thus be defined by using \eqref{eq:plug_in} such that,
\begin{equation}\label{eq:thetap_extest1}
  \tilde \theta_p^{\rm lim}\left(1-\varepsilon'_n\right) = \left( \frac{\widehat{\Pi}_{p,q} \varepsilon_n}{\varepsilon'_n} \right)^{\hat\gamma_H} \hat \theta_q(1-\varepsilon_n).
\end{equation}
We here use superscripts ``{\rm int}", ``{\rm ext}" and ``{\rm lim}" to mark these three estimators.

Beyond that of \cite{Daouia2019}, the above three estimators are all embedded in the estimators of TRELT, and their uncertainty are new and unknown.  In practice, the TRELT-based extrapolation may better characterize the uncertainty of risks when risk measure varies from $L_q$-quantile to $L_p$-quantile. Hence, \eqref{eq:thetap_extest2} - \eqref{eq:thetap_extest1} allow the estimation of $\theta_p(1-\varepsilon'_n)$ using different $\theta_q(1-\varepsilon_n)$, rather than just $p$ and 1. Finally, under some mild conditions, the asymptotic properties of $\tilde \theta_p^{\rm int}\left(1-\varepsilon'_n\right)$, $\tilde \theta_p^{\rm ext}\left(1-\varepsilon'_n\right)$ and $\tilde \theta_p^{\rm lim}\left(1-\varepsilon'_n\right)$ are well established in Theorems \ref{th:asynorm_thetap2}, \ref{th:asynorm_thetap3}, and \ref{th:asynorm_thetap1} respectively.

\begin{theorem}\label{th:asynorm_thetap2}
  Suppose the conditions of Theorem \ref{th:asynorm_pi_inter} and Assumption \ref{ass:cond_2o_expansion} hold. Then, we have that as $n\to\infty$,
  \begin{equation*}
    \frac{\sqrt{n\varepsilon_n}}{\log[\varepsilon_n/\varepsilon_n^{\prime}]} \left( \frac{\tilde \theta_p^{\rm int}\left(1-\varepsilon'_n\right)}{\theta_p(1-\varepsilon_n^{\prime})} -1 \right) \xrightarrow{d} \mathcal{N}\left( \frac{\lambda}{1-\rho},\gamma^2 \right),
  \end{equation*}
  provided $\varepsilon_n$ and  $\varepsilon'_n$ satisfy $1 - \varepsilon_n \uparrow 1$, $n\varepsilon_n \to \infty$ and $1 - \varepsilon'_n \uparrow 1$, $n\varepsilon'_n \to a < \infty$, as well as
  \begin{equation*}
  \lim_{n \to \infty} \frac{\sqrt{n\varepsilon_n}}{\log [\varepsilon_n / \varepsilon^{\prime}_n]} = \infty, \quad \lim_{n \to \infty}\sqrt{n\varepsilon_n}A\left( \frac{1}{\varepsilon_n} \right) = \lambda < \infty.
  \end{equation*}
\end{theorem}

\begin{theorem}\label{th:asynorm_thetap3}
  Suppose the conditions of Theorem \ref{th:asy_pi_ext} and Assumption \ref{ass:cond_2o_expansion} hold. Then, we have that as $n\to\infty$,
  \begin{equation*}
  \frac{\tilde \theta_p^{\rm ext}\left(1-\varepsilon'_n\right)}{\theta_p(1-\varepsilon_n^{\prime})} \xrightarrow{\mathbb{P}}
  \begin{cases}
  \Delta^{\gamma}, & \mbox{if } a > 0, \\
  \left[\frac{B(q,\gamma^{-1}-q+1)B(p,(2\gamma)^{-1}-p+1)}{B(p,\gamma^{-1}-p+1)B(q,(2\gamma)^{-1}-q+1)}\right]^{\gamma}, & \mbox{if } a = 0,
  \end{cases}
  \end{equation*}
  provided $\varepsilon_n$ and $\varepsilon'_n$ satisfy $1 - \varepsilon_n \uparrow 1$, $n\varepsilon_n \to \infty$ and $1 - \varepsilon'_n \uparrow 1$, $n\varepsilon'_n \to a < \infty$, as well as
  \begin{equation*}
  \lim_{n \to \infty} \frac{\sqrt{n\varepsilon_n}}{\log [\varepsilon_n / \varepsilon^{\prime}_n]} = \infty, \quad \lim_{n \to \infty}\sqrt{n\varepsilon_n}A\left( \frac{1}{\varepsilon_n} \right) = \lambda < \infty.
  \end{equation*}
Here, $\Delta$ is defined in Theorem \ref{th:asy_pi_ext}.
\end{theorem}

\begin{theorem}\label{th:asynorm_thetap1}
  Suppose the conditions of Theorem \ref{th:asynorm_plugin} and Assumption \ref{ass:cond_2o_expansion} hold. Then, we have that as $n\to\infty$,
  \begin{equation*}
    \frac{\sqrt{n\varepsilon_n}}{\log[\varepsilon_n/\varepsilon_n^{\prime}]} \left( \frac{\tilde\theta_p^{\rm lim}\left(1-\varepsilon'_n\right)}{\theta_p(1-\varepsilon_n^{\prime})} -1 \right) \xrightarrow{d} \mathcal{N}\left( \frac{\lambda}{1-\rho},\gamma^2 \right),
  \end{equation*}
  provided $\varepsilon_n$ and $\varepsilon'_n$ satisfy $1 - \varepsilon_n \uparrow 1$, $n\varepsilon_n \to \infty$ and $1 - \varepsilon'_n \uparrow 1$, $n\varepsilon'_n \to a < \infty$, as well as
  \begin{equation*}
  \lim_{n \to \infty} \frac{\sqrt{n\varepsilon_n}}{\log [\varepsilon_n / \varepsilon^{\prime}_n]} = \infty, \quad \lim_{n \to \infty}\sqrt{n\varepsilon_n}A\left( \frac{1}{\varepsilon_n} \right) = \lambda < \infty.
  \end{equation*}
\end{theorem}

Note that $\tilde \theta_p^{\rm int}\left(1-\varepsilon'_n\right)$, $\tilde \theta_p^{\rm lim}\left(1-\varepsilon'_n\right)$ and $\tilde\theta^{\rm sta}_p(1-\varepsilon'_n)$, $\tilde\theta^{\rm qua}_p(1-\varepsilon'_n)$ all share the same limit distribution, which is exactly the asymptotical distribution of Hill estimator $\hat{\gamma}_H$. This also implies that these four estimators are all asymptotically unbiased when $\lambda = 0$. Yet, $\tilde \theta_p^{\rm ext}\left(1-\varepsilon'_n\right)$ also fails to have an asymptotic normality, primarily due to the inability of $\widetilde\Pi_{p,q}(\varepsilon'_n)$ to achieve asymptotic Gaussian.



\section{Simulation Study}\label{sec5}

In this section, we implement some simulation studies to examine the finite sample performance of the estimators proposed in Sections \ref{sec3} and \ref{sec4}. The following three distributions, all with an extreme value index $\gamma$, will be considered in the experiments:
\begin{itemize}
  \item Pareto distribution with distribution function $F(x) = 1-x^{-1/\gamma},~x>1$;
  \item Fr$\acute{e}$chet distribution with distribution function $F(x) = \exp\{ -x^{-1/\gamma} \},~x>0$;
  \item Student-$t$ distribution with density function $f(x) = \frac{\Gamma\left(\frac{1/\gamma+1}{2}\right)}{\sqrt{\pi/\gamma}\Gamma\left(\frac{1/\gamma}{2}\right)}(1+\gamma x^2)^{-\frac{1/\gamma+1}{2}}$ and degree of freedom $1/\gamma$.
\end{itemize}
To characterize the heavy tails, we set $\gamma = 1/3$ and 0.45. The selection of orders $p,q$ cannot be arbitrary; it must meet the constraints $1 \leq q < p$ and $1-q < p - \frac{1}{2\gamma} < 1$. The parameters we use are listed in Table \ref{tab:para_conf}.

\begin{table}
  \centering
  \caption{The parameter configuration for this simulation.}
  \label{tab:para_conf}
  \begin{tabular}{@{}clll@{}}
  \hline
   & Pareto & Fr$\acute{e}$chet & Student-$t$ \\
  \hline
  $\gamma$ & \multicolumn{3}{c}{The values of pair $(p,q)$}  \\
  \hline
  {1/3}  & (2.4,~1.8)  & (2.4,~1.8) & (2.4,~1.8)  \\
         & (2.4,~2.0)  & (2.4,~2.0) & (2.4,~2.0)  \\
  {0.45} & (2.0,~1.5)  & (2.0,~1.5) & (2.0,~1.5)  \\
         & (2.0,~1.8)  & (2.0,~1.8) & (2.0,~1.8)  \\
  \hline
  \end{tabular}
\end{table}

We now need to determine $\tau_0$ empirically with $p,q$ given in Table \ref{tab:para_conf}. To do this, we plot the curves of $L_p$-quantiles, which are depicted against $\tau$ varying from 0 to 1 by step size 0.001. It shows that, for Pareto and Fr$\acute{e}$chet distributions, $\theta_p(\tau) > \theta_q(\tau)$ always holds on $(0,1)$. As for Student-$t$ distribution, due to symmetry, $\theta_p(\tau) > \theta_q(\tau)$ holds on $[0.5,1)$ while $\theta_p(\tau) < \theta_q(\tau)$ holds on $(0,0.5)$. Therefore, it is sufficient to take $\tau_0 = 0.5$ for the construction of CTRELT. Moreover, a tractable approach to choosing an intermediate $k (= n\varepsilon_n)$ is plotting the Hill estimator $\hat{\gamma}_H$ against $k$ and then choosing a suitable $k$ according to the first stable parts. We summarize the chosen values of $k$ in Table \ref{tab:k_values}.


\begin{table}
  \centering
  \caption{The chosen values of $k$ for Pareto, Fr$\acute{e}$chet and Student-$t$ with $\gamma=1/3,0.45$.}
  \label{tab:k_values}
  \begin{tabular}{@{}ccccc@{}}
  \hline
  & & Pareto & Fr$\acute{e}$chet & Student-$t$ \\
  \hline
   $n$ & $\gamma$ & \multicolumn{3}{c}{The values of $k$} \\
  \hline
  {2000} & 1/3   & 58 & 77 & 53 \\
         & 0.45  & 58 & 77 & 57 \\
  {5000} & 1/3   & 55 & 80 & 60 \\
         & 0.45  & 54 & 80 & 55 \\
  \hline
  \end{tabular}
\end{table}

In our study, we set the sample size $n = 2000, 5000$, extreme level $\varepsilon_n^{\prime} = 0.005$, intermediate level $\varepsilon_n = k/n$ and repeat the simulation $N = 1000$ times. We will implement the following methods to compare the finite-sample performance:
\begin{itemize}
  \item LimTRELT-I: the plug-in estimator $\widehat{\Pi}_{p,q}$ (see \eqref{eq:plug_in}) for $\Pi_{p,q}(\varepsilon_n)$ at intermediate level $\varepsilon_n$.
  \item LimTRELT-II: the plug-in estimator $\widehat{\Pi}_{p,q}$ (see \eqref{eq:plug_in}) for $\Pi_{p,q}(\varepsilon'_n)$ at extreme level $\varepsilon'_n$.
  \item IntTRELT: the empirical estimator $\widetilde\Pi_{p,q}(\varepsilon_n)$ (see \eqref{eq:pi_inter_est}) for $\Pi_{p,q}(\varepsilon_n)$ at intermediate level $\varepsilon_n$.
  \item ExtTRELT: the empirical estimator $\widetilde\Pi_{p,q}(\varepsilon'_n)$ (see \eqref{eq:pi_ext_est}) for $\Pi_{p,q}(\varepsilon'_n)$ at extreme level $\varepsilon'_n$.
  \item BM: the standard extrapolative estimator $\tilde\theta^{\rm sta}_p(1-\varepsilon'_n)$ (see \eqref{eq:exm1_2019}) for $\theta_p(1-\varepsilon'_n)$, served as a benchmark.
  \item ExtraM-I: the proposed extrapolative estimator $\tilde \theta_p^{\rm int}\left(1-\varepsilon'_n\right)$ (see \eqref{eq:thetap_extest2}) for $\theta_p(1-\varepsilon'_n)$.
  \item ExtraM-II: the proposed extrapolative estimator $\tilde \theta_p^{\rm ext}\left(1-\varepsilon'_n\right)$ (see \eqref{eq:thetap_extest3}) for $\theta_p(1-\varepsilon'_n)$.
  \item ExtraM-III: the proposed extrapolative estimator $\tilde \theta_p^{\rm lim}\left(1-\varepsilon'_n\right)$ (see \eqref{eq:thetap_extest1}) for $\theta_p(1-\varepsilon'_n)$.
\end{itemize}

\begin{figure}[htbp]
\centering
\begin{minipage}[b]{0.32\textwidth}
\includegraphics[width=\textwidth,height = 0.18\textheight]{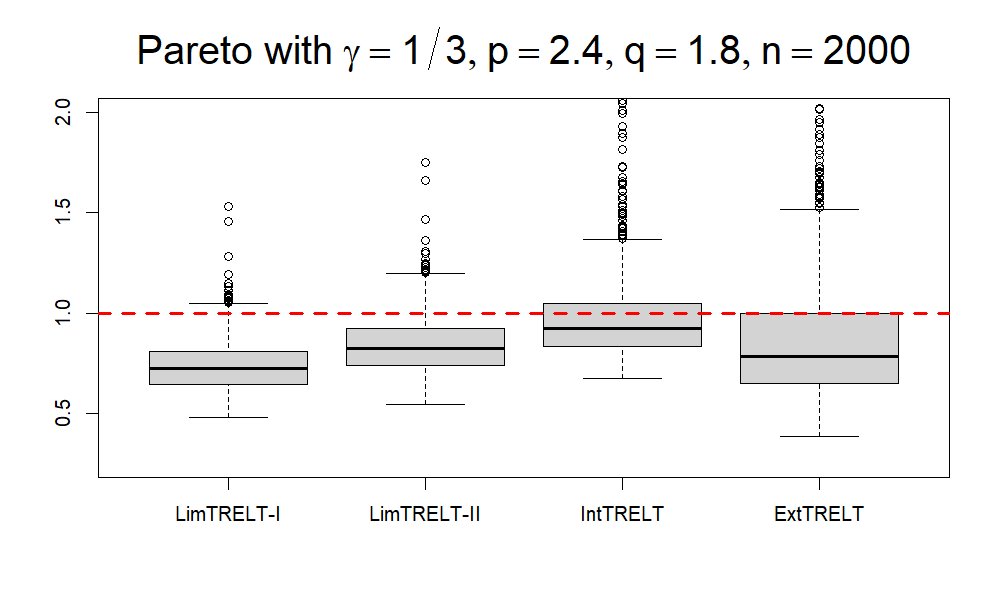}
\end{minipage}
\begin{minipage}[b]{0.32\textwidth}
\includegraphics[width=\textwidth,height = 0.18\textheight]{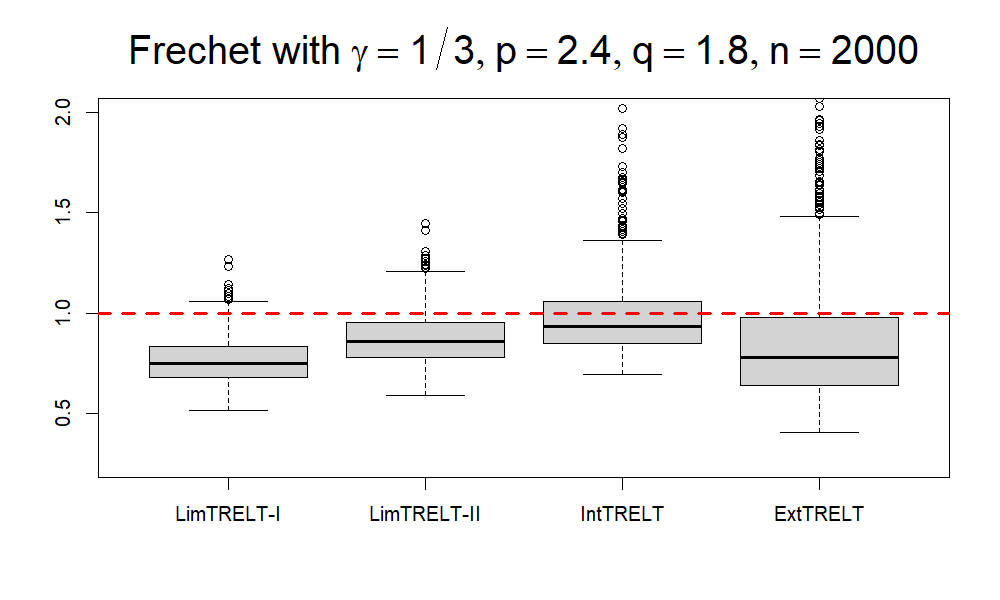}
\end{minipage}
\begin{minipage}[b]{0.32\textwidth}
\includegraphics[width=\textwidth,height = 0.18\textheight]{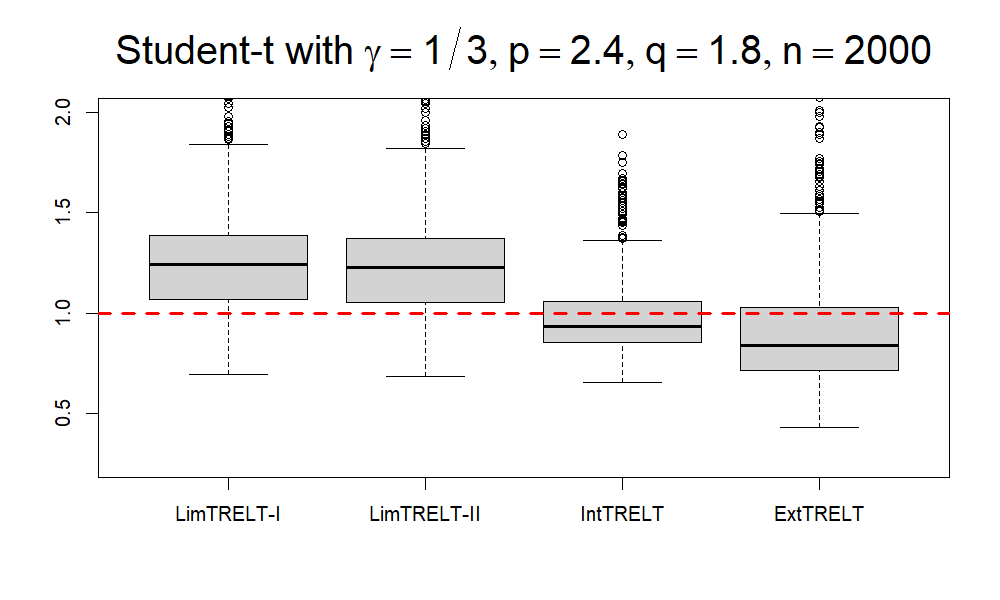}
\end{minipage}
\\
\begin{minipage}[b]{0.32\textwidth}
\includegraphics[width=\textwidth,height = 0.18\textheight]{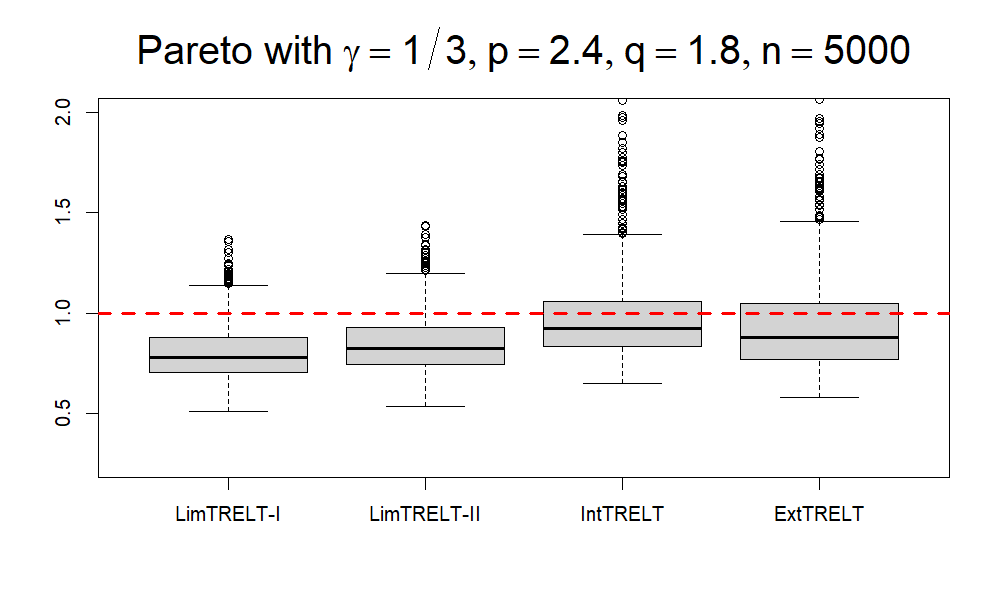}
\end{minipage}
\begin{minipage}[b]{0.32\textwidth}
\includegraphics[width=\textwidth,height = 0.18\textheight]{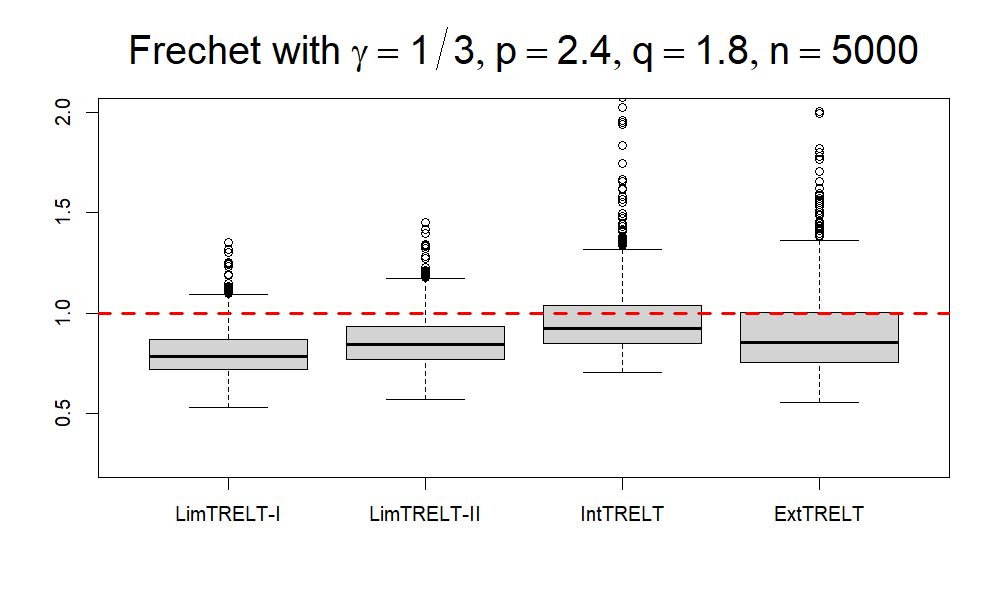}
\end{minipage}
\begin{minipage}[b]{0.32\textwidth}
\includegraphics[width=\textwidth,height = 0.18\textheight]{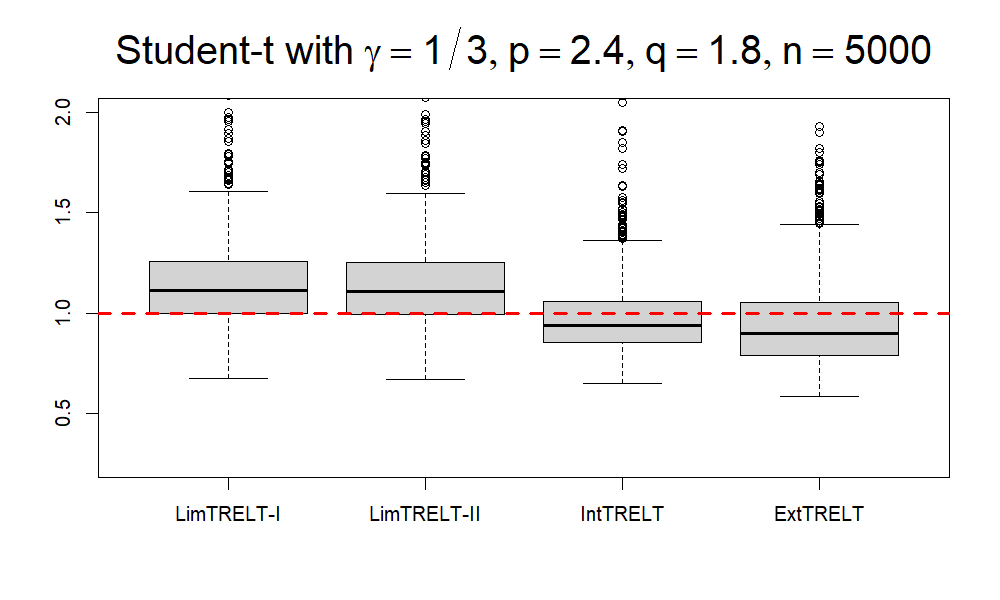}
\end{minipage}
\\
\begin{minipage}[b]{0.32\textwidth}
\includegraphics[width=\textwidth,height = 0.18\textheight]{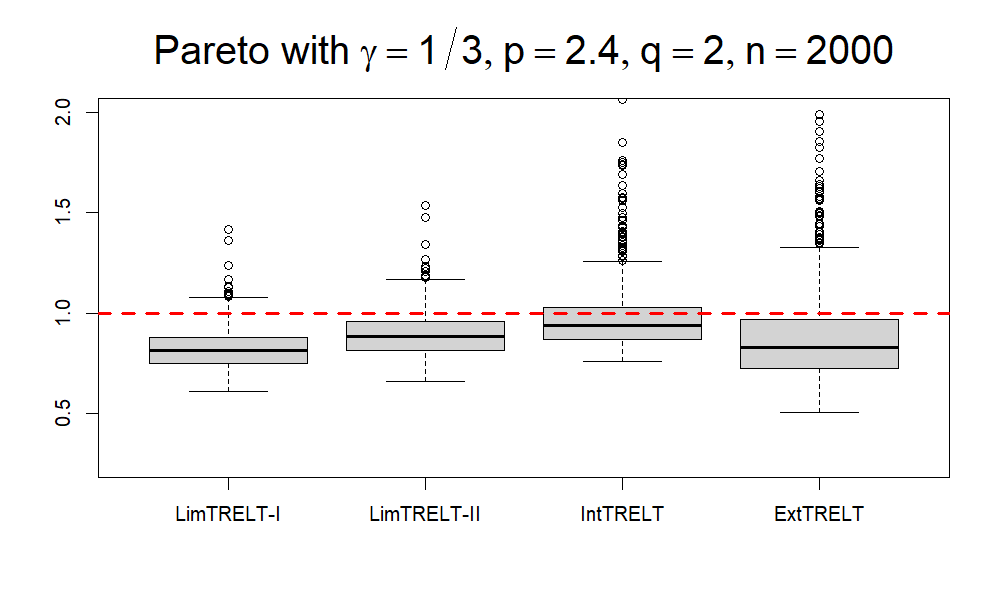}
\end{minipage}
\begin{minipage}[b]{0.32\textwidth}
\includegraphics[width=\textwidth,height = 0.18\textheight]{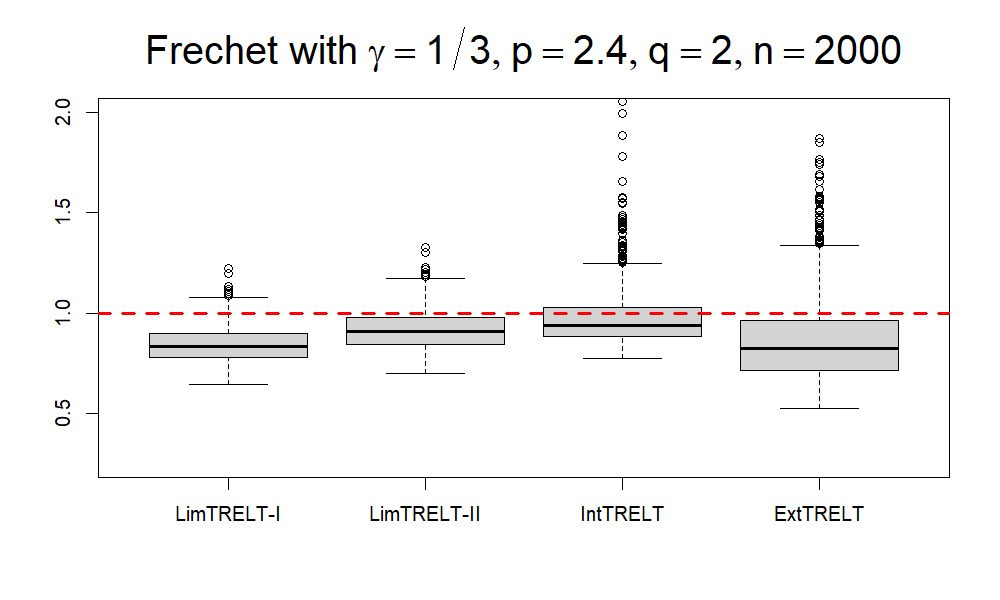}
\end{minipage}
\begin{minipage}[b]{0.32\textwidth}
\includegraphics[width=\textwidth,height = 0.18\textheight]{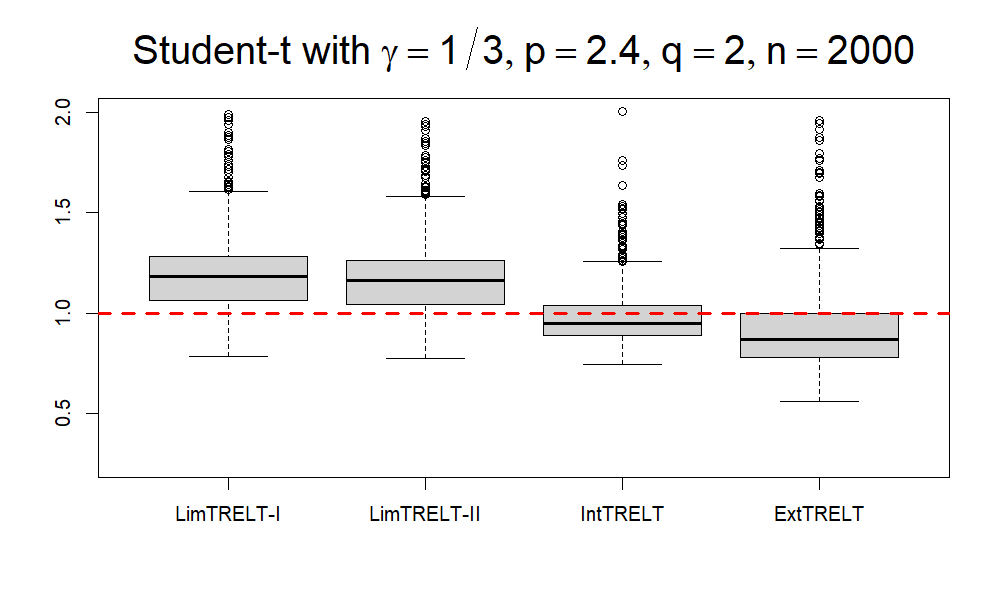}
\end{minipage}
\\
\begin{minipage}[b]{0.32\textwidth}
\includegraphics[width=\textwidth,height = 0.18\textheight]{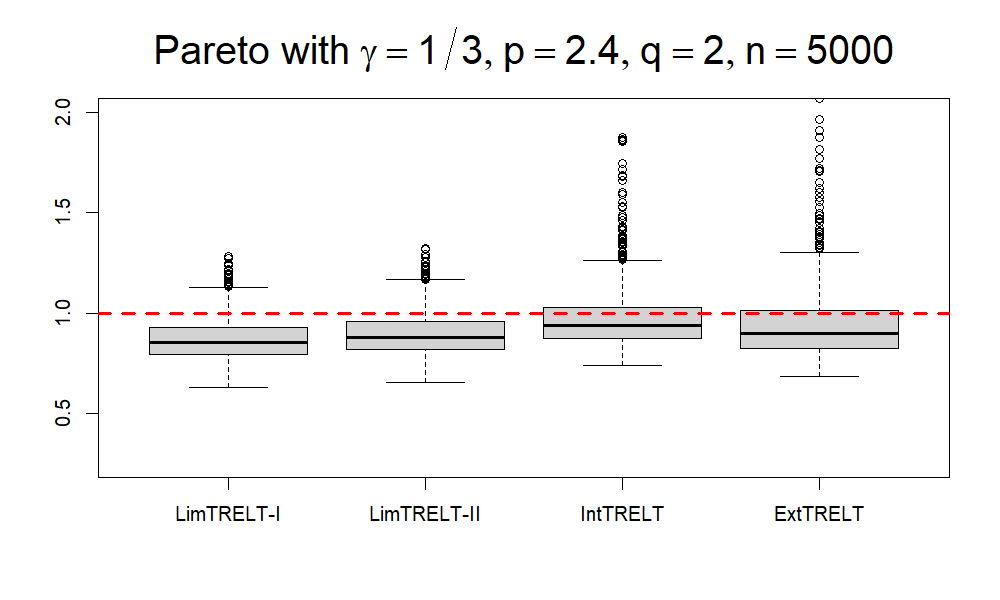}
\end{minipage}
\begin{minipage}[b]{0.32\textwidth}
\includegraphics[width=\textwidth,height = 0.18\textheight]{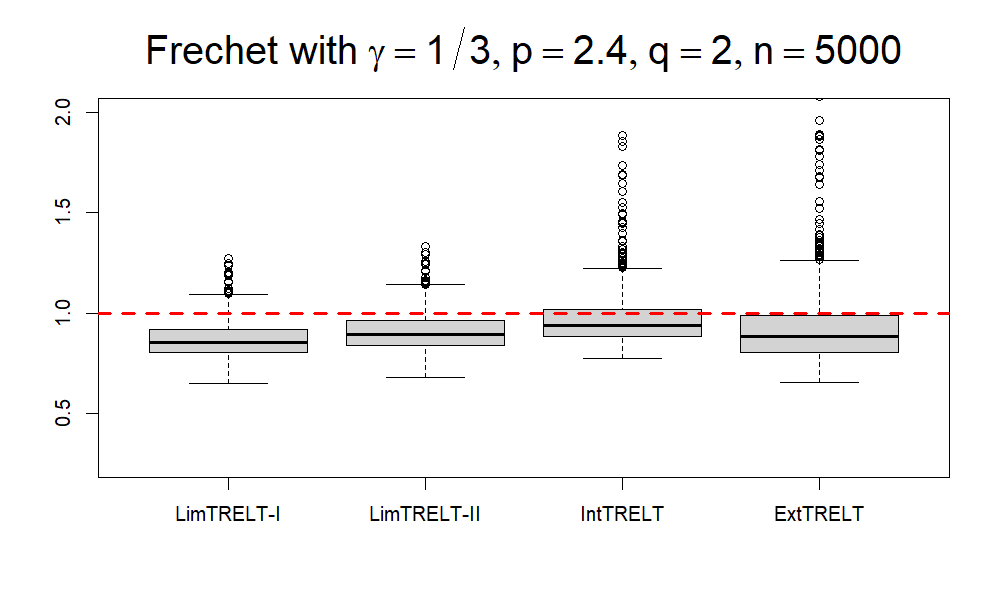}
\end{minipage}
\begin{minipage}[b]{0.32\textwidth}
\includegraphics[width=\textwidth,height = 0.18\textheight]{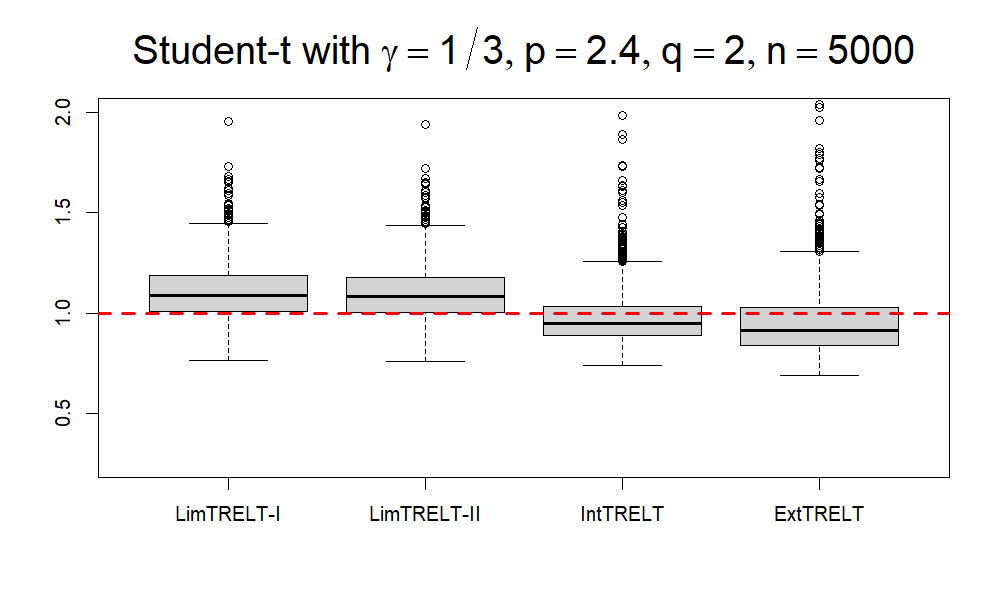}
\end{minipage}
\caption{The boxplots of LimTRELT-I, LimTRELT-II, IntTRELT and ExtTRELT for Pareto (left column), Fr$\acute{e}$chet (middle column) and Student-$t$ (right column) distributions with $\gamma = 1/3$. The boxplots in the top two lines are drawn for $p=2.4,q=1.8$ with $n=2000,5000$ while the boxplots in the bottom two lines are drawn for $p=2.4,q=2$ with $n=2000,5000$ respectively.}
\label{Fig:trelt_gam3333}
\end{figure}

\begin{figure}[htbp]
\centering
\begin{minipage}[b]{0.32\textwidth}
\includegraphics[width=\textwidth,height = 0.18\textheight]{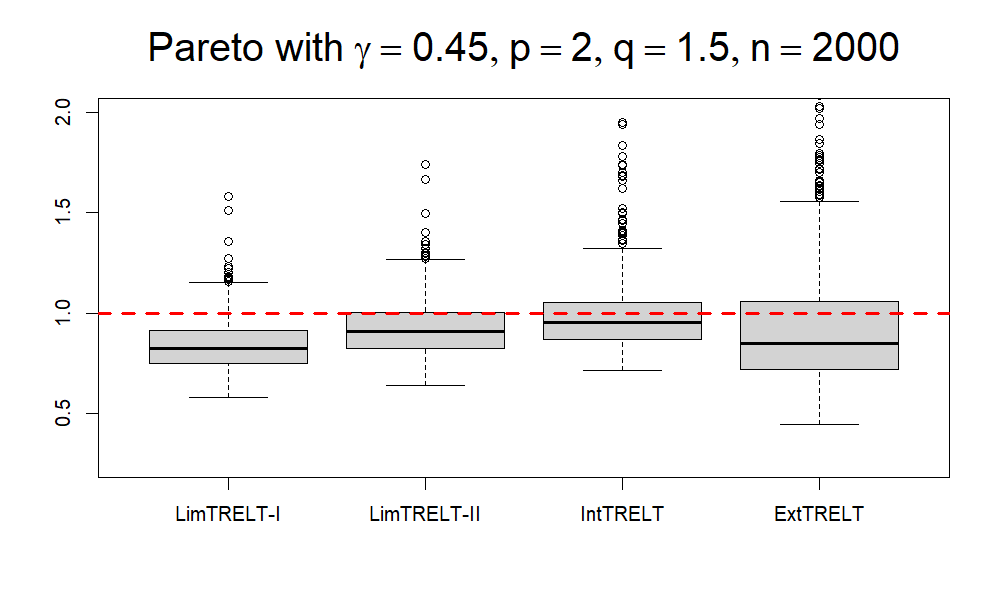}
\end{minipage}
\begin{minipage}[b]{0.32\textwidth}
\includegraphics[width=\textwidth,height = 0.18\textheight]{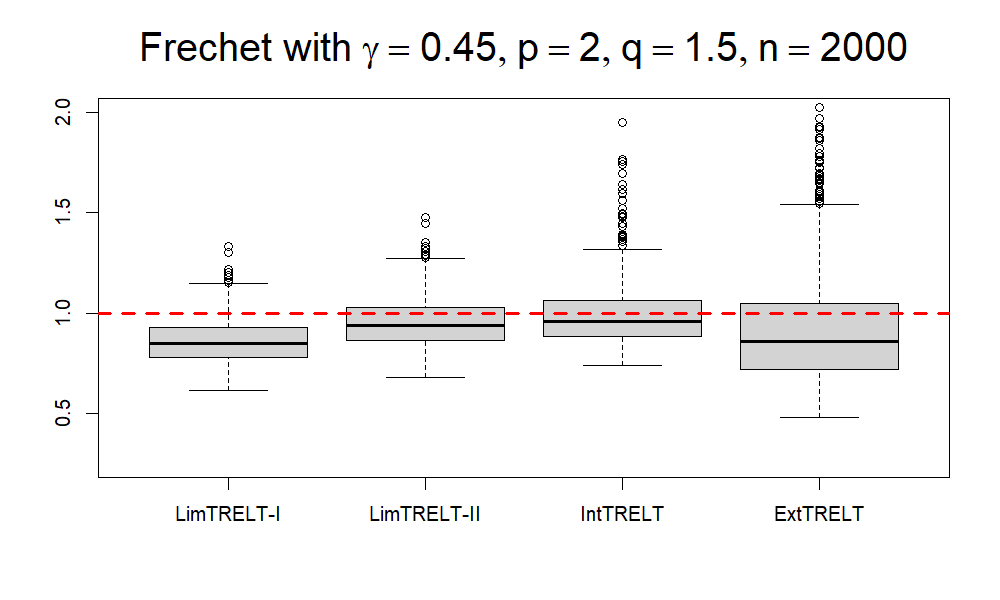}
\end{minipage}
\begin{minipage}[b]{0.32\textwidth}
\includegraphics[width=\textwidth,height = 0.18\textheight]{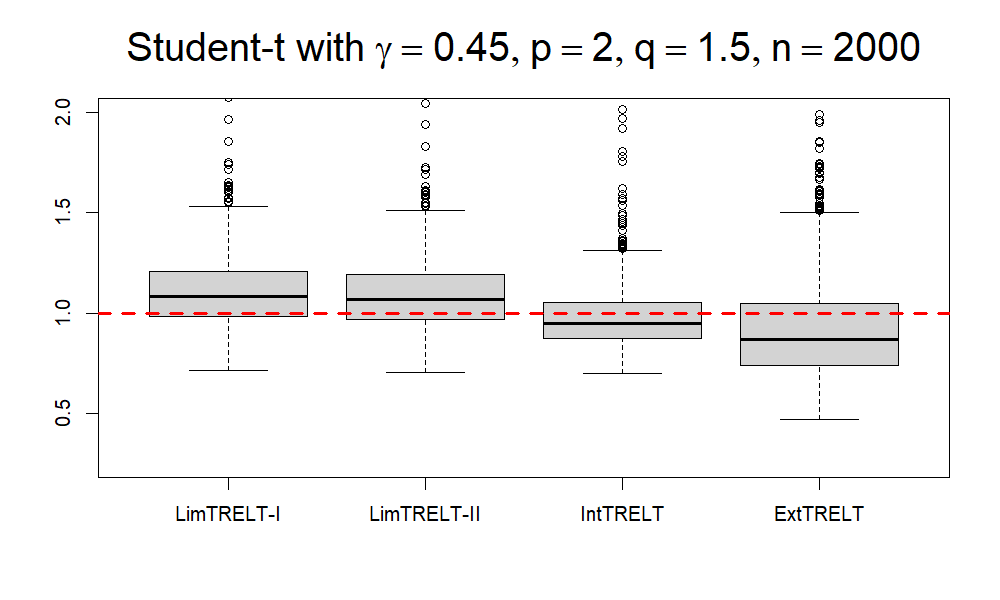}
\end{minipage}
\\
\begin{minipage}[b]{0.32\textwidth}
\includegraphics[width=\textwidth,height = 0.18\textheight]{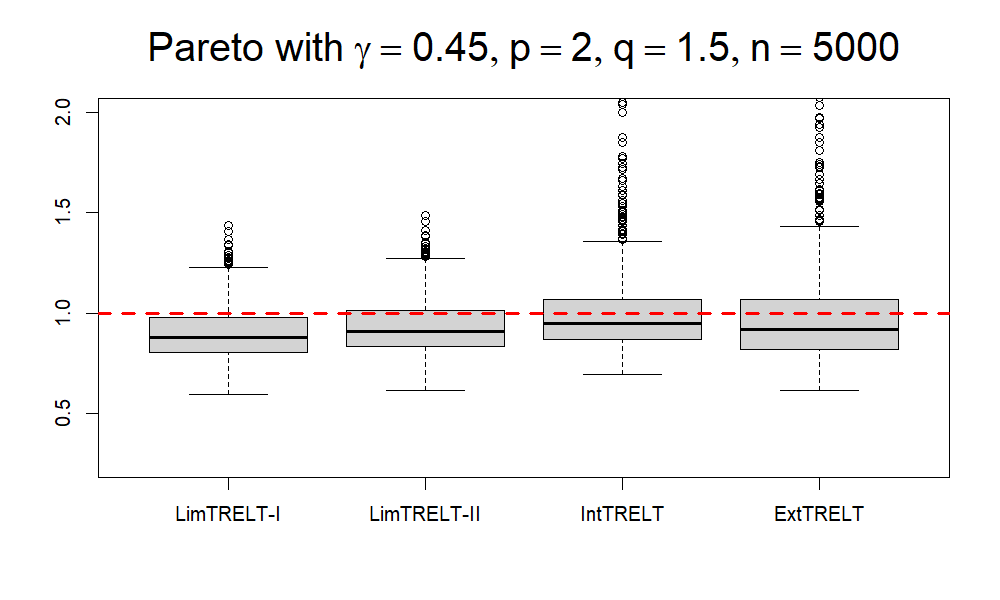}
\end{minipage}
\begin{minipage}[b]{0.32\textwidth}
\includegraphics[width=\textwidth,height = 0.18\textheight]{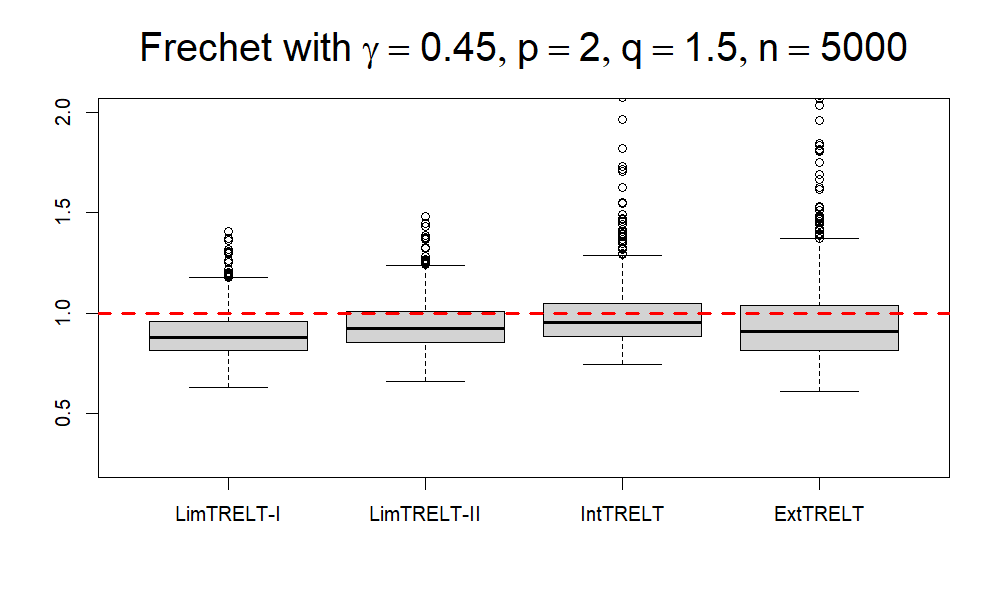}
\end{minipage}
\begin{minipage}[b]{0.32\textwidth}
\includegraphics[width=\textwidth,height = 0.18\textheight]{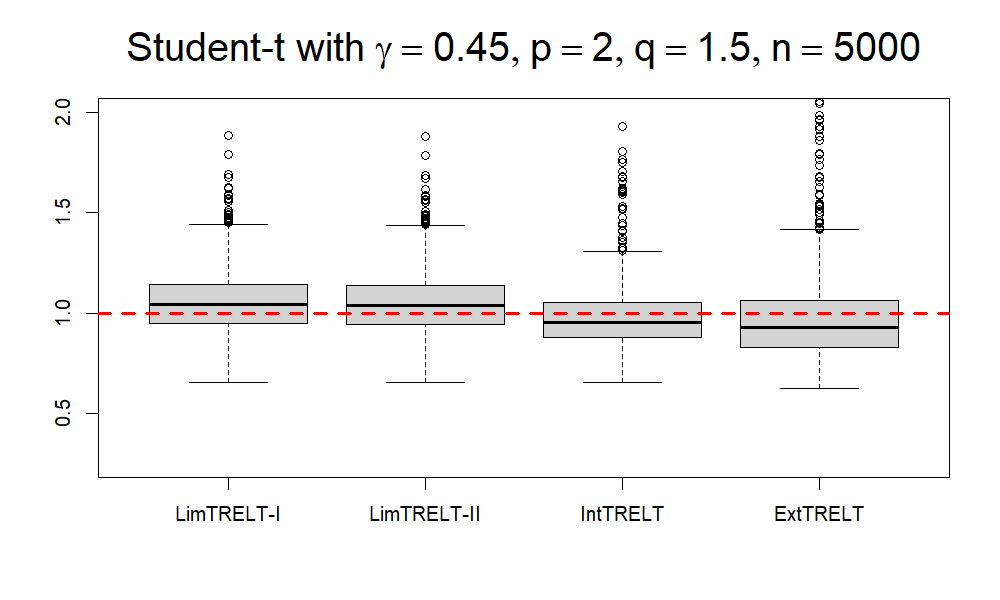}
\end{minipage}
\\
\begin{minipage}[b]{0.32\textwidth}
\includegraphics[width=\textwidth,height = 0.18\textheight]{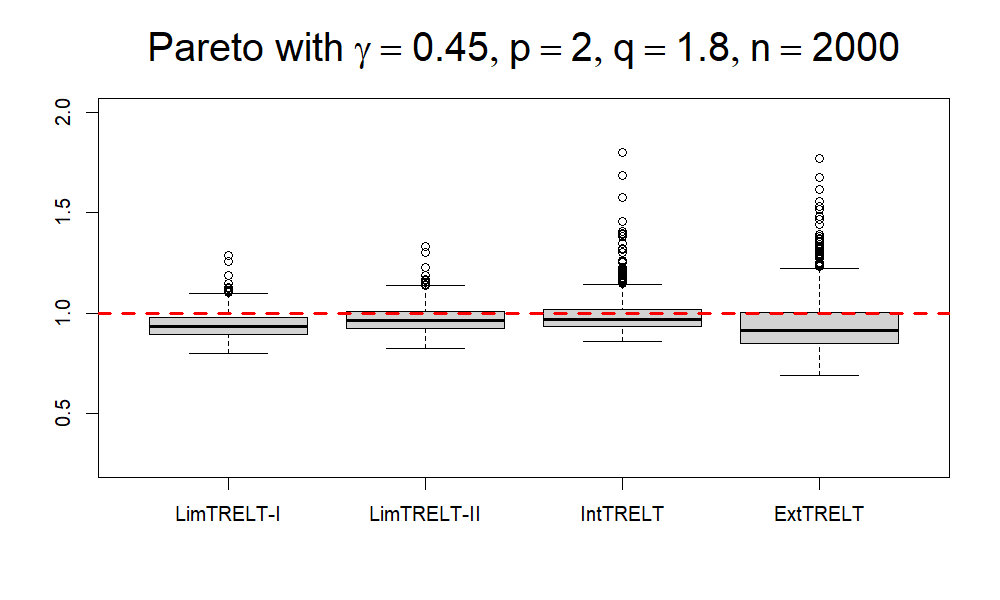}
\end{minipage}
\begin{minipage}[b]{0.32\textwidth}
\includegraphics[width=\textwidth,height = 0.18\textheight]{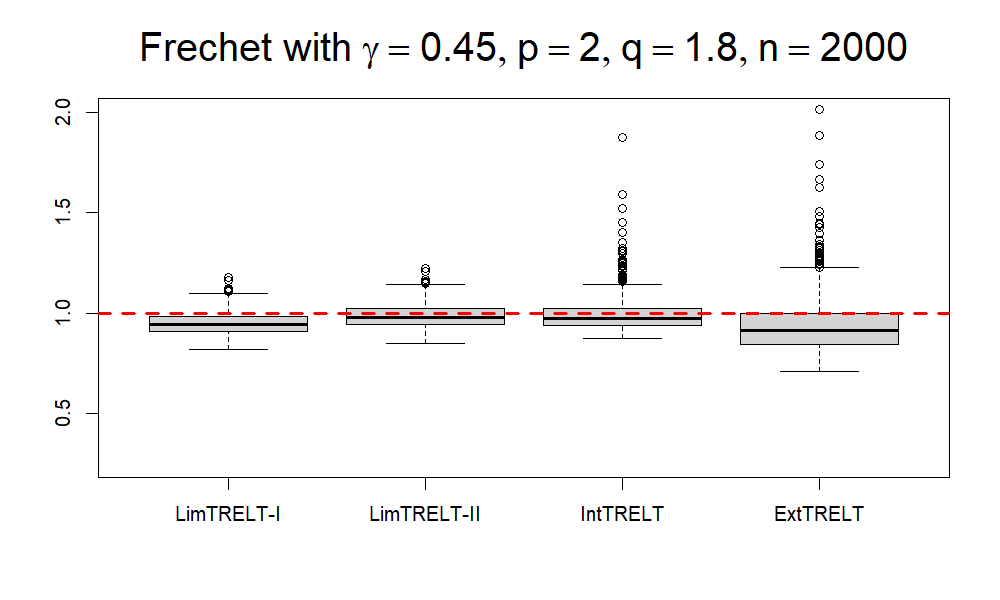}
\end{minipage}
\begin{minipage}[b]{0.32\textwidth}
\includegraphics[width=\textwidth,height = 0.18\textheight]{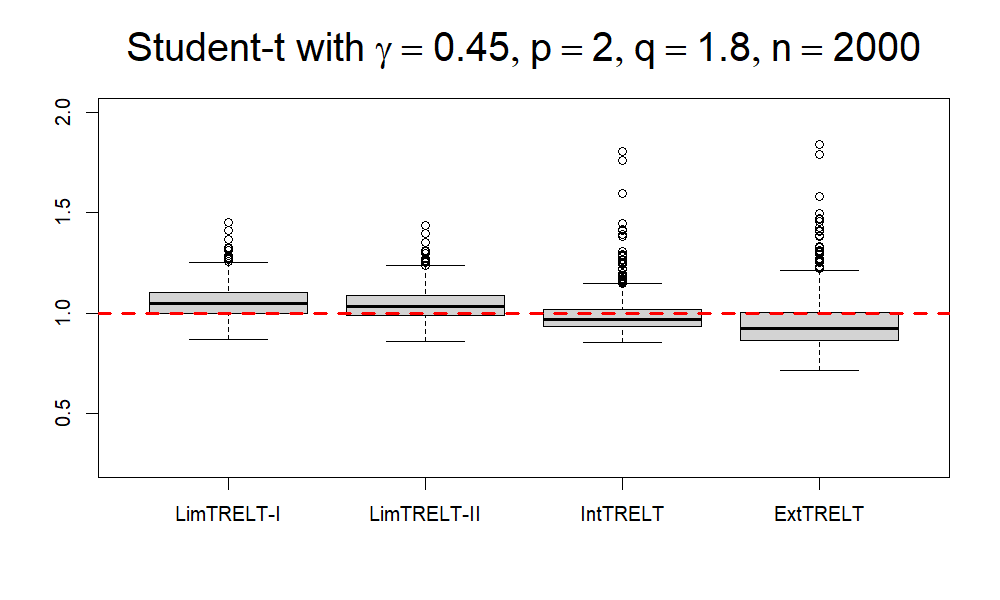}
\end{minipage}
\\
\begin{minipage}[b]{0.32\textwidth}
\includegraphics[width=\textwidth,height = 0.18\textheight]{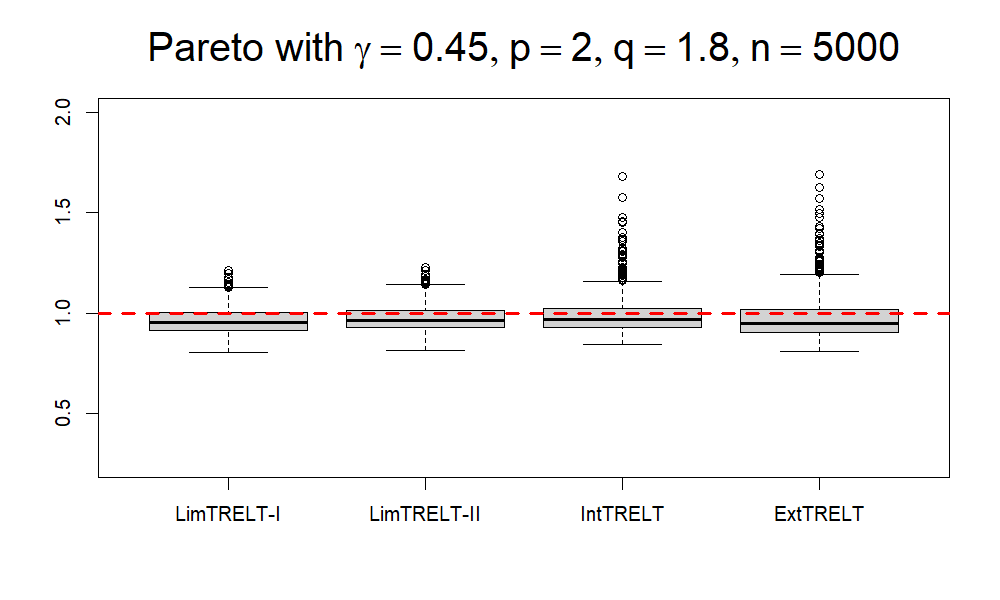}
\end{minipage}
\begin{minipage}[b]{0.32\textwidth}
\includegraphics[width=\textwidth,height = 0.18\textheight]{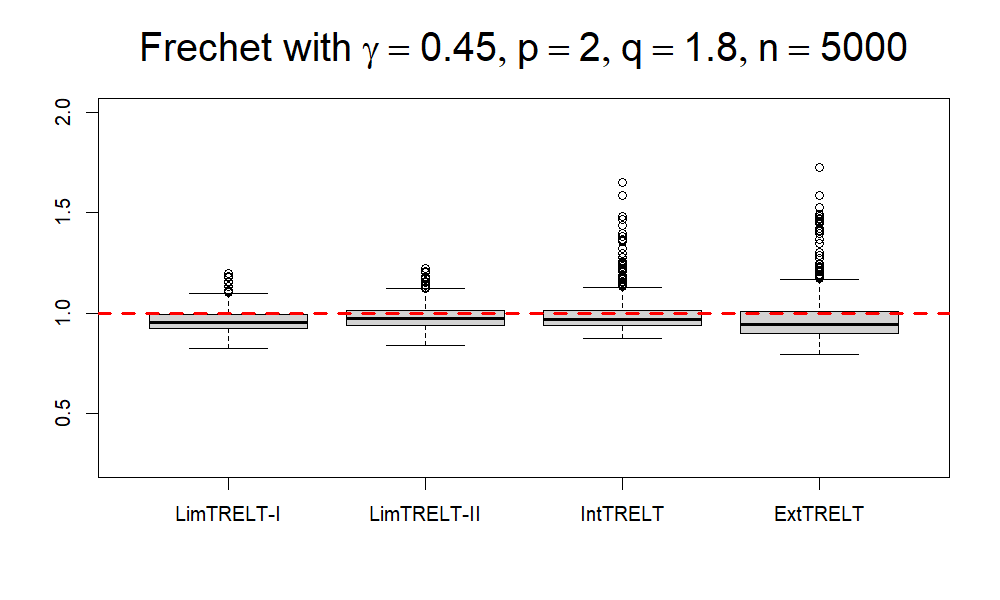}
\end{minipage}
\begin{minipage}[b]{0.32\textwidth}
\includegraphics[width=\textwidth,height = 0.18\textheight]{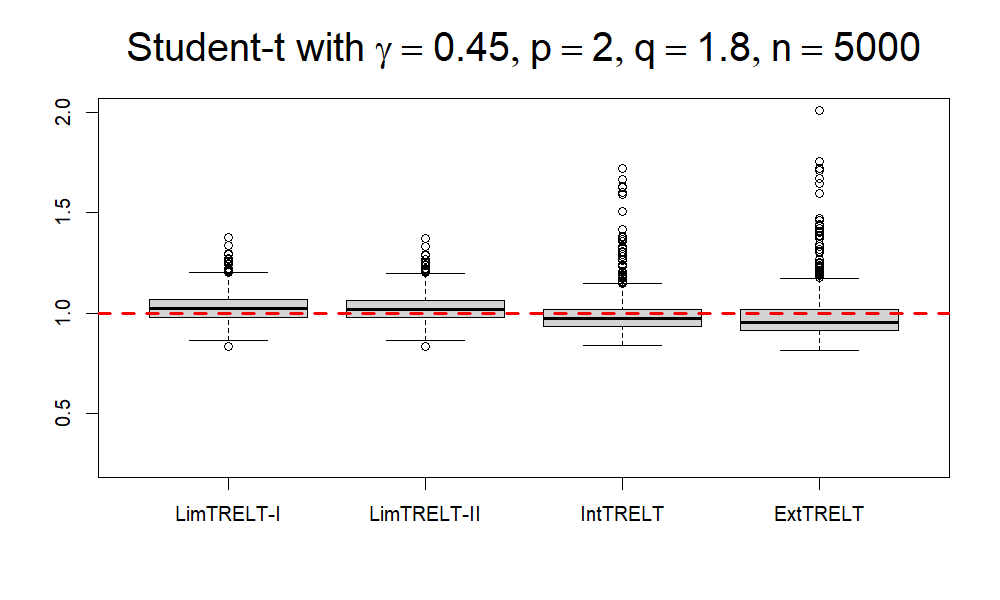}
\end{minipage}
\caption{The boxplots of LimTRELT-I, LimTRELT-II, IntTRELT and ExtTRELT for Pareto (left column), Fr$\acute{e}$chet (middle column) and Student-$t$ (right column) distributions with $\gamma = 0.45$. The boxplots in the top two lines are drawn for $p=2,q=1.5$ with $n=2000,5000$ while the boxplots in the bottom two lines are drawn for $p=2,q=1.8$ with $n=2000,5000$ respectively.}
\label{Fig:trelt_gam45}
\end{figure}

\begin{figure}[htbp]
\centering
\begin{minipage}[b]{0.32\textwidth}
\includegraphics[width=\textwidth,height = 0.18\textheight]{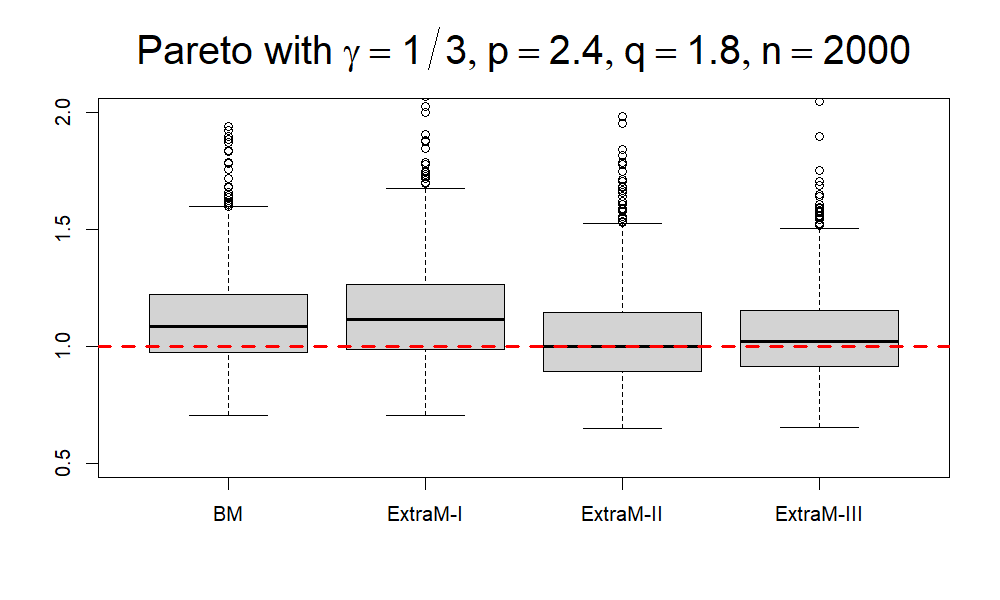}
\end{minipage}
\begin{minipage}[b]{0.32\textwidth}
\includegraphics[width=\textwidth,height = 0.18\textheight]{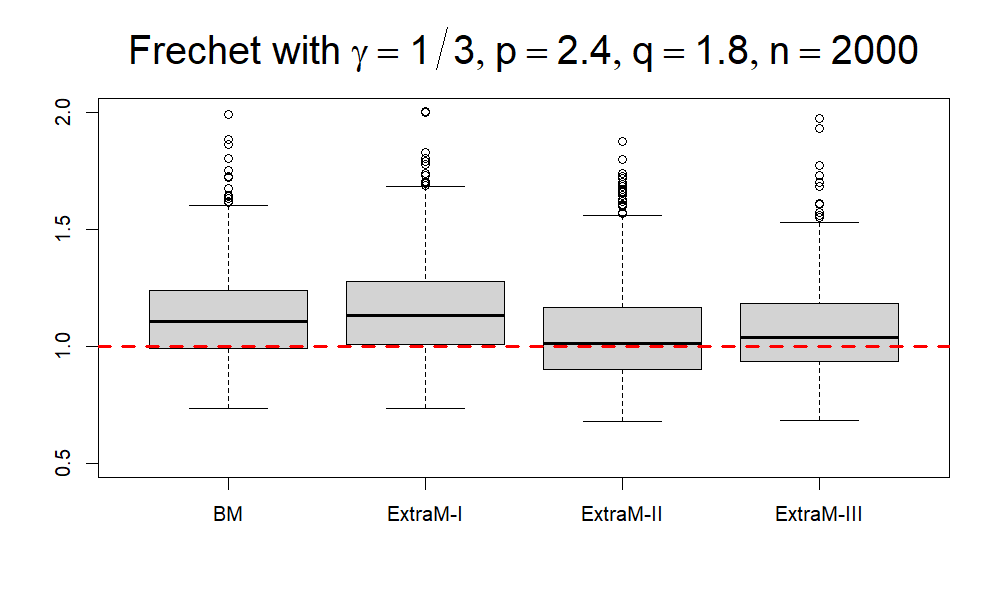}
\end{minipage}
\begin{minipage}[b]{0.32\textwidth}
\includegraphics[width=\textwidth,height = 0.18\textheight]{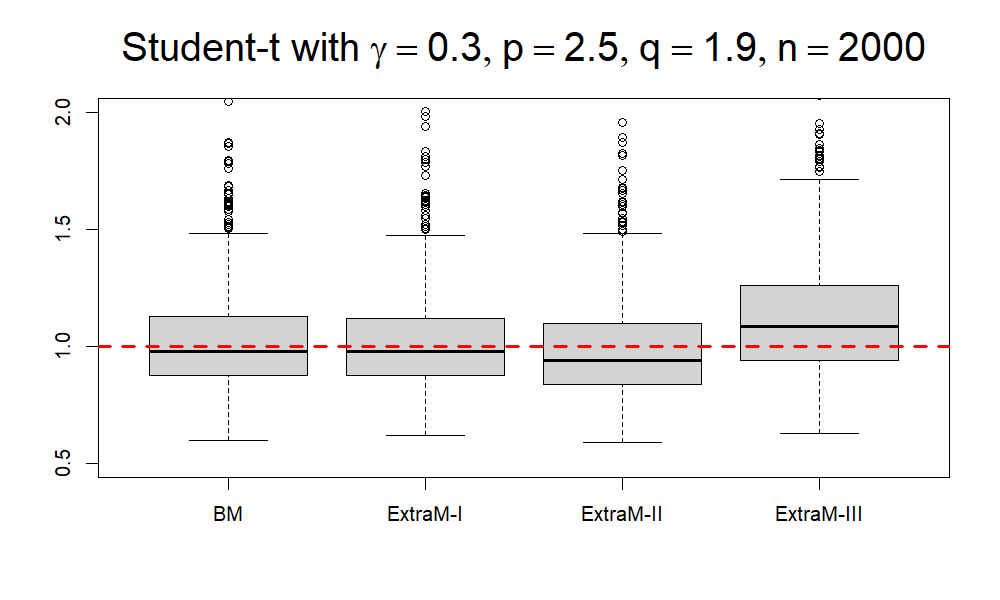}
\end{minipage}
\\
\begin{minipage}[b]{0.32\textwidth}
\includegraphics[width=\textwidth,height = 0.18\textheight]{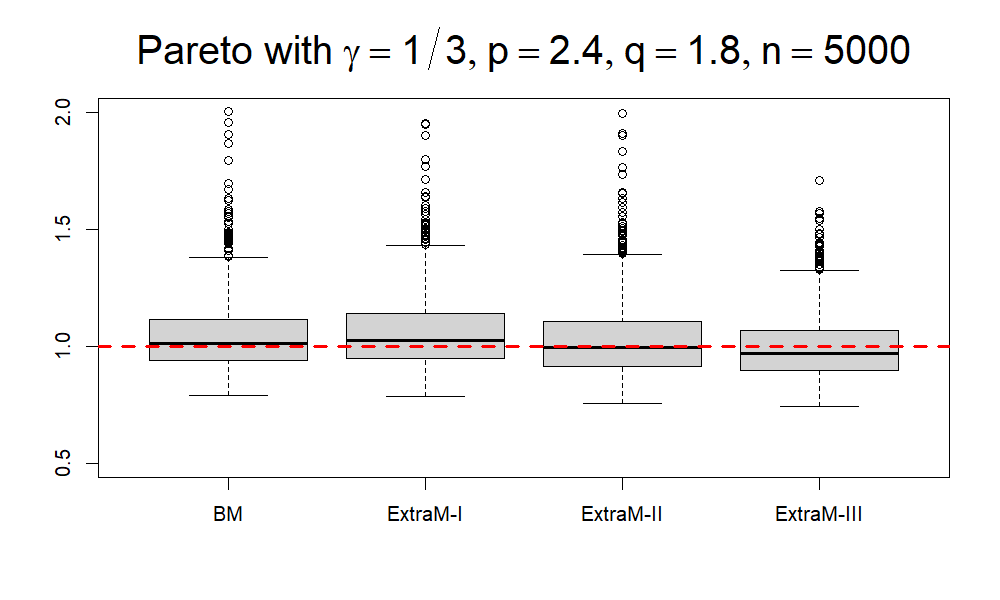}
\end{minipage}
\begin{minipage}[b]{0.32\textwidth}
\includegraphics[width=\textwidth,height = 0.18\textheight]{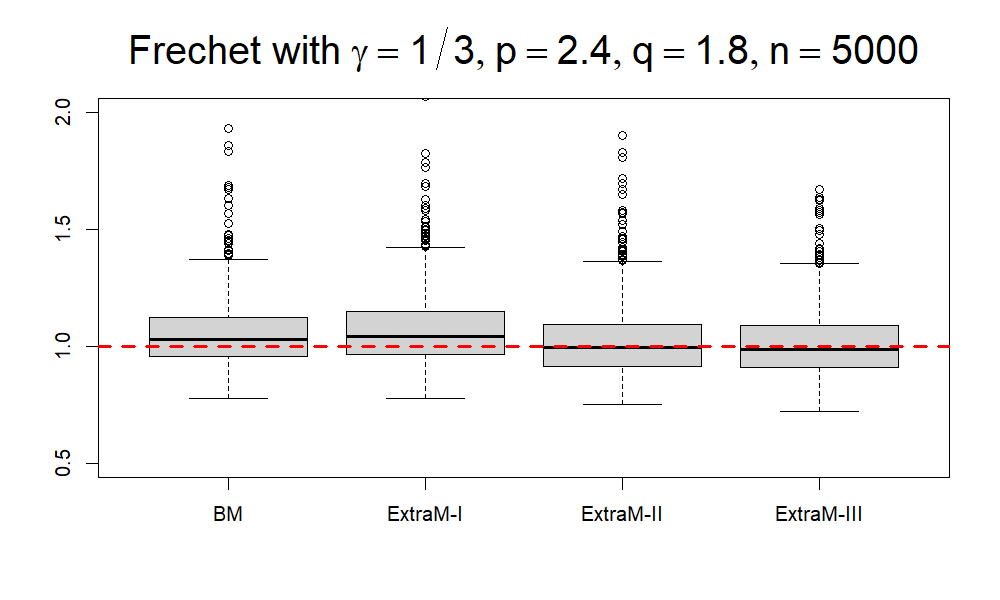}
\end{minipage}
\begin{minipage}[b]{0.32\textwidth}
\includegraphics[width=\textwidth,height = 0.18\textheight]{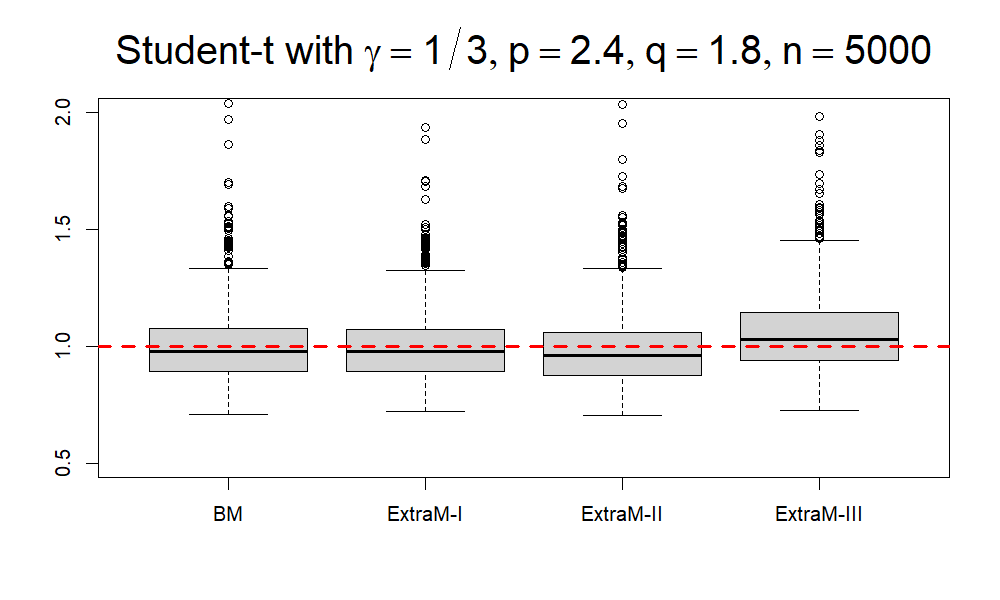}
\end{minipage}
\\
\begin{minipage}[b]{0.32\textwidth}
\includegraphics[width=\textwidth,height = 0.18\textheight]{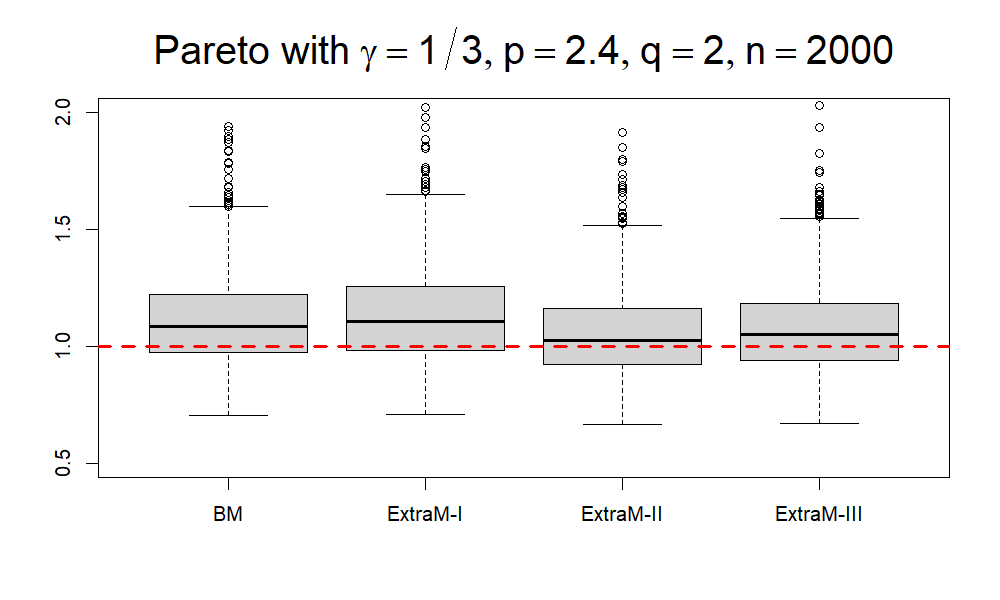}
\end{minipage}
\begin{minipage}[b]{0.32\textwidth}
\includegraphics[width=\textwidth,height = 0.18\textheight]{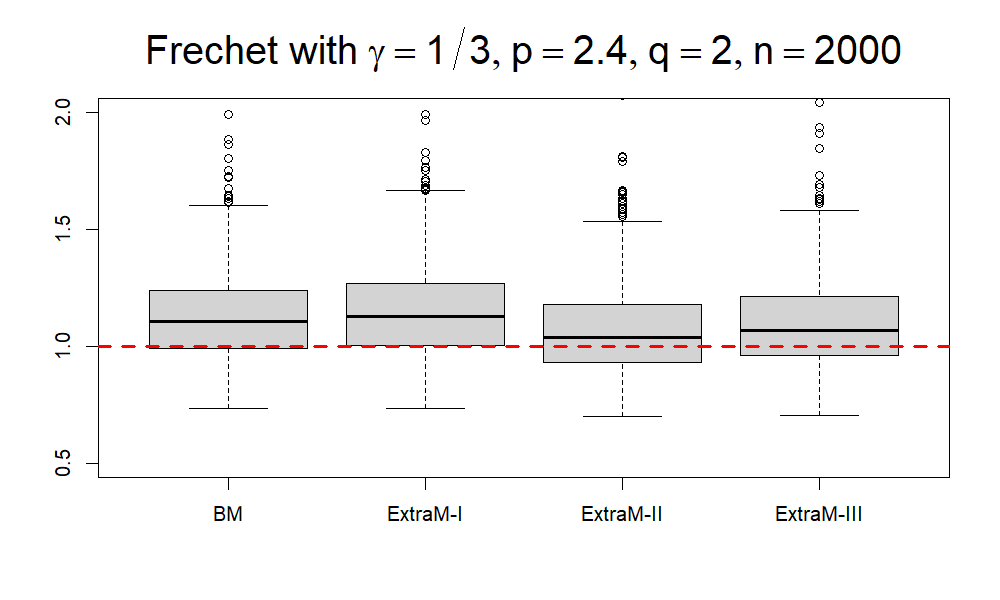}
\end{minipage}
\begin{minipage}[b]{0.32\textwidth}
\includegraphics[width=\textwidth,height = 0.18\textheight]{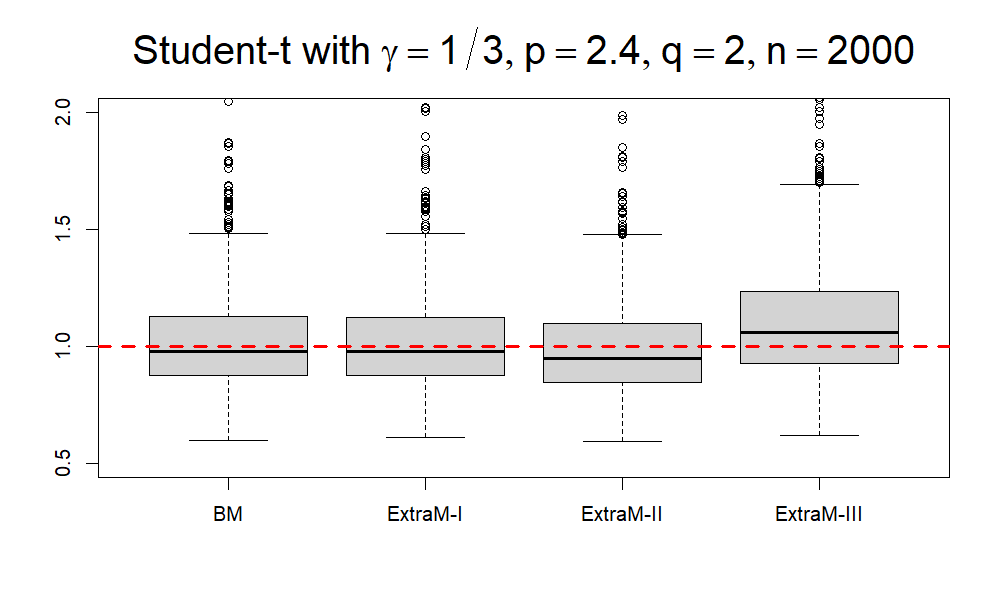}
\end{minipage}
\\
\begin{minipage}[b]{0.32\textwidth}
\includegraphics[width=\textwidth,height = 0.18\textheight]{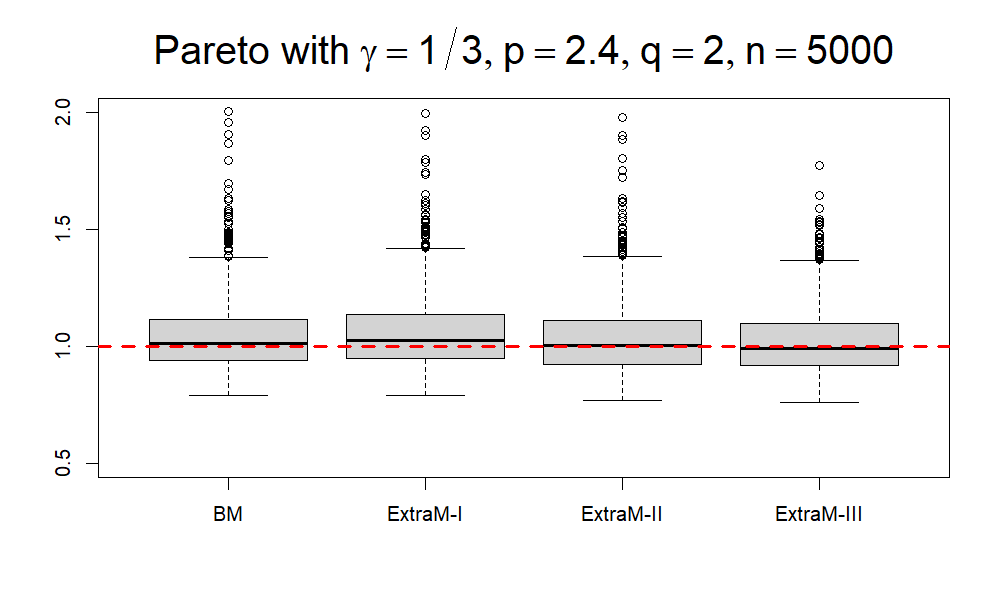}
\end{minipage}
\begin{minipage}[b]{0.32\textwidth}
\includegraphics[width=\textwidth,height = 0.18\textheight]{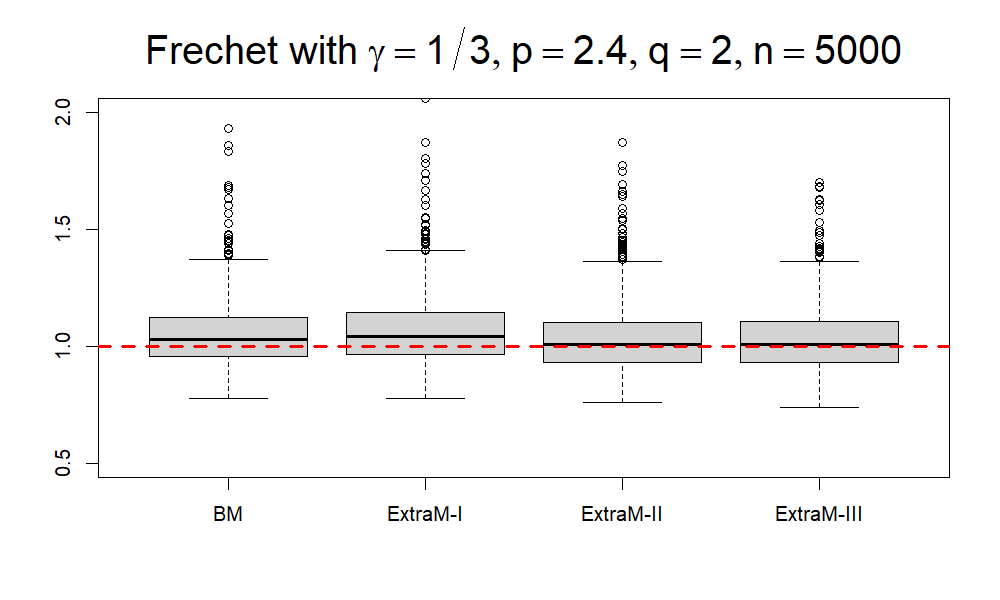}
\end{minipage}
\begin{minipage}[b]{0.32\textwidth}
\includegraphics[width=\textwidth,height = 0.18\textheight]{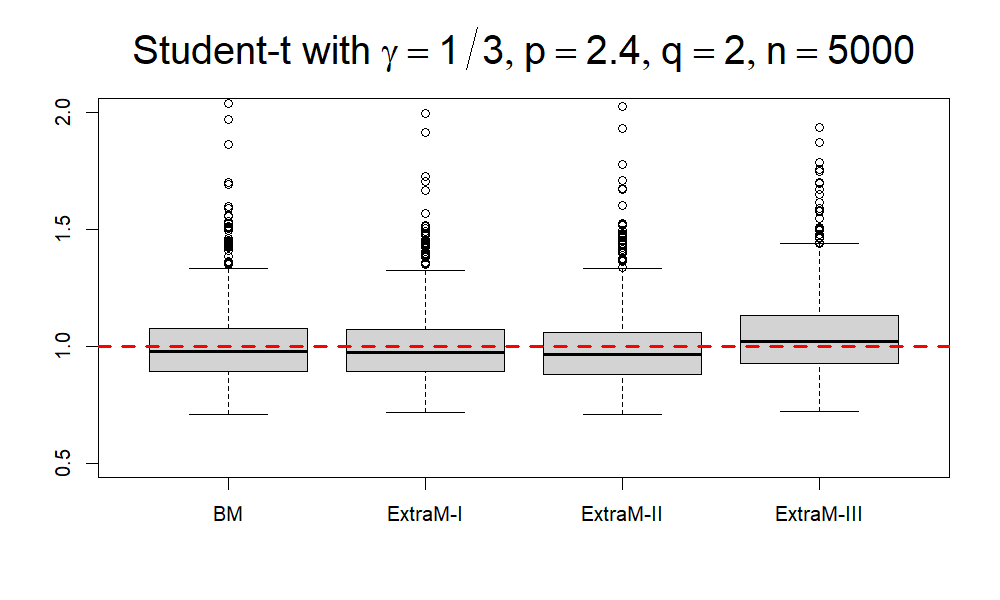}
\end{minipage}
\caption{The boxplots of BM, ExtraM-I, ExtraM-II and ExtraM-III for Pareto (left column), Fr$\acute{e}$chet (middle column) and Student-$t$ (right column) distributions with $\gamma = 1/3$. The boxplots in the top two lines are drawn for $p=2.4,q=1.8$ with $n=2000,5000$ while the boxplots in the bottom two lines are drawn for $p=2.4,q=2$ with $n=2000,5000$ respectively.}
\label{Fig:theta_gam3333}
\end{figure}

\begin{figure}[htbp]
\centering
\begin{minipage}[b]{0.32\textwidth}
\includegraphics[width=\textwidth,height = 0.18\textheight]{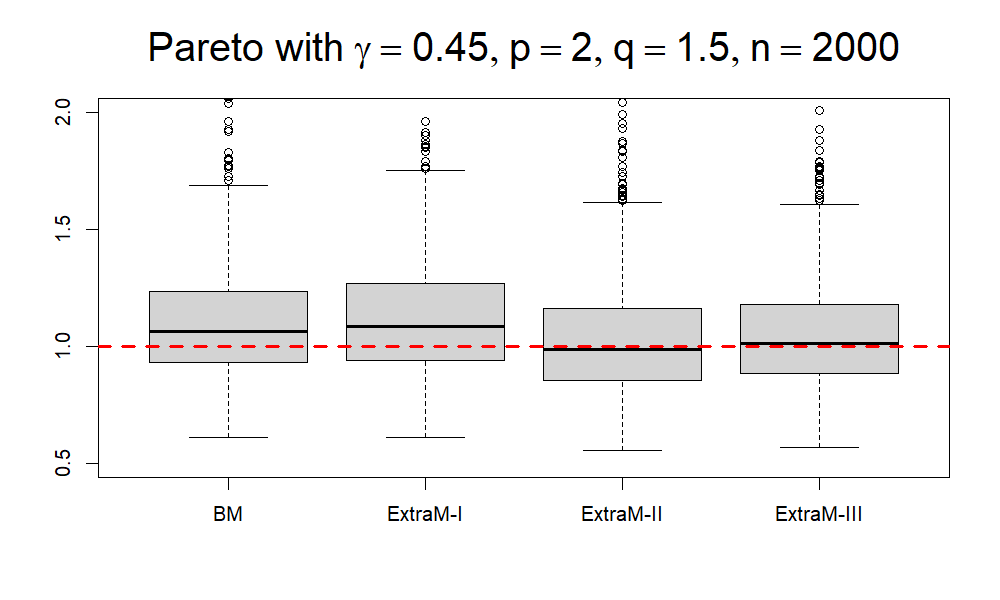}
\end{minipage}
\begin{minipage}[b]{0.32\textwidth}
\includegraphics[width=\textwidth,height = 0.18\textheight]{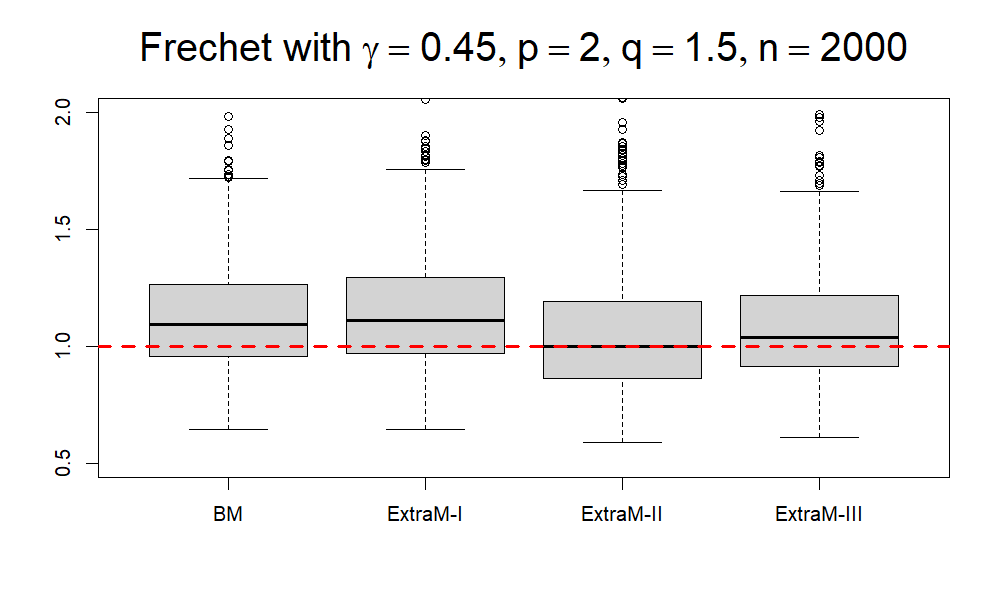}
\end{minipage}
\begin{minipage}[b]{0.32\textwidth}
\includegraphics[width=\textwidth,height = 0.18\textheight]{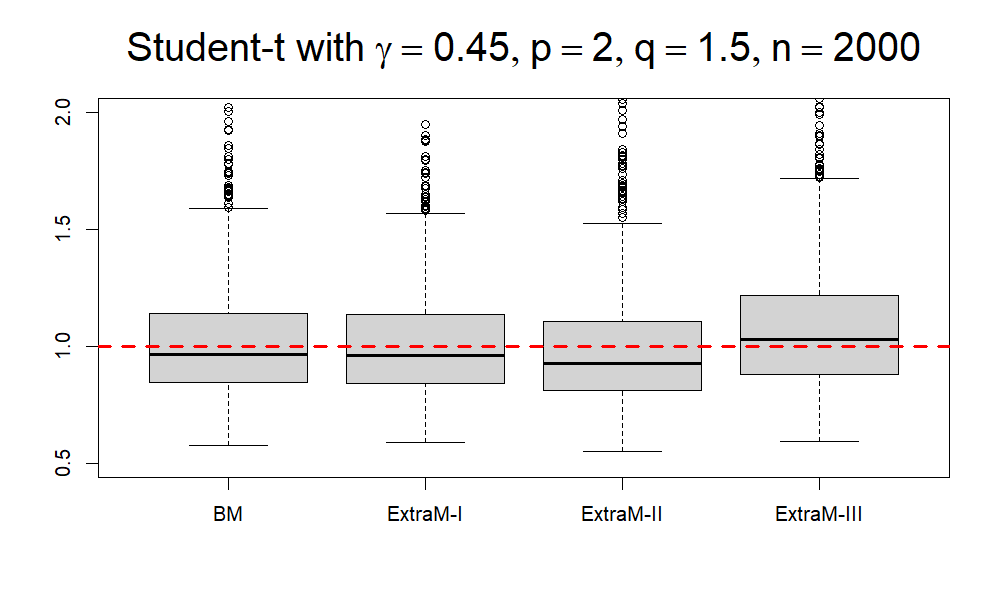}
\end{minipage}
\\
\begin{minipage}[b]{0.32\textwidth}
\includegraphics[width=\textwidth,height = 0.18\textheight]{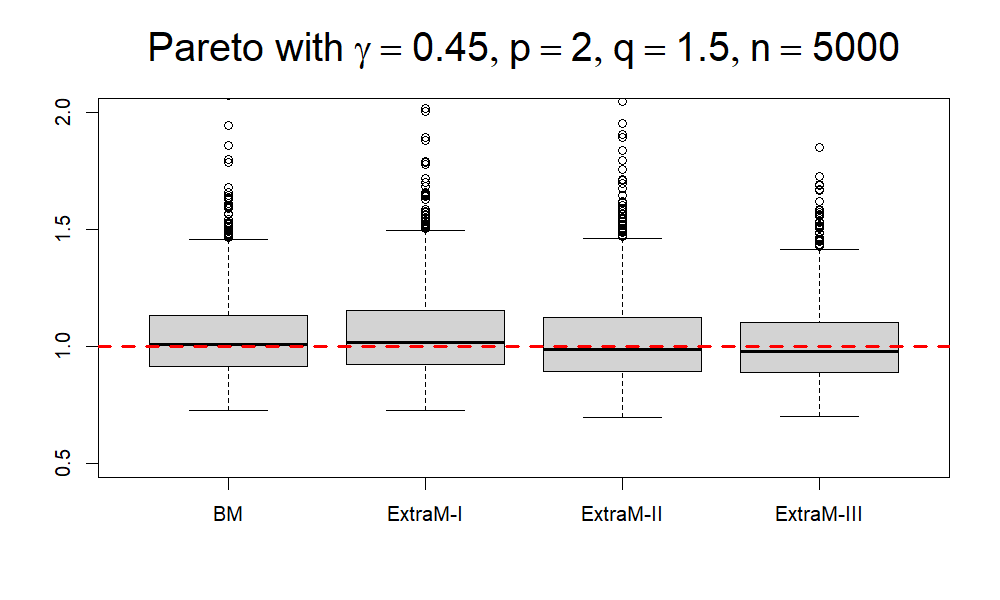}
\end{minipage}
\begin{minipage}[b]{0.32\textwidth}
\includegraphics[width=\textwidth,height = 0.18\textheight]{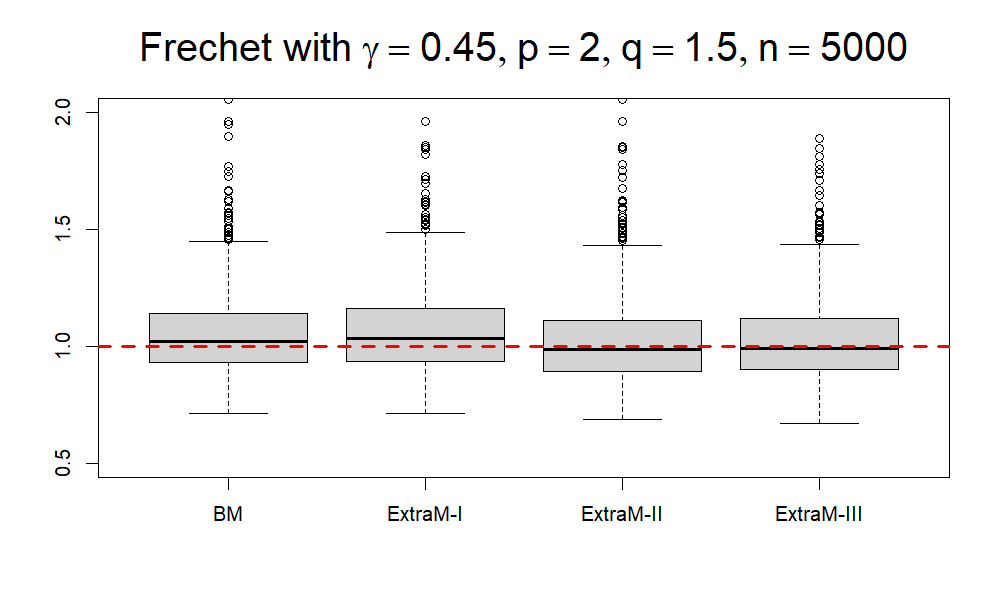}
\end{minipage}
\begin{minipage}[b]{0.32\textwidth}
\includegraphics[width=\textwidth,height = 0.18\textheight]{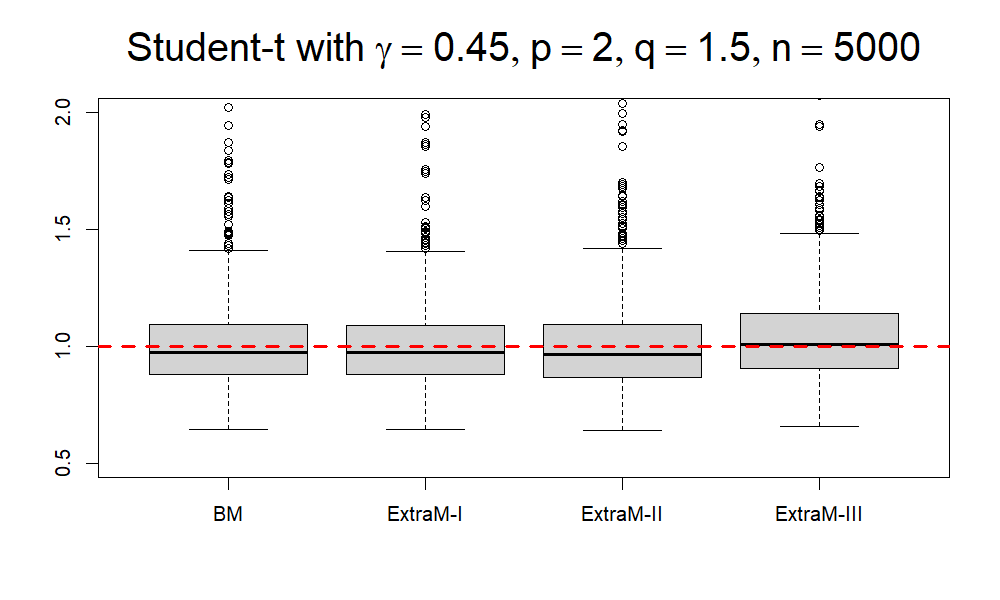}
\end{minipage}
\\
\begin{minipage}[b]{0.32\textwidth}
\includegraphics[width=\textwidth,height = 0.18\textheight]{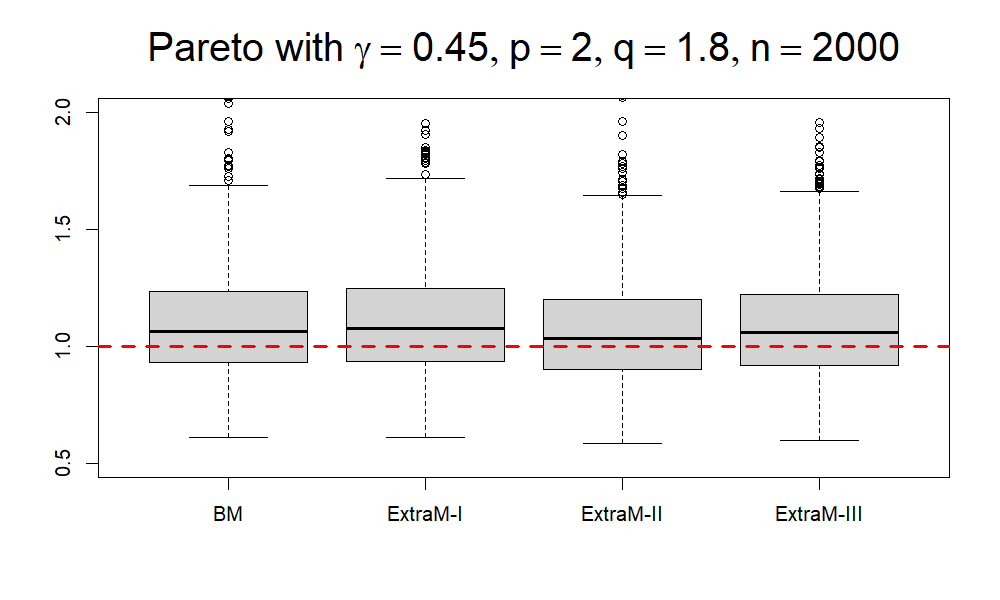}
\end{minipage}
\begin{minipage}[b]{0.32\textwidth}
\includegraphics[width=\textwidth,height = 0.18\textheight]{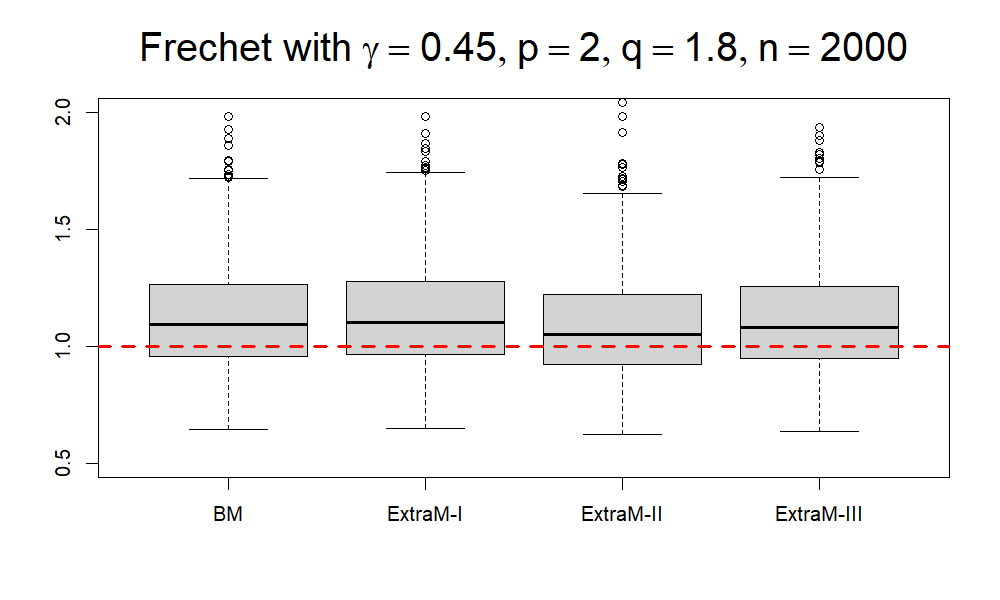}
\end{minipage}
\begin{minipage}[b]{0.32\textwidth}
\includegraphics[width=\textwidth,height = 0.18\textheight]{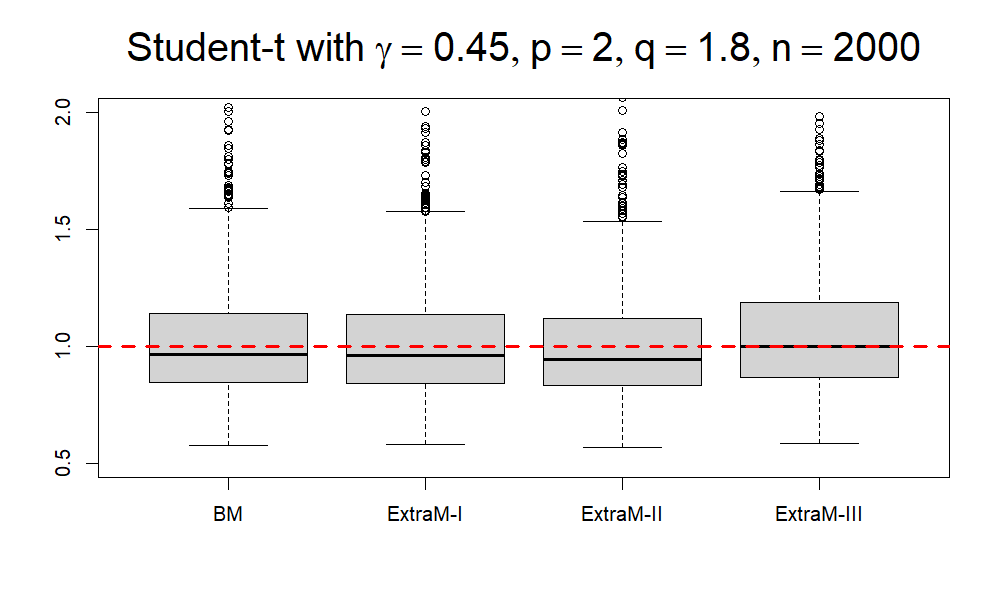}
\end{minipage}
\\
\begin{minipage}[b]{0.32\textwidth}
\includegraphics[width=\textwidth,height = 0.18\textheight]{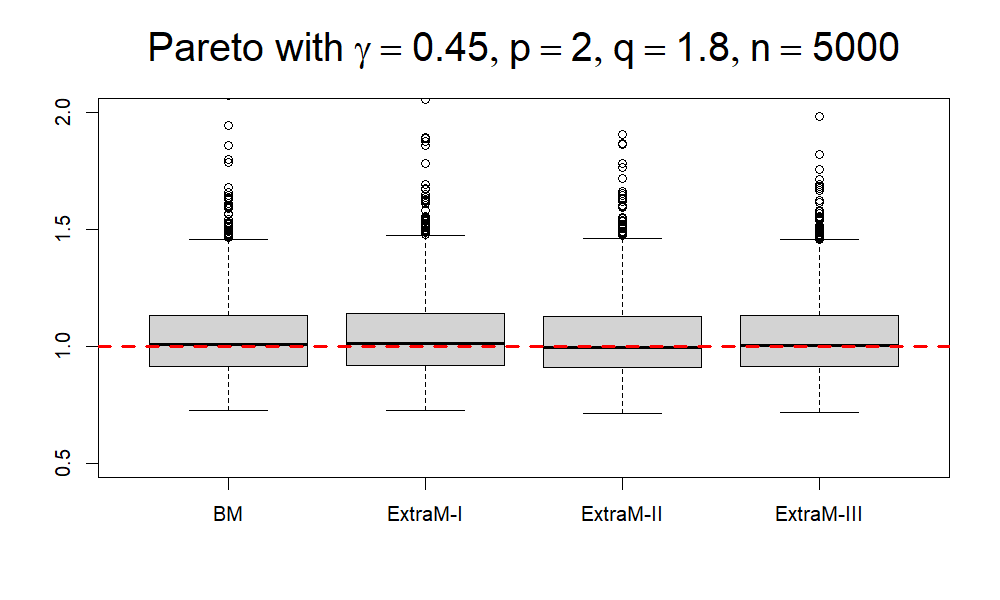}
\end{minipage}
\begin{minipage}[b]{0.32\textwidth}
\includegraphics[width=\textwidth,height = 0.18\textheight]{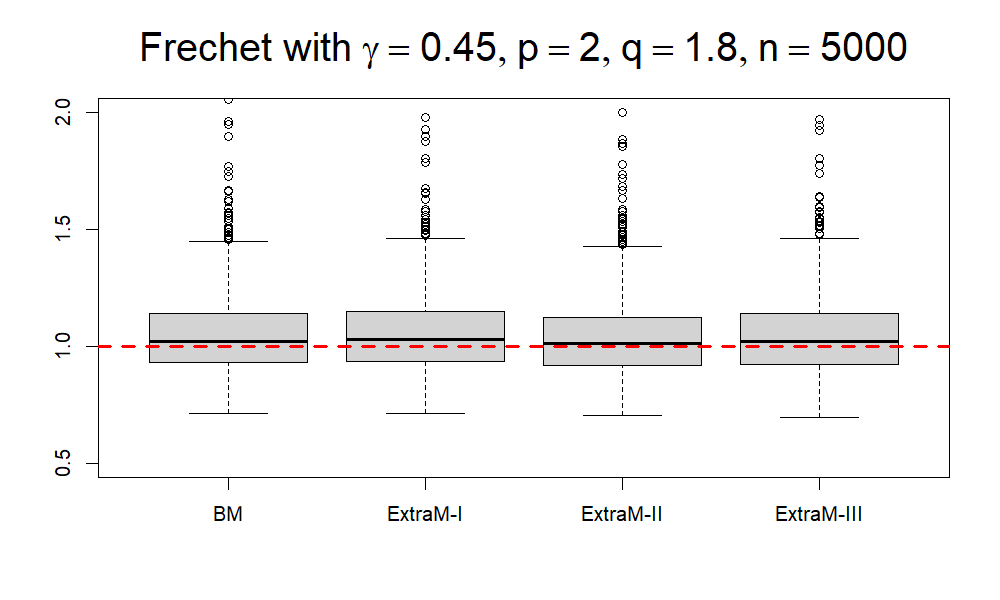}
\end{minipage}
\begin{minipage}[b]{0.32\textwidth}
\includegraphics[width=\textwidth,height = 0.18\textheight]{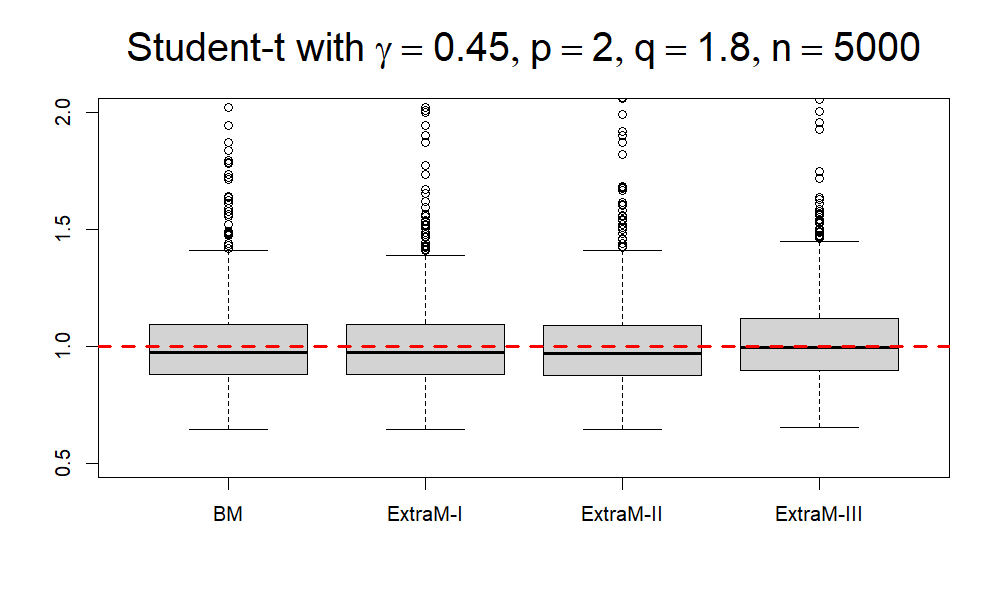}
\end{minipage}
\caption{The boxplots of BM, ExtraM-I, ExtraM-II and ExtraM-III for Pareto (left column), Fr$\acute{e}$chet (middle column) and Student-$t$ (right column) distributions with $\gamma = 0.45$. The boxplots in the top two lines are drawn for $p=2,q=1.5$ with $n=2000,5000$ while the boxplots in the bottom two lines are drawn for $p=2,q=1.8$ with $n=2000,5000$ respectively.}
\label{Fig:theta_gam45}
\end{figure}

To evaluate the performance of all estimators, we calculate the {\it Mean Squared Relative Error} (MSRE) based on $N$ replications, which is given by,
\begin{equation*}
  {\rm MSRE} = \frac{1}{N} \sum_{i=1}^{N} \left( \frac{\hat \theta_n^{(i)}}{\theta} - 1 \right)^2,
\end{equation*}
where $\hat \theta_n^{(i)}$ is the estimator we are interested in given the simulated data of the $i$-th replication, and $\theta$ is the true value. We use \texttt{uniroot} function in \texttt{RStudio} to compute $\theta_p(1-\varepsilon)$ via relationship \eqref{eq:lp-quantile2} and to compute $\Pi_{p,q}(\varepsilon)$ via relationship \eqref{eq:ctrelt}. We summarize the values of MSREs in Tables \ref{tab:msre_c3333}, \ref{tab:msre_c45}, \ref{tab:msre_theta3333} and \ref{tab:msre_theta45}. To reflect the performance intuitively, we additionally display the comparable boxplots in Figures \ref{Fig:trelt_gam3333}, \ref{Fig:trelt_gam45}, \ref{Fig:theta_gam3333} and \ref{Fig:theta_gam45}. It is immediately visually apparent that as the sample size increases, all these estimators become more and more concentrated, whose MSREs also show an overall decreasing trend with increasing samples.

\textbf{The analyses of methods: LimTRELT-I, LimTRELT-II, IntTRELT and ExtTRELT.} In the intermediate case, it is evident from an intuitive observation of Figure \ref{Fig:trelt_gam3333} that LimTRELT-I exhibits significantly more bias compared to IntTRELT. This may be attributed to the fact that LimTRELT-I is more suitable for smaller levels, whereas the intermediate level $\varepsilon_n$ is not sufficiently small. Furthermore, the MSREs of LimTRELT-I and IntTRELT corroborate this observation, especially for a smaller $\gamma$. The extreme case is somewhat more complex. When compared to LimTRELT-II, both boxplots and MSREs indicate that ExtTRELT presents more dispersion in most cases. This could be due to two main reasons: firstly, the extreme level is small enough to render LimTRELT-II more appropriate and effective, resulting in lower MSREs; secondly, as observed, the BM method does indeed display some bias, which may contribute to the unsatisfactory performance of ExtTRELT, since $\widetilde{\Pi}_{p,q}(\varepsilon'_n)$ is exactly established by substituting BM into \eqref{eq:pi_ext_est}. Of course, there are also some drawbacks for LimTRELT-II, such as bias, especially for Student-$t$ distribution, for example, see plots with titles ``Student-$t$ with $\gamma=1/3, p=2.4, q=1.8, n=2000$" and ``Student-$t$ with $\gamma=1/3, p=2.4, q=2, n=2000$" in Figure \ref{Fig:trelt_gam3333}. Additionally, all four methods demonstrate increased robustness with heavier-tailed populations.

\textbf{The analyses of methods: BM, ExtraM-I, ExtraM-II and ExtraM-III.} First, the empirical performance of ExtraM-I is on par with that of the standard extrapolative method BM, which is not only evident in the boxplots but also reflected in the accurate values of MSREs. However, some bias are observed within the context of small sample size. Second, it seems that the most robust one is the method ExtraM-II, which exhibits minimal bias and relatively smaller MSREs for both large and small sample size. Besides, another well-performed method appears to be ExtraM-III, which has the lowest MSREs in most cases, albeit with some bias for Student-$t$ distribution. It may be because the extreme level $\varepsilon'_n$ is sufficiently small to make estimator $\widehat{\Pi}_{p,q}$ more reasonable, which in turn results in smaller MSREs of ExtraM-III. To sum up, compared to the standard extrapolative estimator BM, our proposed methods ExtraM-I, ExtraM-II and ExtraM-III via TRELT indeed demonstrate some merits in quantifying extreme risks via $L_p$-quantiles. They not only enjoy lower MSREs, but also show more robustness than BM.

\begin{table}
\centering
\caption{The MSREs of LimTRELT-I, LimTRELT-II, IntTRELT and ExtTRELT for Pareto, Fr$\acute{e}$chet and Student-$t$ distributions with $\gamma = 1/3$, $\varepsilon_n = k/n$ and $\varepsilon'_n = 0.005$. The bold numbers are the smallest values in each row.}
\label{tab:msre_c3333}
\setlength{\tabcolsep}{9pt}
\begin{tabular}{@{}ccccccc@{}}
\hline
&  & & LimTRELT-I & LimTRELT-II & IntTRELT & ExtTRELT \\
\cline{4-7}
& $\varepsilon_n$ & $(p,q)$ & \multicolumn{4}{c}{$n=2000$} \\
\hline
{Pareto}  & 0.0290 & (2.4,1.8) & 0.08621 & \textbf{0.04659} & 0.05564 & 0.14720  \\
          &        & (2.4,2.0) & 0.04212 & \textbf{0.02356} & 0.02800 & 0.06712   \\
\hline
{Fr$\acute{e}$chet}  &0.0385 & (2.4,1.8) & 0.06887 & \textbf{0.03260} & 0.07509  & 0.19160  \\
                     &       & (2.4,2.0) & 0.03247 & \textbf{0.01620} & 0.02952  & 0.07440  \\
\hline
{Student-$t$}  &0.0265 & (2.4,1.8) & 0.14574 & 0.13605 & \textbf{0.05006}  & 0.12116  \\
               &       & (2.4,2.0) & 0.07289 & 0.06461 & \textbf{0.02530}  & 0.05273  \\
\hline
&  &  & \multicolumn{4}{c}{$n=5000$} \\
\hline
{Pareto}  &0.0110 &(2.4,1.8)   & 0.05923 & \textbf{0.04620} & 0.05719  & 0.08763  \\
          &       &(2.4,2.0)   & 0.02959 & \textbf{0.02345} & 0.02841  & 0.04048   \\
\hline
{Fr$\acute{e}$chet}  &0.0160 & (2.4,1.8) & 0.05337 & \textbf{0.03591} & 0.07362  & 0.12345  \\
                     &       & (2.4,2.0) & 0.02590 & \textbf{0.01788} & 0.02886  & 0.04643  \\
\hline
{Student-$t$}  &0.0120 & (2.4,1.8) & 0.06626 & 0.06415 & \textbf{0.05096}  & 0.08526  \\
               &       & (2.4,2.0) & 0.03317 & 0.03131 & \textbf{0.02589}  & 0.03842  \\
\hline
\end{tabular}
\end{table}

\begin{table}
\centering
\caption{The MSREs of LimTRELT-I, LimTRELT-II, IntTRELT and ExtTRELT for Pareto, Fr$\acute{e}$chet and Student-$t$ distributions with $\gamma = 0.45$, $\varepsilon_n = k/n$ and $\varepsilon'_n = 0.005$. The bold numbers are the smallest values in each row.}
\label{tab:msre_c45}
\setlength{\tabcolsep}{9pt}
\begin{tabular}{@{}ccccccc@{}}
\hline
&  & & LimTRELT-I & LimTRELT-II & IntTRELT  & ExtTRELT \\
\cline{4-7}
& $\varepsilon_n$ & $(p,q)$ & \multicolumn{4}{c}{$n=2000$} \\
\hline
{Pareto}  & 0.0290 & (2.0,1.5) & 0.04157 & \textbf{0.02456} & 0.03483  & 0.12541   \\
          &        & (2.0,1.8) & 0.00787 & \textbf{0.00516} & 0.00832  & 0.02145 \\
\hline
{Fr$\acute{e}$chet}  &0.0385 & (2.0,1.5) & 0.03193 & \textbf{0.01658} & 0.05719  & 0.20204  \\
                     &       & (2.0,1.8) & 0.00559 & \textbf{0.00342} & 0.00855  & 0.02331  \\
\hline
{Student-$t$}  &0.0285 & (2.0,1.5) & 0.04222 & 0.03816 & \textbf{0.03614}  & 0.12617  \\
               &       & (2.0,1.8) & 0.00896 & \textbf{0.00757} & 0.00862  & 0.01888  \\
\hline
&  &  & \multicolumn{4}{c}{$n=5000$} \\
\hline
{Pareto}  & 0.0108 &(2.0,1.5)   & 0.02949 & \textbf{0.02531} & 0.03804  & 0.06494   \\
          &        &(2.0,1.8)   & 0.00607 & \textbf{0.00537} & 0.00861  & 0.01252   \\
\hline
{Fr$\acute{e}$chet}  & 0.0160 & (2.0,1.5) & 0.02460 & \textbf{0.01813} & 0.04939  & 0.10478  \\
                     &        & (2.0,1.8) & 0.00469 & \textbf{0.00369} & 0.00819  & 0.01375  \\
\hline
{Student-$t$}  & 0.0110 &(2.0,1.5) & 0.03075 & \textbf{0.03003} & 0.04876  & 0.09202  \\
               &        & (2.0,1.8) & 0.00628 & \textbf{0.00604} & 0.01036  & 0.01431  \\
\hline
\end{tabular}
\end{table}

\begin{table}
\centering
\caption{The MSREs of BM, ExtraM-I, ExtraM-II and ExtraM-III for Pareto, Fr$\acute{e}$chet and Student-$t$ distributions with $\gamma=1/3$, $\varepsilon_n = k/n$ and $\varepsilon'_n = 0.005$. The bold numbers are the smallest values in each row.}
\label{tab:msre_theta3333}
\setlength{\tabcolsep}{14pt}
\begin{tabular}{@{}ccccccc@{}}
\hline
&  & & BM & ExtraM-I & ExtraM-II & ExtraM-III  \\
\cline{4-7}
& $\varepsilon_n$ & $(p,q)$ & \multicolumn{4}{c}{$n=2000$} \\
\hline
{Pareto}  & 0.0290 & (2.4,1.8) & 0.05596 & 0.06796 & 0.04594 & \textbf{0.03977}  \\
          &        & (2.4,2.0) & 0.05596 & 0.06335 & \textbf{0.04295} & 0.04459   \\
\hline
{Fr$\acute{e}$chet}  &0.0385 & (2.4,1.8) & 0.11180 & 0.07991 & 0.05771 & \textbf{0.03851}  \\
                     &       & (2.4,2.0) & 0.11180 & 0.08533 & 0.06004 & \textbf{0.04956}  \\
\hline
{Student-$t$}  &0.0265 & (2.4,1.8) & 0.06039 & \textbf{0.05114} & 0.05528 & 0.09832  \\
               &       & (2.4,2.0) & 0.06039 & 0.05319 & \textbf{0.05109} & 0.08580  \\
\hline
&  & & \multicolumn{4}{c}{$n=5000$} \\
\hline
{Pareto}  &0.0110 &(2.4,1.8) & 0.02995  & 0.03341 & 0.03056 & \textbf{0.02170} \\
          &       &(2.4,2.0) & 0.02995  & 0.03177 & 0.02869 & \textbf{0.02307} \\
\hline
{Fr$\acute{e}$chet}  &0.0160 & (2.4,1.8) & 0.04911 & 0.03612 & 0.03270 & \textbf{0.01974}  \\
                     &       & (2.4,2.0) & 0.04911 & 0.03757 & 0.03262 & \textbf{0.02273}  \\
\hline
{Student-$t$}  &0.0120 & (2.4,1.8) & 0.03313 & \textbf{0.02638} & 0.02997 & 0.03596 \\
               &       & (2.4,2.0) & 0.03313 & \textbf{0.02742} & 0.02866 & 0.03301 \\
\hline
\end{tabular}
\end{table}

\begin{table}
\centering
\caption{The MSREs of BM, ExtraM-I, ExtraM-II and ExtraM-III for Pareto, Fr$\acute{e}$chet and Student-$t$ distributions with $\gamma=0.45$, $\varepsilon_n = k/n$ and $\varepsilon'_n = 0.005$. The bold numbers are the smallest values in each row.}
\label{tab:msre_theta45}
\setlength{\tabcolsep}{14pt}
\begin{tabular}{@{}ccccccc@{}}
\hline
&  & & BM & ExtraM-I & ExtraM-II & ExtraM-III  \\
\cline{4-7}
& $\varepsilon_n$ & $(p,q)$ & \multicolumn{4}{c}{$n=2000$} \\
\hline
{Pareto}  & 0.0290 & (2.0,1.5) & 0.07669 & 0.08439 & 0.07160 & \textbf{0.06319}  \\
          &        & (2.0,1.8) & 0.07669 & 0.07788 & \textbf{0.06347} & 0.06985  \\
\hline
{Fr$\acute{e}$chet}  &0.0385 & (2.0,1.5) & 0.24294 & 0.10273 & 0.09877 & \textbf{0.05934}  \\
                     &       & (2.0,1.8) & 0.24294 & 0.14290 & 0.11562 & \textbf{0.09251}  \\
\hline
{Student-$t$}  &0.0285 & (2.0,1.5) & 0.07961 & \textbf{0.06417} & 0.08161 & 0.08649  \\
               &       & (2.0,1.8) & 0.07961 & 0.06928 & \textbf{0.06721} & 0.07611  \\
\hline
&  & & \multicolumn{4}{c}{$n=5000$} \\
\hline
{Pareto}  &0.0108 &(2.0,1.5)  & 0.04262 & 0.04314 & 0.04350 & \textbf{0.03327} \\
          &       &(2.0,1.8)  & 0.04262 & 0.04148 & 0.03982 & \textbf{0.03630} \\
\hline
{Fr$\acute{e}$chet}  &0.0160 & (2.0,1.5) & 0.08966 & 0.04772 & 0.05037 & \textbf{0.03106}  \\
                     &       & (2.0,1.8) & 0.08966 & 0.05812 & 0.05373 & \textbf{0.04015}  \\
\hline
{Student-$t$}  &0.0110  & (2.0,1.5) & 0.05667 & \textbf{0.03583} & 0.04275 & 0.03921  \\
               &        & (2.0,1.8) & 0.05667 & 0.04150 & 0.04194 & \textbf{0.03789}  \\
\hline
\end{tabular}
\end{table}

\section{Real Data Analysis}\label{sec6}

In this section, we further illustrate the empirical performance of the proposed estimators through a real data analysis. The theories for TRELT and extreme $L_p$-quantile estimations are derived for independent and identically distributed samples. To reduce the potential serial dependence, we utilize the weekly historical adjusted closing prices of the S\&P500 index\footnote{The S\&P500 index data are obtained from Yahoo Finance (\url{https://finance.yahoo.com/}).} for our empirical analysis. The data spans from May 22th, 1967, to November 18th, 2024, consisting of 3001 trading records. We use weekly log-loss (negative log-return) data over the observation period with a moving window of 1800 trading days for weekly estimators of $\gamma$, $\Pi_{p,q}$ and $\theta_p(1-\varepsilon'_n)$. For instance, an estimation of $\gamma$ on November 18th, 2024, will use 1800 pieces of data from May 21st, 1990, to November 11st, 2024. This rolling method typically covers a long period and can be used to see the dynamics of the tail heaviness $\gamma$.

\begin{figure}[htbp]
\centering
\begin{subfigure}[b]{0.35\textwidth}
\includegraphics[width=\textwidth,height = 0.2\textheight]{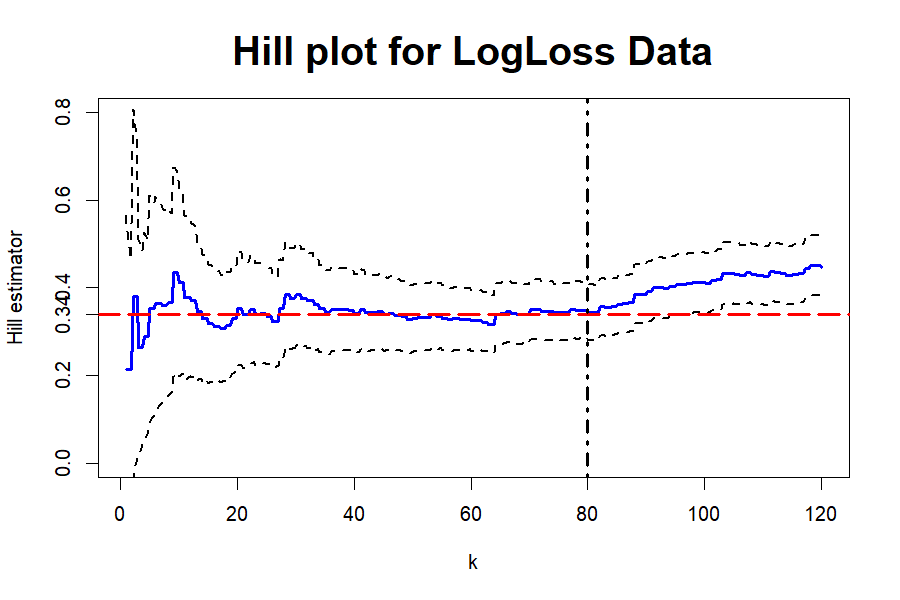}
\caption{}
\end{subfigure}
\begin{subfigure}[b]{0.35\textwidth}
\includegraphics[width=\textwidth,height = 0.2\textheight]{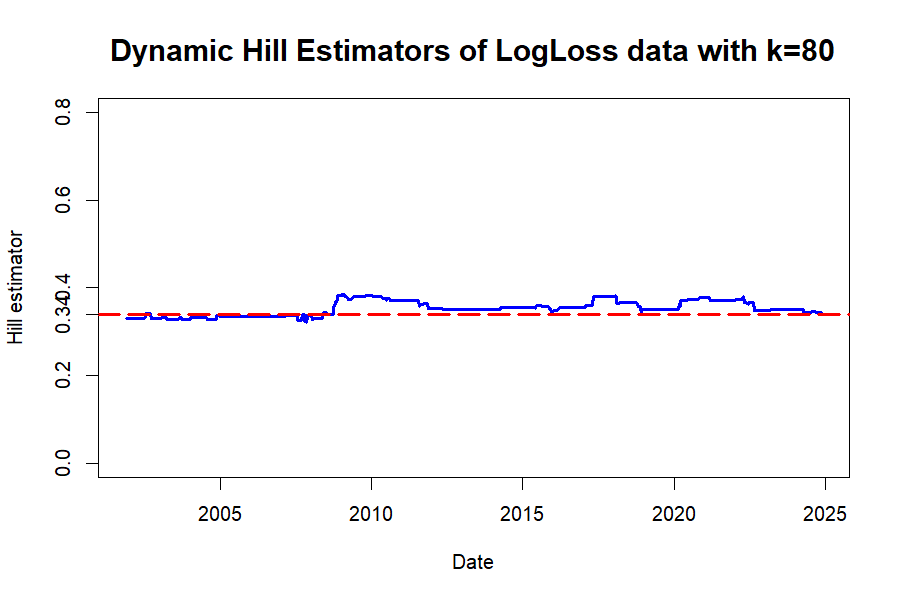}
\caption{}
\end{subfigure}
\caption{The Hill plots for selection of $k$. Plot (a) is drawn for choosing a suitable $k$, where the blue line is the Hill estimator against $k$,  the upper and lower dashed lines are the 90\% confidence bounds, and the vertical line shows the chosen $k = 80$. Plot (b) shows the dynamic Hill estimators with chosen $k$ by rolling method.}
\label{Fig:hillplots}
\end{figure}

Before proceeding with the data analysis, we need to determine the parameters $p, q$, and $\gamma$. We first draw the Hill plots for weekly log-loss data by using the recent 1800 pieces of data. As depicted in Figure \ref{Fig:hillplots}, we choose $k = 80$, which stabilize the Hill estimators around 0.34. We additionally draw the dynamic Hill plot with the chosen $k$ using rolling method. As we can see (b) in Figure \ref{Fig:hillplots}, the Hill estimators tend to stabilize around 0.34, suggesting that the choice of $k$ is appropriate and the weekly log-loss distribution is empirically heavy-tailed. Based on these observations, we now determine the values of $p,q$ under constraint $1-q < p - \frac{1}{2\gamma} < 1$ according to $\gamma=0.34$. We pick three different sets of $(p,q) = \{(2,1), (2.2,1.5), (2.4,2) \}$ and always set $\varepsilon'_n = 0.005$.

\begin{figure}[htbp]
\centering
\begin{subfigure}[b]{0.32\textwidth}
\includegraphics[width=\textwidth,height = 0.18\textheight]{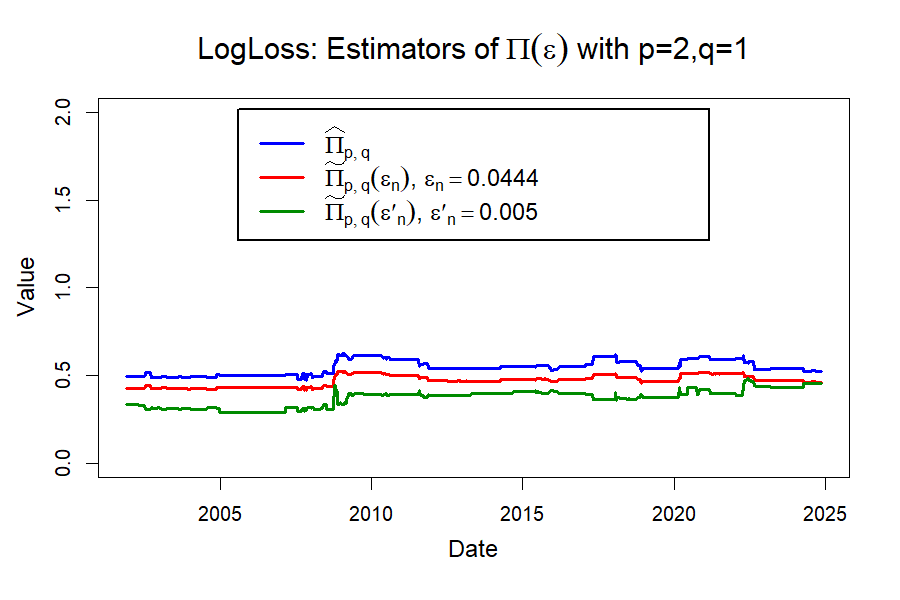}
\end{subfigure}
\begin{subfigure}[b]{0.32\textwidth}
\includegraphics[width=\textwidth,height = 0.18\textheight]{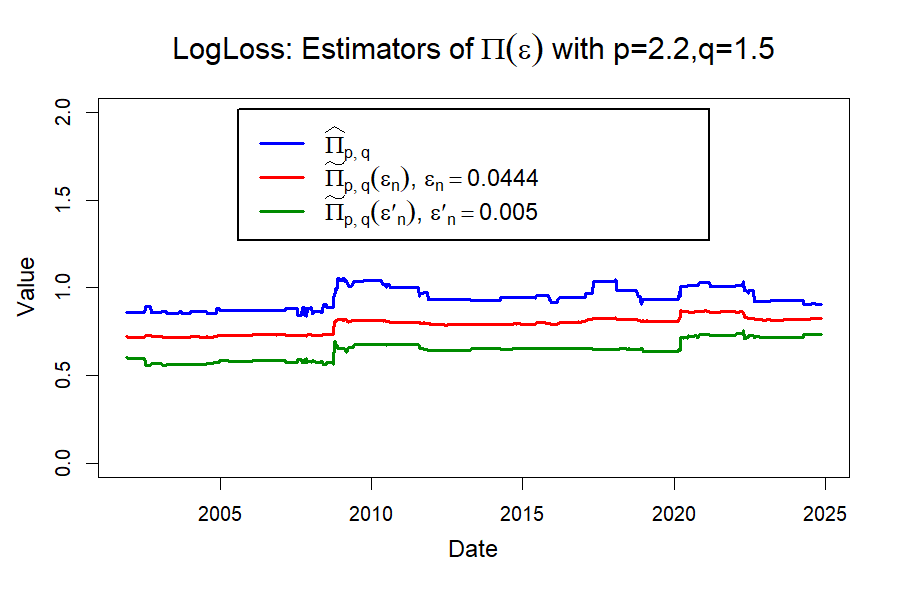}
\end{subfigure}
\begin{subfigure}[b]{0.32\textwidth}
\includegraphics[width=\textwidth,height = 0.18\textheight]{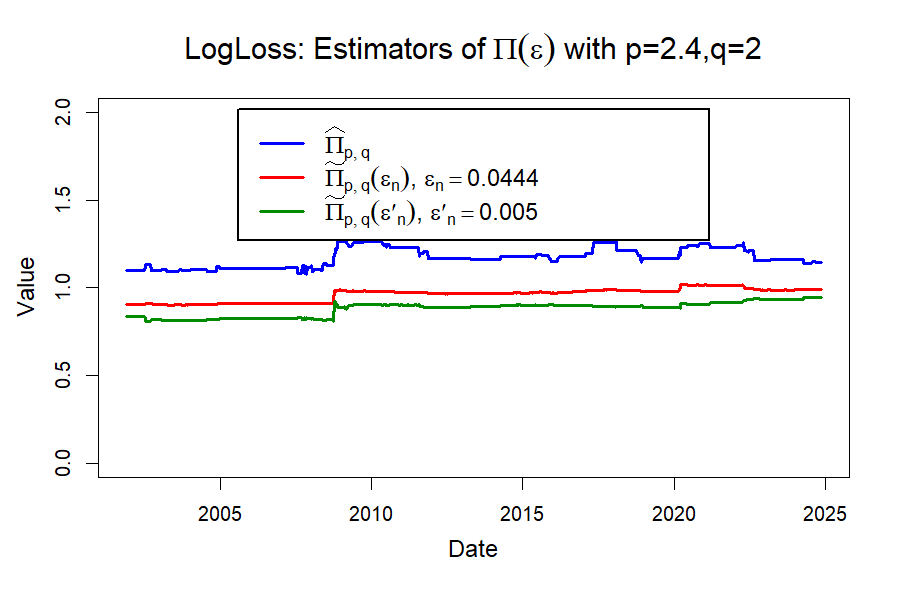}
\end{subfigure}
\\
\begin{subfigure}[b]{0.32\textwidth}
\includegraphics[width=\textwidth,height = 0.18\textheight]{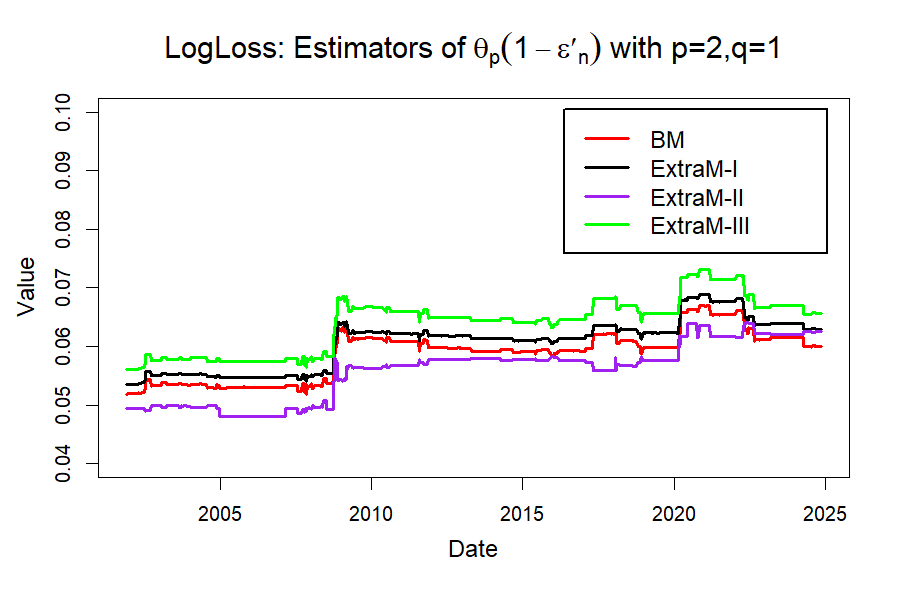}
\end{subfigure}
\begin{subfigure}[b]{0.32\textwidth}
\includegraphics[width=\textwidth,height = 0.18\textheight]{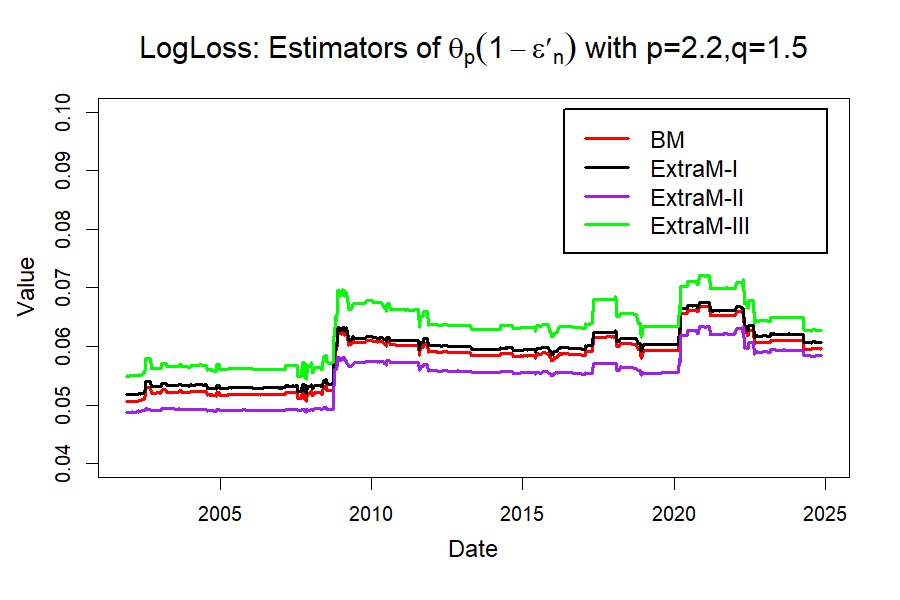}
\end{subfigure}
\begin{subfigure}[b]{0.32\textwidth}
\includegraphics[width=\textwidth,height = 0.18\textheight]{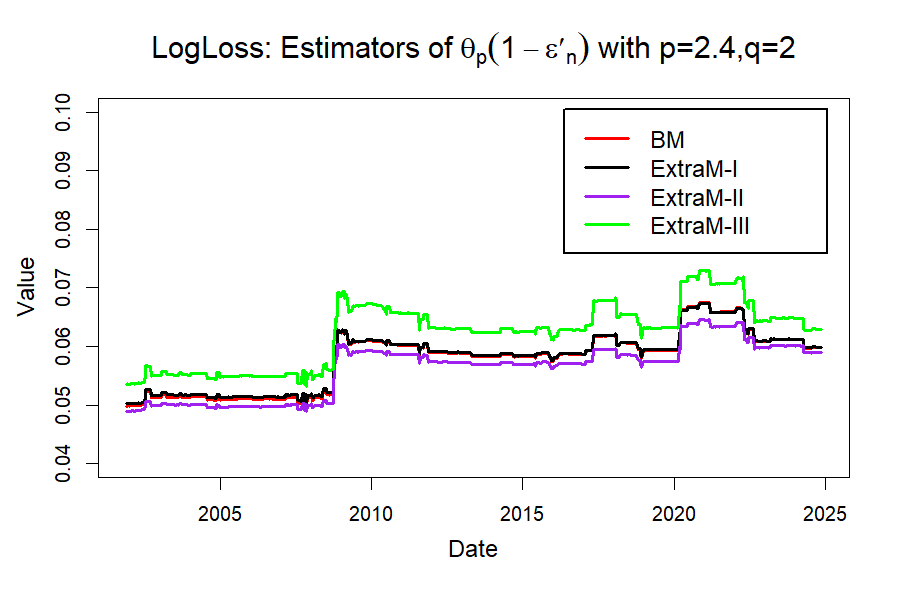}
\end{subfigure}
\caption{The dynamic estimations of $\Pi_{p,q}(\varepsilon)$ and $\theta_p(1-\varepsilon'_n)$ by rolling method with a 1800-moving window. The plots in top line correspond to the estimations of $\Pi_{p,q}(\varepsilon)$. The plots in bottom line correspond to the estimations of $\theta_p(1-\varepsilon'_n)$.}
\label{Fig:plots_c_theta}
\end{figure}

To demonstrate the performance of these methods, we further present the curves of the estimations of both $\Pi_{p,q}(\varepsilon)$ and $\theta_p(1-\varepsilon'_n)$ in Figure \ref{Fig:plots_c_theta} by the rolling method with a 1800 moving-window. We make several observations by comparing their empirical performances.

\begin{itemize}
  \item As expected, the estimator $\widehat{\Pi}_{p,q}$ and Hill estimator \eqref{eq:hill} share the same trend since the $\widehat{\Pi}_{p,q}$ is entirely dependent on the Hill estimator. Recall that $\widehat{\Pi}_{p,q}$ is established by substituting the Hill estimator in the limit $\ell(\gamma,p,q)$ directly.
  \item Overall, the lines of $\widetilde{\Pi}_{p,q}(\varepsilon_n)$ and $\widetilde{\Pi}_{p,q}(\varepsilon'_n)$ remain quite stable during the past three decades. Additionally, it is noteworthy that two significant changes, occurring around 2009 and 2020, are also well-reflected. This observation aligns with the widely held belief that both the financial crisis (2009) and the COVID-19 pandemic (2020) have had substantial impacts on financial markets.
  \item As a risk measure, the empirical values of $\theta_p(1-\varepsilon'_n)$ display the time-varying volatility observed over the past three decades. It is evident that the trends of the four methods—BM, ExtraM-I, ExtraM-II, and ExtraM-III—are nearly identical. The two significant fluctuations, occurring around 2009 and 2020, are also clearly depicted in Figure \ref{Fig:plots_c_theta}.
  \item The ranking of size for these four methods is as follows: ExtraM-III $>$ ExtraM-I $\approx$ BM $>$ ExtraM-II. This outcome is highly consistent with the simulation results presented in Section \ref{sec5}. Due to lower risk preference, the regulators may opt for the ExtraM-II method as the most preferred choice, followed by either ExtraM-I or BM as the second-best option for practical application of $L_p$-quantiles in quantifying extreme risks. This preference is justified because a smaller value indicates a lower capital requirement.
\end{itemize}

In summary, based on the substantial simulations and empirical studies, we can assert that TRELT serves as a tail-based measure of variability. Its values mirror the stability of the financial market and are capable of effectively identifying abnormal fluctuations. Moreover, these empirical studies also provide compelling evidence that our methods ExtraM-I, ExtraM-II and ExtraM-III are more efficient for predicting extreme risks via $L_p$-quantiles.

\section{Conclusion}

In this paper, we propose the concept of tail risk equivalent level transition (TRELT) between $L_p$-quantiles, which is motivated by the PELVE in \cite{Li2023}. The TRELT (and its dual) is developed under the extreme value theory to bridge different risk measures given tail equivalence of risks, which is novel in tail risk measurement. We study the theoretical properties, such as existence, uniqueness, and limiting properties, of the coefficient of TRELT, and further propose the estimation approach for it. In addition, to predict the extreme $L_p$-quantiles, we propose new extrapolative estimators based on the TRELT approach. Simulation studies show that our proposed estimators are effective for predicting extreme risks. As for further studies, it is of theoretical interest to study the TRELT between more general risk measures, as well as of practical interest to propose real applications of TRELT in tail risk measurement.

\appendix

\section{Auxiliary results}\label{sec:par}

\begin{lemma}\label{lem:finite_mom}
Let $X$ be a random variable with distribution function $F$ satisfying Assumption \ref{ass:forv} for some $\gamma > 0$, then for all $-\infty < x < U(\infty)$($U(\infty)$ is the right endpoint of $F$),
\begin{equation*}
\mathbb{E}[|X|^{\iota} \one_{\{X>x\}}]<\infty,
\end{equation*}
if $0<\iota < \frac{1}{\gamma}$ and $\mathbb{E}[|X|^{\iota} \one_{\{X>x\}}]=\infty$ if $\iota > \frac{1}{\gamma}$.
\end{lemma}

\begin{proposition}\label{pro:mp_lpquant}
Let $X$ be a random variable with continuous distribution function $F$, $x_* = \inf_{x \in \mathbb{R}}\{ ~x~ \big| ~F(x)>0 \}$ and $x^* = \inf_{x \in \mathbb{R}}\{ ~x~ \big| ~F(x) \ge 1 \}$ denote the left and right endpoints of $X$ respectively. For $p>1$, we have the following statements.
\begin{enumerate}
  \item \label{mp_lpquant1} (Existence and Uniqueness) For all $\tau \in (0,1)$, \eqref{eq:lp-quantile2} has a unique solution $\theta_{p}(\tau) \in (x_* , x^*)$;
  \item \label{mp_lpquant2} (Monotonicity in level) The mapping $\tau \in (0,1) \mapsto \theta_{p}(\tau) \in \mathbb{R}$ is strictly increasing, and
  $$
  \lim_{\tau \uparrow 1} \theta_{p}(\tau) = x^*, ~ \lim_{\tau \downarrow 0} \theta_{p}(\tau) = x_*;
  $$
  \item \label{mp_lpquant3} (Continuity) The mapping $\tau \in (0,1) \mapsto \theta_{p}(\tau) \in \mathbb{R}$ is continuous.
\end{enumerate}
\end{proposition}

\begin{proposition}\label{pro:daouiath1}
  Under the conditions of Theorem \ref{th:asynorm_pi_inter}, for all $q > 1$, we have,
  \begin{equation}\label{eq:th1daouia_tep}
  \begin{split}
  \sqrt{n \varepsilon_n} \left( \frac{\hat{\theta}_q(1-\varepsilon_n)}{\theta_q(1-\varepsilon_n)} -1 \right) \xrightarrow{d} & \mathcal{N}\left( \frac{\lambda(q-1)}{\gamma^{\rho-1}\rho}\frac{B(q-1,-(\rho-1)/\gamma-q+1)}{[B(q,\gamma^{-1}-q+1)]^{1-\rho}} - \frac{\lambda}{\rho}, \right. \\
  & ~~ \left. \frac{\gamma}{4} \frac{[B(q,(2\gamma)^{-1}-q+1)]^2}{B(q,\gamma^{-1}-q+1)} \right).
  \end{split}
  \end{equation}
\end{proposition}

\begin{proposition}\label{pro:2order_expansion}
Suppose $F$ is strictly increasing and satisfies both Assumption \ref{ass:sorv} for some $\gamma > 0$ and Assumption \ref{ass:mom_negapart} with an order $p \in (1, 1+1/\gamma)$. Then, for all $q \in [1,p)$, we have that,
\begin{equation}\label{eq:2order_expansion}
  \frac{\theta_p(\tau)}{\theta_q(\tau)} = \mathcal{L}(\gamma, p ,q) \left( 1- \gamma R(p,q,\gamma,\tau) + A\left( \frac{1}{\overline{F}(\theta_{q}(\tau))} \right) \left\{ \frac{1}{\rho} \left[ \left[ \frac{B(p,\gamma^{-1}-p+1)}{B(q,\gamma^{-1}-q+1)} \right]^{\rho} -1 \right] +o(1) \right\} \right),
\end{equation}
as $\tau \uparrow 1$, where
\begin{align*}
R(p,q,\gamma,\tau) = & - \frac{\gamma}{B(p,\gamma^{-1}-p+1)} A\left( \frac{1}{\overline{F}(\theta_q(\tau))} \right) K(p, q, \gamma, \rho)(1+o(1))  \\
  - & (p-1)r(\mathcal{L}(\gamma,p,q)\theta_{q}(\tau), p, \gamma, X), \\
K(p, q, \gamma, \rho) = &
\begin{cases}
\begin{aligned}
& \left[ \frac{B(p,\gamma^{-1}-p+1)}{B(q,\gamma^{-1}-q+1)} \right]^{\rho}\frac{1}{\gamma^2 \rho}[(1-\rho)B(p,(1-\rho)/\gamma-p+1) \\
& - B(p,\gamma^{-1}-p+1)], & \mbox{if } \rho < 0, \\
& \frac{p-1}{\gamma^2} \int_{1}^{+\infty} (x-1)^{p-2} x^{-\frac{1}{\gamma}} \log x \,dx , & \mbox{if } \rho = 0,
\end{aligned}
\end{cases}
\\
r(\theta_{q}(\tau), p, \gamma, X) = &
\begin{cases}
\frac{\mathbb{E}\left[X \one_{\left\{0< X < \theta_{q}(\tau)\right\}}\right]}{\theta_{q}(\tau)}(1+o(1)), & \mbox{if } \gamma \leq 1, \\
\overline{F}(\theta_{q}(\tau))B(p-1, 1-\gamma^{-1})(1+o(1)) , & \mbox{if } \gamma > 1.
\end{cases}
\end{align*}
In particular, if $q=1$, then \eqref{eq:2order_expansion} becomes, as $\tau \uparrow 1$,
\begin{equation}\label{eq:2order_expansion_p1}
\frac{\theta_p(\tau)}{\theta_1(\tau)} = \mathcal{L}(\gamma, p ,1) \left( 1- \gamma R(p,1,\gamma,\tau) + A\left( \frac{1}{1-\tau} \right) \left\{ \frac{1}{\rho} \left[ \left[\frac{B(p,\gamma^{-1}-p+1)}{\gamma} \right]^{\rho} -1 \right] +o(1) \right\} \right).
\end{equation}
\end{proposition}

\end{document}